\documentclass[twocolumn,times]{aastex631}

\usepackage{float}
\usepackage{moresize}

\newcommand{\sextractor}{\texttt{SExtractor}}
\newcommand{\astropy}{\texttt{AstroPy}}

\newcommand{\eazy}{\texttt{EAZY}}
\newcommand{\cigale}{\texttt{CIGALE}}
\newcommand{\trilogy}{\texttt{Trilogy}}
\newcommand{\galfit}{\texttt{GalFit}}
\newcommand{\Gaia}      {Gaia}%

{\relax}%
\renewcommand{\added}[1]{#1}%
{\relax}%
\newlength{\txw}
\newlength{\txh}

\definecolor{blue}    {rgb}{0.0,0.7,1.0}

\newcommand{\DELETED}[1]{\relax}%

\newcommand\degree       {{\ifmmode^\circ\else$^\circ$\fi}}
\newcommand\arcm         {{\ifmmode {'\ }\else$'     $\fi}}
\newcommand\arcs         {{\ifmmode{''\ }\else$''    $\fi}}
\newcommand\cge          {{$\gtrsim$}}

\newcommand\eg           {{e.g.},}
\newcommand\ie           {{i.e.},}

\newcommand{\HST}        {\emph{HST}}
\newcommand{\JWST}       {\emph{JWST}}

\newcommand\mAB          {{$m_{\rm AB}$}}

\newcommand\mum          {{\micron}}

\newcommand\VLA          {\emph{VLA}}

\newcommand{\nd}{\nodata}

\defcitealias{Driver2016b}{D16}
\defcitealias{Windhorst2011}{W11}
\defcitealias{Carleton2022}{SKYSURF-2}
\defcitealias{Windhorst2022}{SKYSURF-1}

\def\ltsima{$\; \buildrel < \over \sim \;$}
\def\lsim{\lower.5ex\hbox{\ltsima}}
\def\gtsima{$\; \buildrel > \over \sim \;$}
\def\gsim{\lower.5ex\hbox{\gtsima}}

\shorttitle{Central Point-Source Galaxies Within the JWST NEP--TDF}
\shortauthors{Ortiz et al.}

\begin{document}
\title{PEARLS: Discovery of Point-Source Features Within Galaxies in the North Ecliptic Pole Time Domain Field}

\author[0000-0002-6150-833X]{Rafael Ortiz III} 
\affiliation{School of Earth and Space Exploration, Arizona State University,
Tempe, AZ 85287-1404, USA}

\author[0000-0001-8156-6281]{Rogier A. Windhorst} 
\affiliation{School of Earth and Space Exploration, Arizona State University,
Tempe, AZ 85287-1404, USA}

\author[0000-0003-3329-1337]{Seth H. Cohen} 
\affiliation{School of Earth and Space Exploration, Arizona State University,
Tempe, AZ 85287-1404, USA}

\author[0000-0002-9895-5758]{Steven P. Willner} 
\affiliation{Center for Astrophysics \textbar\ Harvard \& Smithsonian, 60 Garden St., Cambridge, MA 02138 USA
}

\author[0000-0003-1268-5230]{Rolf A. Jansen} 
\affiliation{School of Earth and Space Exploration, Arizona State University,
Tempe, AZ 85287-1404, USA}

\author[0000-0001-6650-2853]{Timothy Carleton} 
\affiliation{School of Earth and Space Exploration, Arizona State University,
Tempe, AZ 85287-1404, USA}

\author[0000-0001-9394-6732]{Patrick S. Kamieneski}
\affiliation{School of Earth and Space Exploration, Arizona State University,
Tempe, AZ 85287-1404, USA}

\author[0000-0001-7016-5220]{Michael J. Rutkowski}
\affiliation{Department of Physics and Astronomy, Minnesota State University, Mankato, MN 56001, USA}

\author[0000-0002-0648-1699]{Brent M. Smith}
\affiliation{School of Earth and Space Exploration, Arizona State University,
Tempe, AZ 85287-1404, USA}

\author[0000-0002-7265-7920]{Jake Summers} 
\affiliation{School of Earth and Space Exploration, Arizona State University,
Tempe, AZ 85287-1404, USA}


\author[0000-0003-0202-0534]{Cheng Cheng}
\affiliation{Chinese Academy of Sciences South America Center for
Astronomy, National Astronomical Observatories, CAS, Beijing 100101,
China}

\author[0000-0001-7410-7669]{Dan Coe} 
\affiliation{AURA for the European Space Agency (ESA), Space Telescope Science
Institute, 3700 San Martin Drive, Baltimore, MD 21210, USA}

\author[0000-0003-1949-7638]{Christopher J. Conselice} 
\affiliation{Jodrell Bank Centre for Astrophysics, Alan Turing Building, 
University of Manchester, Oxford Road, Manchester M13 9PL, UK}

\author[0000-0001-9065-3926]{Jose M. Diego} 
\affiliation{Instituto de F\'isica de Cantabria (CSIC-UC). Avda. Los Castros s/n. 39005 Santander, Spain}

\author[0000-0001-9491-7327]{Simon P. Driver} 
\affiliation{International Centre for Radio Astronomy Research (ICRAR) and the
International Space Centre (ISC), The University of Western Australia, M468,
35 Stirling Highway, Crawley, WA 6009, Australia}

\author[0000-0002-9816-1931]{Jordan C. J. D'Silva} 
\affiliation{International Centre for Radio Astronomy Research (ICRAR) and the
International Space Centre (ISC), The University of Western Australia, M468,
35 Stirling Highway, Crawley, WA 6009, Australia}
\affiliation{ARC Centre of Excellence for All Sky Astrophysics in 3 Dimensions
(ASTRO 3D), Australia}

\author[0000-0003-1625-8009]{Brenda L. Frye} 
\affiliation{Steward Observatory, University of Arizona, 933 N Cherry Ave,
Tucson, AZ, 85721-0009, USA}

\author[0000-0003-1436-7658]{Hansung B. Gim} 
\affiliation{Department of Physics, Montana State University, Bozeman, MT 59717, US}

\author[0000-0001-9440-8872]{Norman A. Grogin} 
\affiliation{Space Telescope Science Institute, 3700 San Martin Drive, 
Baltimore, MD 21210, USA}

\author[0000-0001-8751-3463]{Heidi B. Hammel}
\affiliation{Association of Universities for Research in Astronomy, Washington, D.C. 20004, USA}

\author[0000-0001-6145-5090]{Nimish P. Hathi} 
\affiliation{Space Telescope Science Institute, 3700 San Martin Drive, Baltimore, MD 21218, USA.}

\author[0000-0002-4884-6756]{Benne W. Holwerda} 
\affiliation{Department of Physics, University of Louisville, Natural Science Building 102, Louisville KY 40292, USA}

\author[0000-0003-4738-4251]{Minhee Hyun}
\affiliation{Korea Astronomy and Space Science Institute, Yuseong-gu, Daejeon 34055, Republic of Korea}

\author[0000-0002-8537-6714]{Myungshin Im}
\affiliation{SNU Astronomy Research Center, Dept. of Physics \& Astronomy, Seoul 08826, Republic of Korea}

\author[0000-0002-6131-9539]{William C. Keel} 
\affiliation{Department of Physics and Astronomy, University of Alabama, Box 870324, Tuscaloosa, AL\,35404, USA}

\author[0000-0002-6610-2048]{Anton M. Koekemoer} 
\affiliation{Space Telescope Science Institute, 3700 San Martin Drive,
Baltimore, MD 21210, USA}

\author[0000-0002-8184-5229]{Juno Li}
\affiliation{International Centre for Radio Astronomy Research (ICRAR) and the
International Space Centre (ISC), The University of Western Australia, M468,
35 Stirling Highway, Crawley, WA 6009, Australia}
\affiliation{ARC Centre of Excellence for All Sky Astrophysics in 3 Dimensions
(ASTRO 3D), Australia}

\author[0000-0001-6434-7845]{Madeline A. Marshall} 
\affiliation{National Research Council of Canada, Herzberg Astronomy \&
Astrophysics Research Centre, 5071 West Saanich Road, Victoria, BC V9E 2E7, 
Canada}
\affiliation{ARC Centre of Excellence for All Sky Astrophysics in 3 Dimensions
(ASTRO 3D), Australia}

\author[0000-0002-5506-3880]{Tyler J. McCabe} 
\affiliation{School of Earth and Space Exploration, Arizona State University,
Tempe, AZ 85287-1404, USA}

\author[0009-0000-5821-4325]{Noah J. McLeod} 
\affiliation{School of Earth and Space Exploration, Arizona State University,
Tempe, AZ 85287-1404, USA}

\author[0000-0001-7694-4129]{Stefanie N. Milam}
\affiliation{Astrochemistry Laboratory, NASA Goddard Space Flight Center, Greenbelt, MD 20771, USA}

\author[0000-0003-3351-0878]{Rosalia O'Brien} 
\affiliation{School of Earth and Space Exploration, Arizona State University,
Tempe, AZ 85287-1404, USA}

\author[0000-0003-3382-5941]{Nor Pirzkal} 
\affiliation{Space Telescope Science Institute, 3700 San Martin Drive,
Baltimore, MD 21210, USA}

\author[0000-0003-0429-3579]{Aaron S. G. Robotham} 
\affiliation{International Centre for Radio Astronomy Research (ICRAR) and the
International Space Centre (ISC), The University of Western Australia, M468,
35 Stirling Highway, Crawley, WA 6009, Australia}

\author[0000-0003-0894-1588]{Russell E. Ryan, Jr.} 
\affiliation{Space Telescope Science Institute, 3700 San Martin Drive, 
Baltimore, MD 21210, USA}

\author[0000-0001-9262-9997]{Christopher N. A. Willmer} 
\affiliation{Steward Observatory, University of Arizona, 933 N Cherry Ave,
Tucson, AZ, 85721-0009, USA}

\author[0000-0001-7592-7714]{Haojing Yan} 
\affiliation{Department of Physics and Astronomy, University of Missouri,
Columbia, MO 65211, USA}

\author[0000-0001-7095-7543]{Min S. Yun} 
\affiliation{Department of Astronomy, University of Massachusetts, Amherst, MA 01003, USA}

\author[0000-0002-0350-4488]{Adi Zitrin}
\affiliation{Department of Physics, Ben-Gurion University of the Negev, P.O. Box 653, Be'er-Sheva 84105, Israel}


\begin{abstract}
The first public 0.9--4.4~$\mu$m NIRCam images of the North Ecliptic Pole (NEP) Time Domain Field (TDF) uncovered galaxies displaying point-source features in their cores as seen in the longer wavelength filters. We visually identified a sample of 66 galaxies ($\sim$1 galaxy per arcmin$^2$) with point-like cores and \added{have modeled their two-dimensional light profiles with \galfit{}, identifying 16 galactic nuclei with measurable point-source components.} \added{\galfit{} suggests the visual sample is a mix of both compact stellar bulge and point-source galaxy cores.  This core classification is complemented by}  spectral energy distribution (SED) modeling to infer the sample's active galactic nucleus (AGN) and host-galaxy parameters. 
For galaxies with measurable point-source components, the median fractional AGN contribution to their 0.1--30.0~$\mu$m flux is 0.44, and 14/16 are color-classified AGN\null.
We conclude that near-infrared point-source  galaxy cores are signatures of AGN\null. In addition, we define an automated sample-selection criterion to identify these point-source features.
These criteria can be used in other extant and future NIRCam images to streamline the search for galaxies with unresolved IR-luminous AGN\null.
The James Webb Space Telescope's superb angular resolution and sensitivity at infrared wavelengths is resurrecting the morphological identification of AGN.

\end{abstract}

\correspondingauthor{Rafael Ortiz III}
\email{rortizii@asu.edu}

\section{Introduction}

More than eighty years ago, Carl \citet[]{Seyfert1943} drew attention to 6 ``extragalactic nebulae'' with broad emission lines and ``exceedingly luminous stellar or semistellar'' nuclei.  Galaxies of this type became known as ``Seyfert galaxies'' \citep{Burbidge1963}, and they are now recognized \citep[\eg][]{Osterbrock1993} as members of the low-luminosity end of the population of active galactic nuclei (AGN).

AGN are now thought to consist of an accretion disk around a supermassive black hole (SBMH) within a wider, optically thick dust torus \citep[\eg][]{Antonucci1985}. This central engine can emit enormous luminosities and can even outshine the entire host galaxy \citep[\eg][]{Padovani2017}.  Modern studies of AGN  probe the relationship with their host galaxies and intergalactic environments along with properties of the central SMBHs \citep[\eg][]{bollati2023connection, costasouza2023spatially, Sampaio2023}.  

AGN emit energy across the entire electromagnetic spectrum, and identification and classifications of AGN are based on X-ray \citep[\eg][]{Elvis1978}, UV--visible \citep[\eg][]{Seyfert1943}, infrared \citep[\eg][]{Stern2005}, and radio \citep[\eg][]{Fanaroff1974} observations.
AGN are most often identified via their spectral energy distributions (SED)s  \citep[\eg][]{li2023epochs, Lyu2023, CeersYang2023}, their colors \citep[\eg][]{Stern2005,Hwang2021, Joudz2023, FurtakColors2023}, spectroscopy \citep[\eg][]{Mehdipour2024, Burke2024}, variability \citep[\eg][]{Pouliasis2019, OBrien2024}, radio \citep[\eg][]{Hyun2023}, or X-ray emission \citep[\eg][]{Masini2020,Zhao2021}. \added{However, heavily obscured  AGN whose intrinsic UV--visible signatures are hidden by dust may be missed \citep[\eg][]{Glikman2012,Hickox2018}}, and at infrared wavelengths, where the dust extinction is lower, wide-field surveys can only be done from space. Until now, the low angular resolution of infrared space observatories and \added{the ``big data'' era of advanced instrumentation and vast surveys (such as the Sloan Digital Sky Survey \citep[][]{York2000}}) has caused identification of AGN through visual morphologies to have fallen out of favor.

\added{Recent work indicates that a significant population of redder AGN exists. Their UV--visible SEDs suggest intermediate levels of extinction, $A_{\rm V}\sim~1$--3~mag \citep[\eg][]{Wilkes2002, Richards2003, Trump2013, Cales2015, Wang2019}. These objects exhibit hybrid properties, blurring the lines between canonical AGN classifications, and may represent an important evolutionary phase. For instance, modeling suggests the AGN torus geometry evolves as material accretes onto the black hole \citep[\eg][]{Ohsuga2001, Mandal2018}, with early dust-free to intermediate-stage reddened phases necessarily preceding the well-studied mature dust-obscured AGN\null. Disentangling the nature of these moderately-reddened AGN exhibiting a blend of obscured and unobscured traits has proved challenging but vitally important for refining our understanding of the complex coevolution between SMBHs and their host galaxies over cosmic time. Detailed multi-wavelength modeling is critical to robustly characterize the properties of these objects.}

The James Webb Space Telescope (JWST), with its   unprecedented resolution in the infrared \citep[\eg][]{Davies2024, li2023epochs, CeersYang2023}, offers the chance to revisit morphological selection and analysis of AGN. 
This paper is an initial exploration of the possibilities via the characteristic diffraction spikes of \JWST's point-spread function (PSF) at the longer near-infrared wavelengths that suggest pointlike galaxy cores.

The North Ecliptic Pole (NEP) Time Domain Field \citep[TDF;][]{Jansen_2018} is located within \JWST’s continuous viewing zone, making it a prime target for AGN variability and other time-domain science. Because the ultimate goal will be to compare morphological selection with other methods, we used the now-public \JWST\ NEP--TDF images \citep{Windhorst2023} to visually search for galaxies with point-like features in their center, \ie\, potential AGN, following the core morphology and Seyfert class relationship \citep[\eg][]{Cohen2010, Rutkowski2013}. The morphologically selected sample was then analyzed in order to infer physical parameters and AGN presence.

Section~\ref{s:obs} describes the observations and data, along with the construction of a morphologically selected AGN catalog and a means to automate the selection of galaxies with point-source nuclei. Section~\ref{sec:characterization} discusses the results of fitting 2D-light profiles and various galaxy SEDs to the sample, and Section~\ref{sec:discussion} gives a summary.
All magnitudes are in AB units 
\citep{Oke_1983}. Where relevant, we adopt a flat $\Lambda$CDM cosmology with $H_0 = 68$ km s$^{-1}$ Mpc$^{-1}$, $\Omega_M = 0.32$, and $\Omega_{\Lambda} = 0.68$ \citep{PlanckCollaboration2016, PlanckCollaboration2018}.

\section{Data \& Cataloging}
\label{s:obs}
\subsection{Observations}
\label{sec:dataprocessing}

The \JWST\ NEP--TDF is centered at (RA, Decl.)$_{\rm J2000}$ = (17:22:47.896, +65:49:21.54) \citep[][]{Jansen_2018}. The TDF was observed with \JWST\ as part of the Prime Extragalactic Areas for Reionization and Lensing Science (PEARLS) GTO program \citep{Windhorst2023} in eight NIRCam filters: F090W, F115W, F150W, F200W, F277W, F356W, F410M, and F444W\null. The  observations consisted of four orthogonal spokes (see Figure~1 of \citealt{OBrien2024}) observed between 2022 Aug 25 and 2023 May 30. The NIRCam image quality is diffraction-limited  at wavelengths \cge 1.3 \mum\ \citep{rigby2023, Windhorst2023} with point source FWHM values ranging from 60--160 milliarcseconds (mas) at wavelengths of 1.3--4.8$~\mu$m.

\added{The data were retrieved from MAST and post-processed by the PEARLS team using their custom pipeline to mitigate $1/f$-noise, identify and subtract wisps in the NIRCam/SW filters, mask snowball artifacts, and flatten the background across read-out amplifier boundaries. The individual post-processed images were rectified and aligned to \Gaia/DR3. {Full mosaics of the field were created for each filter with a 0\farcs030 pixel scale. The 5$\sigma$ point-source limit is typically between 28.0 and 29.1 mag, depending on the filter, and 29.0 mag in F200W specifically.} \citet{Windhorst2023}, \citet{Robotham_2023}, and Jansen et~al.\ 2024a (in prep.) give more details of the data reduction, calibration, and post-processing.}

\added{While the primary focus of this work is the ability to identify and analyze likely AGN with \JWST/NIRCam photometry, we incorporate ancillary \HST\ observations of the NEP--TDF with the Wide Field Camera~3 (WFC3/UVIS) in the F275W filter ($\lambda_c\simeq0.272$~\micron) and with the Advanced Camera for Surveys (ACS/WFC) in the F435W and F606W filters ($\lambda_c\simeq 0.433$ and 0.592~\micron, respectively) with a total area of $\simeq$194 arcmin$^2$ (ACS/WFC). Observations from \HST\ GO\,15278 were taken between 2017 October 1 and 2019 February 9 and those from GO\,16252+16793 (TREASUREHUNT) between 2020 September 25 and 2022 October 31. Both programs used 4-orbit CVZ visits to reach 2$\sigma$ limiting depths of \mAB\ $\simeq$ 28.0, 28.6, and 29.5 mag in F275W, F435W and F606W, respectively. \citet[][\S2.1]{OBrien2024} and Jansen et~al. (2024b; in prep.)\ give further details of the \HST\ observations.}

For an initial comparison with longer-wavelength observations in the NEP--TDF, we used the VLA 3~GHz radio observations detailed by \citet[their Appendix~A]{Hyun2023}.

\subsection{Catalogs \& Samples}
We used \sextractor{} \citep{sextractor} to generate  catalogs. The detection threshold required 9 contiguous pixels  1.5$\sigma$ above the background. We used 32 deblending sub-thresholds with a 0.06 minimum contrast necessary for  object deblending. We ran \sextractor{} on all eight filters in dual-image mode using F444W as the detection filter to produce position-matched catalogs of the entire \JWST\ NEP--TDF\null. Magnitudes are \texttt{MAG\_AUTO} except where indicated otherwise.

\added{We position-matched objects in both right ascension and declination between \HST\ \citep[]{OBrien2024} and \VLA\ \citep[]{Hyun2023} catalogs using 0\farcs1 and 0\farcs5 separations, respectively, to identify counterparts to the samples used in our analysis.}

\added{We used both the \eazy{}\footnote{\href{https://github.com/gbrammer/eazy-photoz}{https://github.com/gbrammer/eazy-photoz}} \citep[]{eaZy} and \cigale{}\footnote{\href{https://cigale.lam.fr/}{https://cigale.lam.fr/}} \citep{cigale} SED-fitting codes to compute photometric redshifts for our samples, as discussed in Section~\ref{sec:characterization}. 
We used the \cigale{}-computed photometric redshifts for nominal redshifts in this work and for sample construction because the \cigale{} fitting incorporates multiple components, including AGN, whereas \eazy{} was tuned for single-template fitting.}

\subsection{Visual Sample Selection}
\label{sec:sampleselection}

\begin{figure*}[t]
     \centering
\includegraphics[width=0.11\textwidth]{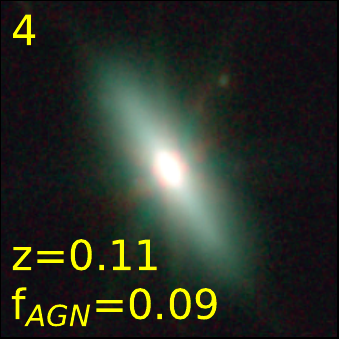}
\includegraphics[width=0.11\textwidth]{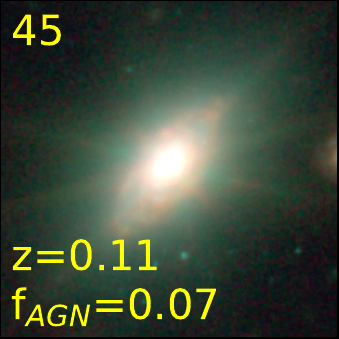}
\includegraphics[width=0.11\textwidth]{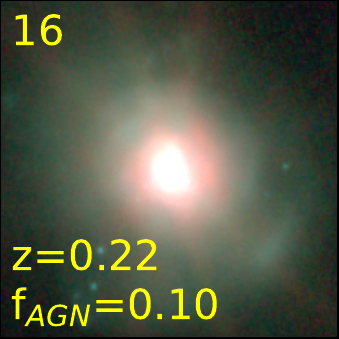}
\includegraphics[width=0.11\textwidth]{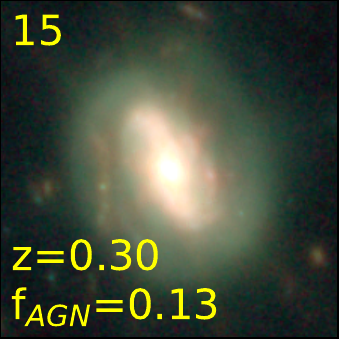}
\includegraphics[width=0.11\textwidth]{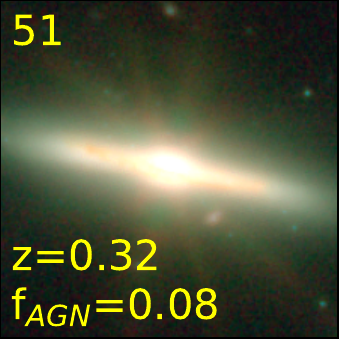}
\includegraphics[width=0.11\textwidth]{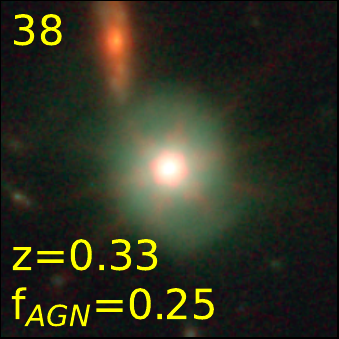} \\
\includegraphics[width=0.11\textwidth]{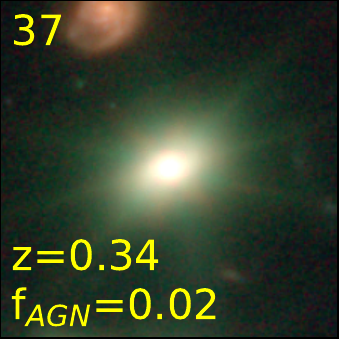}
\includegraphics[width=0.11\textwidth]{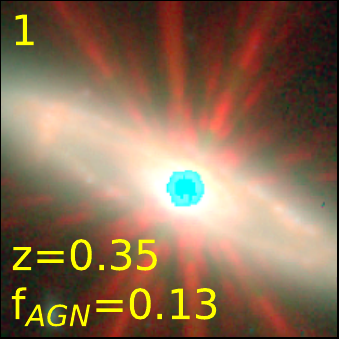}
\includegraphics[width=0.11\textwidth]{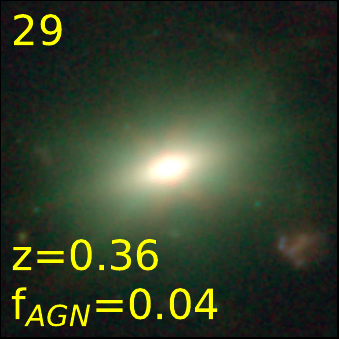}
\includegraphics[width=0.11\textwidth]{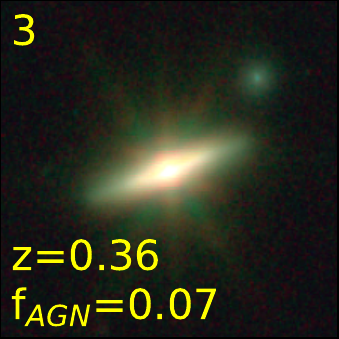}
\includegraphics[width=0.11\textwidth]{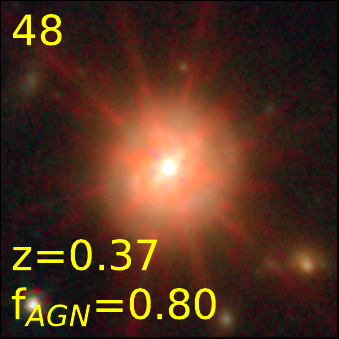}
\includegraphics[width=0.11\textwidth]{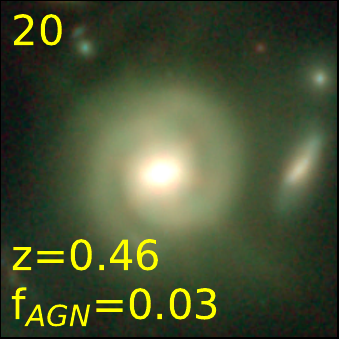} \\
\includegraphics[width=0.11\textwidth]{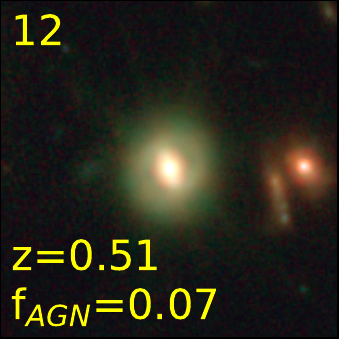}
\includegraphics[width=0.11\textwidth]{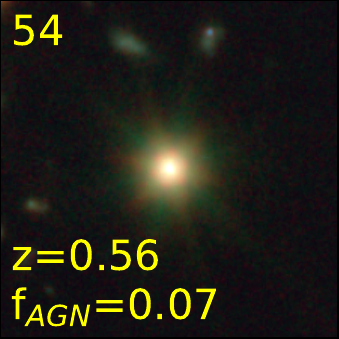}
\includegraphics[width=0.11\textwidth]{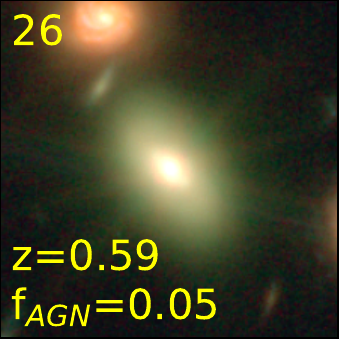}
\includegraphics[width=0.11\textwidth]{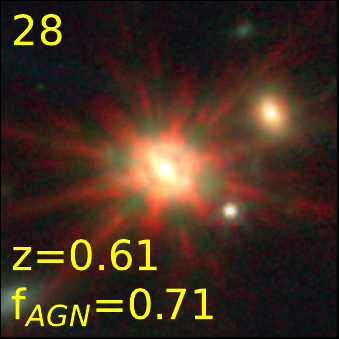}
\includegraphics[width=0.11\textwidth]{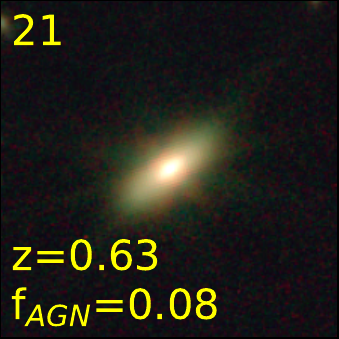}
\includegraphics[width=0.11\textwidth]{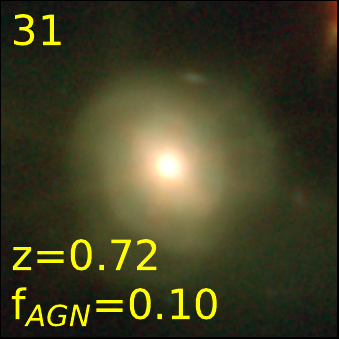} \\
\includegraphics[width=0.11\textwidth]{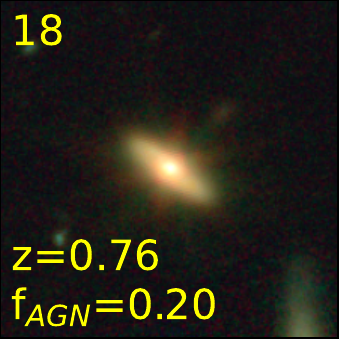}
\includegraphics[width=0.11\textwidth]{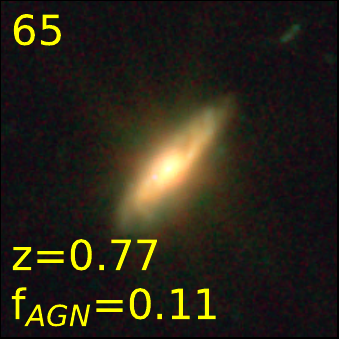}
\includegraphics[width=0.11\textwidth]{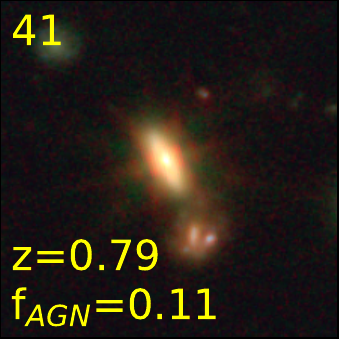}
\includegraphics[width=0.11\textwidth]{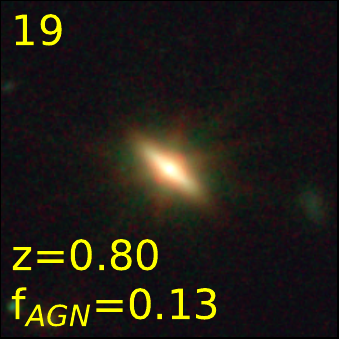}
\includegraphics[width=0.11\textwidth]{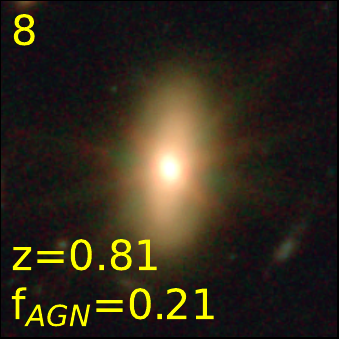}
\includegraphics[width=0.11\textwidth]{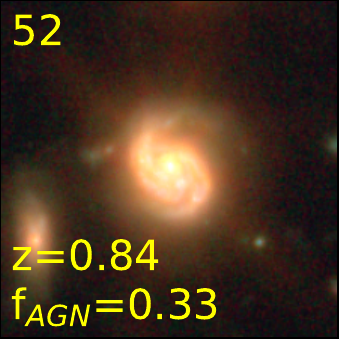} \\
\includegraphics[width=0.11\textwidth]{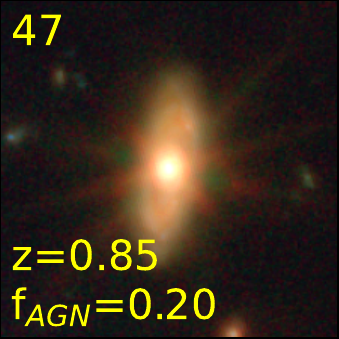}
\includegraphics[width=0.11\textwidth]{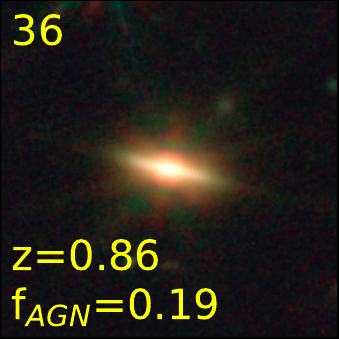}
\includegraphics[width=0.11\textwidth]{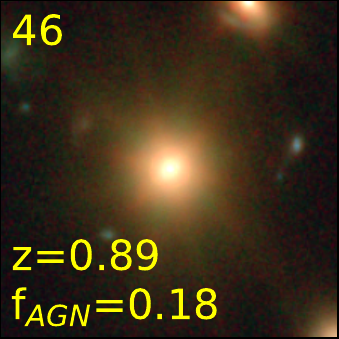}
\includegraphics[width=0.11\textwidth]{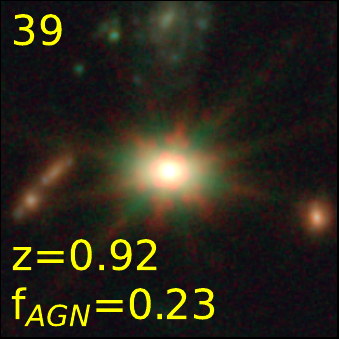}
\includegraphics[width=0.11\textwidth]{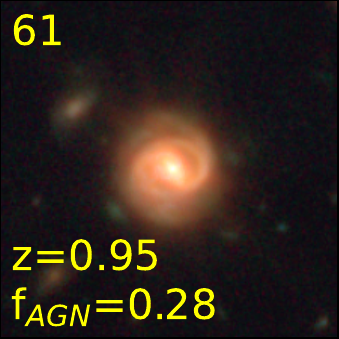}
\includegraphics[width=0.11\textwidth]{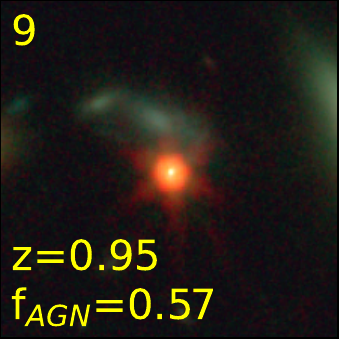} \\
\includegraphics[width=0.11\textwidth]{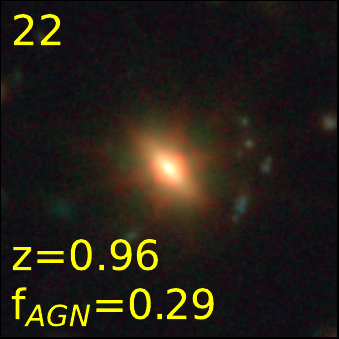}
\includegraphics[width=0.11\textwidth]{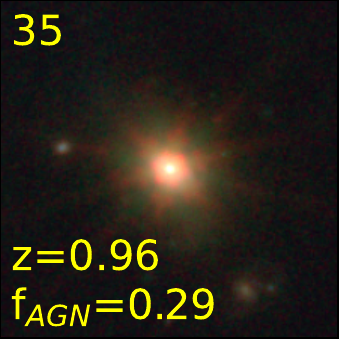}
\includegraphics[width=0.11\textwidth]{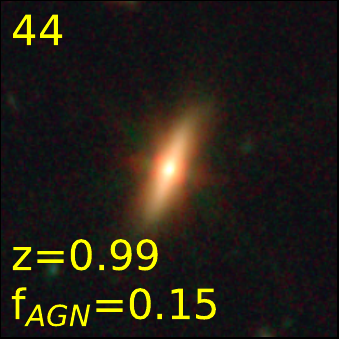}
\includegraphics[width=0.11\textwidth]{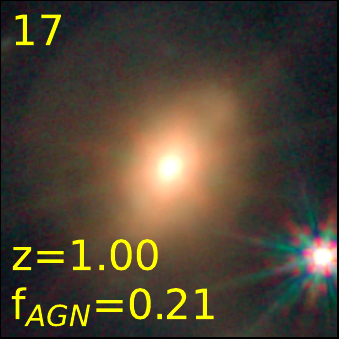}
\includegraphics[width=0.11\textwidth]{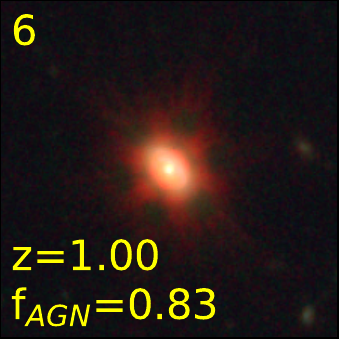}
\includegraphics[width=0.11\textwidth]{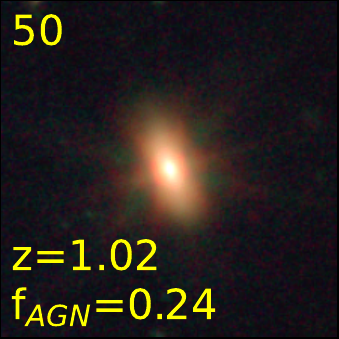} \\
\includegraphics[width=0.11\textwidth]{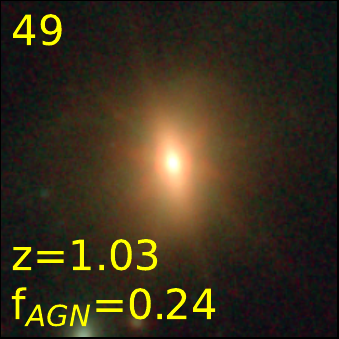}
\includegraphics[width=0.11\textwidth]{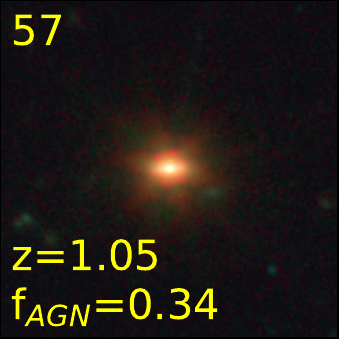}
\includegraphics[width=0.11\textwidth]{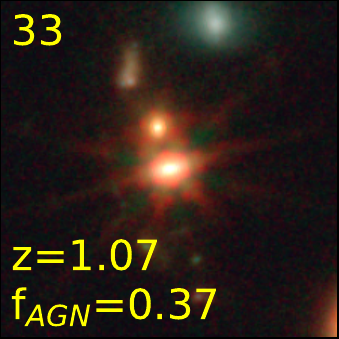}
\includegraphics[width=0.11\textwidth]{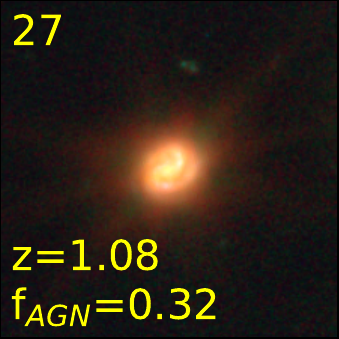}
\includegraphics[width=0.11\textwidth]{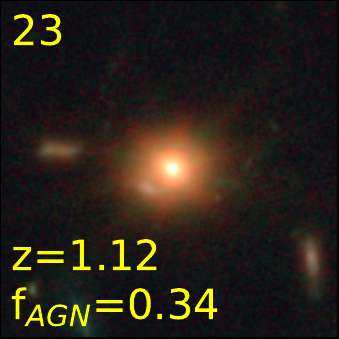}
\includegraphics[width=0.11\textwidth]{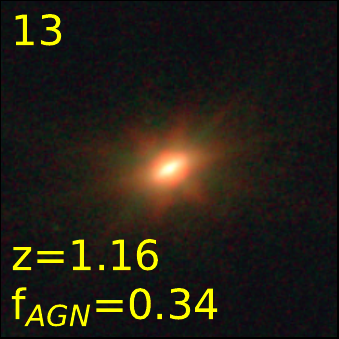} \\
\includegraphics[width=0.11\textwidth]{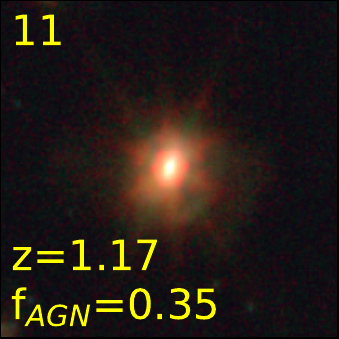}
\includegraphics[width=0.11\textwidth]{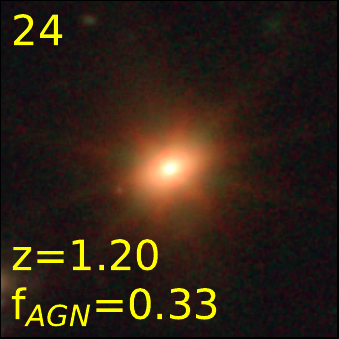}
\includegraphics[width=0.11\textwidth]{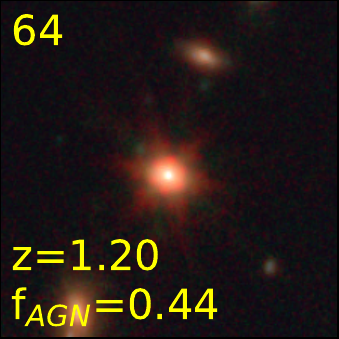}
\includegraphics[width=0.11\textwidth]{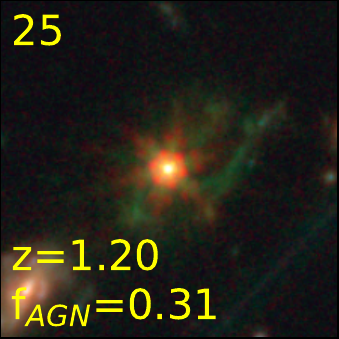}
\includegraphics[width=0.11\textwidth]{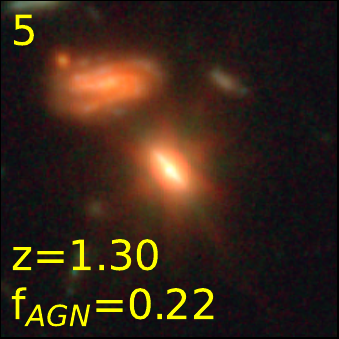}
\includegraphics[width=0.11\textwidth]{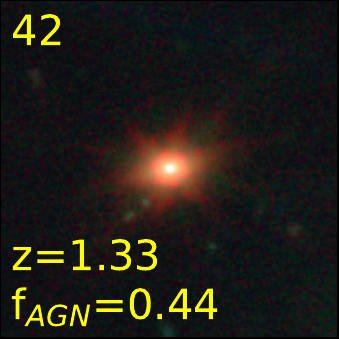} \\
\includegraphics[width=0.11\textwidth]{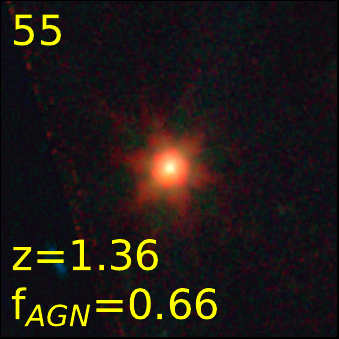}
\includegraphics[width=0.11\textwidth]{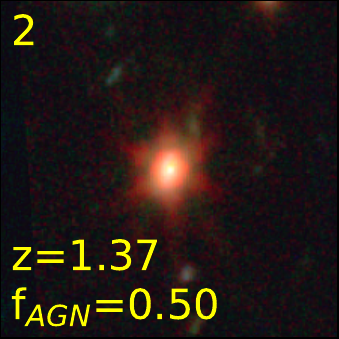}
\includegraphics[width=0.11\textwidth]{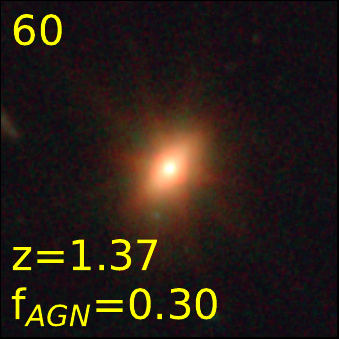}
\includegraphics[width=0.11\textwidth]{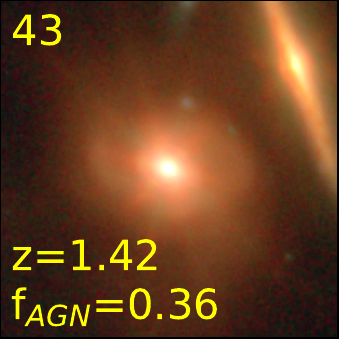}
\includegraphics[width=0.11\textwidth]{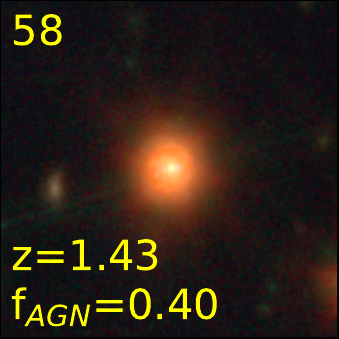}
\includegraphics[width=0.11\textwidth]{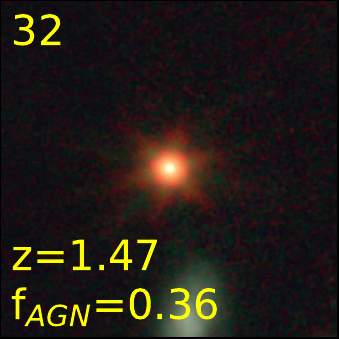} \\
\includegraphics[width=0.11\textwidth]{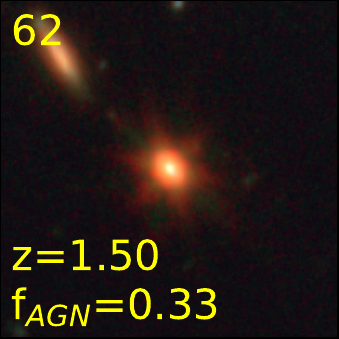}
\includegraphics[width=0.11\textwidth]{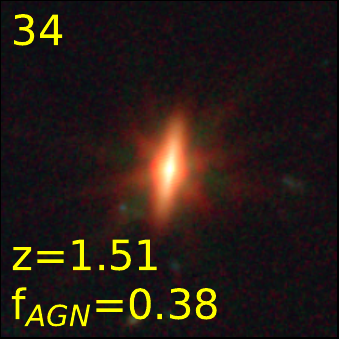}
\includegraphics[width=0.11\textwidth]{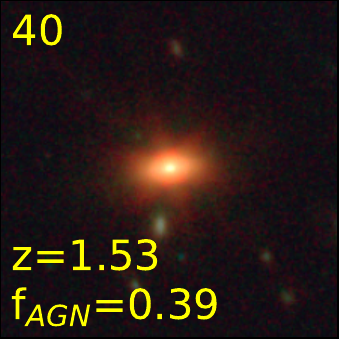}
\includegraphics[width=0.11\textwidth]{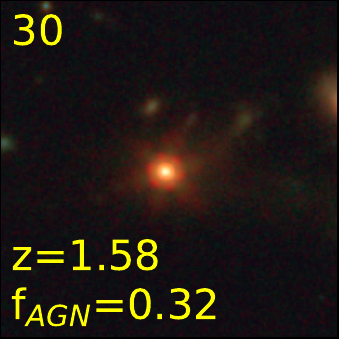}
\includegraphics[width=0.11\textwidth]{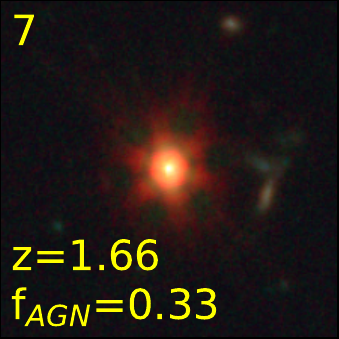}
\includegraphics[width=0.11\textwidth]{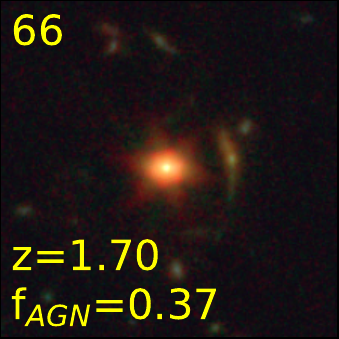} \\
\includegraphics[width=0.11\textwidth]{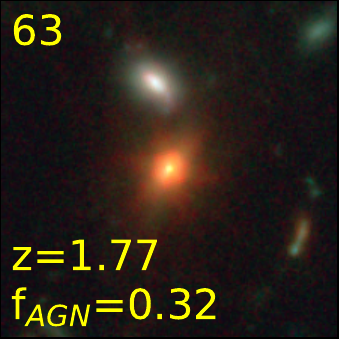}
\includegraphics[width=0.11\textwidth]{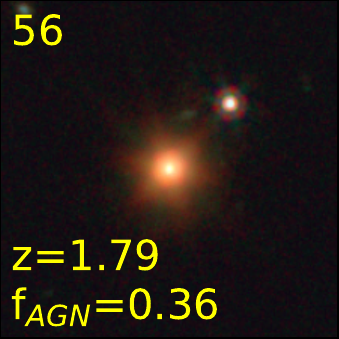}
\includegraphics[width=0.11\textwidth]{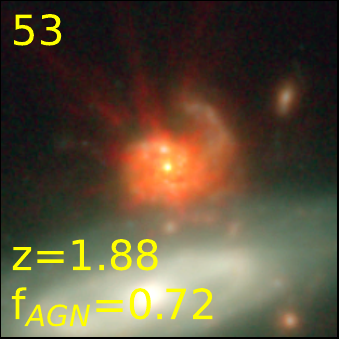}
\includegraphics[width=0.11\textwidth]{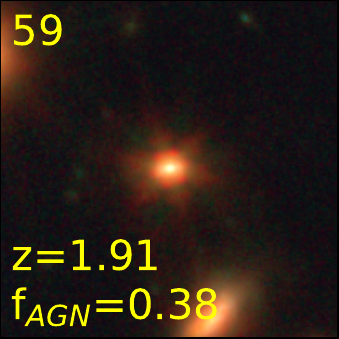}
\includegraphics[width=0.11\textwidth]{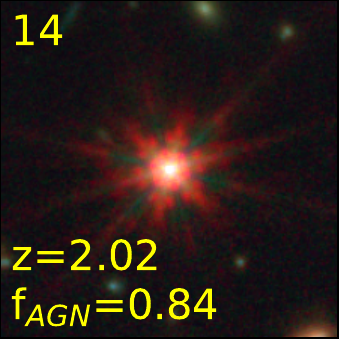}
\includegraphics[width=0.11\textwidth]{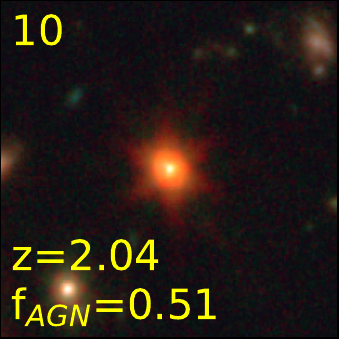}
    \caption{RGB image cutouts for the 66 CPGs sorted by $z_{\rm phot}$ from \cigale{}. All image cutouts are 6\arcsec\ square. The top left corner of each image gives the image ID from Table~1, and $z_{\rm phot}$ and $f_{\rm AGN}$ are in the bottom left. The RGB scaling is R = 0.3$\cdot$F356W + 0.8$\cdot$F410M + 1.0$\cdot$F444W, G = 1$\cdot$F200W + 0.7$\cdot$F277W + 0.5$\cdot$F356W, B = 1$\cdot$F090W + 1$\cdot$F115W + 1$\cdot$F150W + 1$\cdot$F200W + 1$\cdot$F277W + 0.5$\cdot$F356W + 0.5$\cdot$F410M + 0.5$\cdot$F444W.}
    \label{fig:RGB_fig}
\end{figure*}

ROIII visually inspected an 8-filter color-composite of all NIRCam observations within the \JWST\ NEP--TDF to identify resolved galaxies with unresolved point-like features in their cores. The color-composite was constructed according to the \trilogy{}\footnote{\href{https://www.stsci.edu/~dcoe/trilogy}{https://www.stsci.edu/\(\sim \)dcoe/trilogy}} prescription \citep[]{Coe2012}. 
The entire \JWST\ NEP--TDF was visually inspected for galaxies that displayed the characteristic diffraction spikes or PSF effects from compact, luminous galaxy cores. 
This identified 66 galaxies\footnote{We verified that no objects in the sample are stars by applying a FWHM and magnitude cut similar to that used by \citet{Windhorst2023} and by requiring $z_{\rm phot}\neq0$.}, which we refer  as ``CPGs'' (central point-source galaxies) hereafter. \added{Table~\ref{table:catalog} lists the CPGs, and} Figure~\ref{fig:RGB_fig} shows their images and highlights the qualitative criteria, diversity, and similarities of the CPGs. Several objects within the CPG sample display obvious point-like features in their cores (i.e., IDs 1, 14, 28, 48), whereas the vast majority show less distinct features of the characteristic PSF from \JWST.

The NIRCam images of the NEP--TDF sampled an area of 65.4 arcmin$^2$, and hence our sample of 66 CPGs corresponds to $\sim$1 galaxy per arcmin$^{2}$ to $m_{\rm F444W}\lesssim22$~AB mag. This is comparable to the WISE (\textsl{W1}--\textsl{W2}) color-selected AGN density  \citep[\eg][their Table 1]{Assef2013}, $\sim$0.5 AGN per arcmin$^2$ to $m_{\rm W2}\lesssim 20.5$~AB mag.

\subsection{Automated Sample Selection}
\label{sec:procedure}

\begin{figure*}[t]
    \centering
        \includegraphics[width=.8\textwidth]{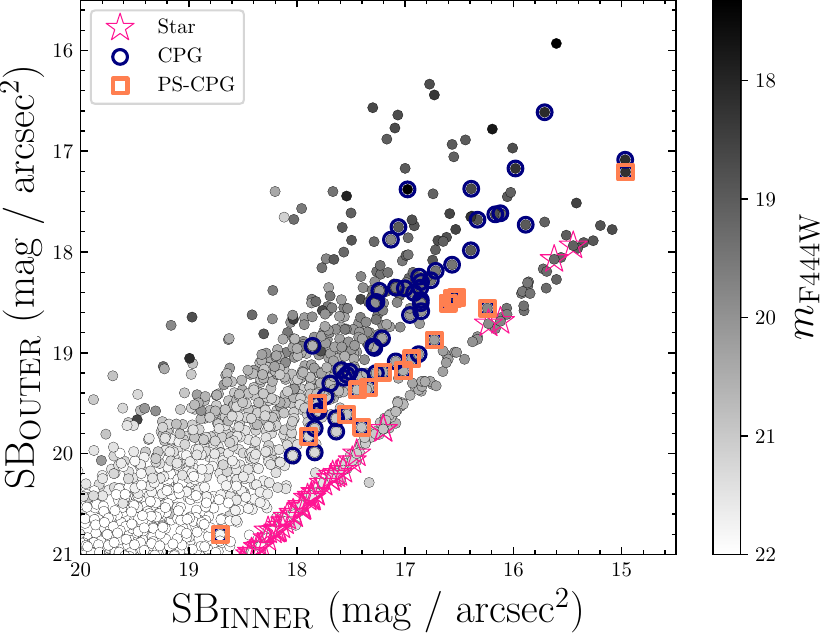}
    \caption{F444W central surface brightness versus the surface brightness in a surrounding annulus. The central surface brightness was measured in a circular aperture with a radius 0\farcs18, and the outer annulus between circular apertures of 0\farcs24 and 0\farcs36. \added{Navy blue circles and orange squares represent CPGs and Point-Source CPGs (see \S~\ref{sec:galfit}), and pink stars represent stars in the field \citep{Windhorst2023}.} Only objects with $m_{\rm F444W}<22$~AB mag, the limiting magnitude of the CPG sample, are plotted. The greyscale bar (right) maps the galaxy total F444W magnitudes (\texttt{MAG\_AUTO}), with brighter galaxies plotted as darker.}
    \label{fig:aperphot_fig}
\end{figure*}

While our initial sample was obtained through qualitative visual selection, a quantitative method is needed to identify objects similar to CPGs, especially for large surveys.
The concentration index \citep[\eg][]{Conselice2003} is one measure, but its basic ratio finds sources that are peaky but not necessarily pointlike.  Instead, we used \sextractor{} to measure F444W magnitudes in circular apertures at the object cores. 
Aperture radii were 6, 8, and 10 pixels, corresponding to radii 0\farcs18, 0\farcs24, and 0\farcs30, respectively. We adopted these apertures to  probe the surface brightness near the cores of the CPG sample and identify objects that are morphologically similar to stars out to a certain radius. We used the smallest aperture  to measure the average surface brightness within an inner area of $\sim$0.1 arcsec$^2$ (SB$_{\rm INNER}$) and used the two other apertures to measure the average surface brightness within an outer annulus (SB$_{\rm OUTER}$), which has an area equal to that within the inner circular aperture. 
Multiple aperture radii for SB$_{\rm INNER}$ and SB$_{\rm OUTER}$ were tested, and by experiment the ones adopted provided the most effective  way to probe point-source features in galaxies. 

Figure~\ref{fig:aperphot_fig}  plots SB$_{\rm OUTER}$  against SB$_{\rm INNER}$.
Stars are identified as straddling the NIRCam diffraction limit with $\rm 150~mas\leq FWHM\leq170~mas$ with $m_{\rm F444W} \leq 28$~mag \citep[][]{Windhorst2023}. CPGs near the stars in Figure~\ref{fig:aperphot_fig} are \added{Point-Source-CPGs (PS-CPGs), characterized and identified in \S\ref{sec:galfit} as having point-source cores}. These objects show that there exists a parameter space where we can probe point sources within galaxies, barring contamination from other stars. 
The longer wavelengths of the near-infrared regime can probe both optically obscured and unobscured AGN \citep[\eg][]{Assef2013} and both star-forming galaxies and weak AGN \citep[\eg][]{Kim2019} via redder colors in the 3.5--3.6~\micron\ and 4.4--4.5~\micron\ bands. Thus, \JWST's aperture photometry---in contrast to shorter wavelength data---can be more effective at probing morphological signatures of AGN at these wavelengths. Obscured AGN in the UV--visible are likely showing up in these images because of the much smaller dust extinction at these wavelengths, \added{and we carried out complementary SED analysis (\S~\ref{sec:characterization}) to clarify whether AGN are the most probable explanation for the} point-like features in our infrared CPGs.

The above classification procedure provides a robust star--CPG separation and a CPG separation from brighter galaxies that automatically recovers our CPG sample and finds objects similar to it. Particularly, this classification procedure can probe the presence of a point source within resolved galaxies. Identifying galaxies that straddle the morphology of stars \added{(such as the orange squares in Figure~\ref{fig:aperphot_fig} representing PS-CPGs)} without contamination would allow a search for galaxies with point-like cores in any \JWST/NIRCam image. The automated method does not pick up all of the weak Seyferts; some will require higher signal-to-noise data to be identified by this procedure. 

\subsection{Control Sample}
\label{sec:controlsample}

\added{Understanding how CPGs differ from other galaxies requires a comprehensive control sample. To create one, we find the four most similar galaxies in the \JWST\ NEP--TDF with respect to brightness and \cigale-computed photometric redshift down to $\rm AB~mag \le 22$, the limiting magnitude for CPG selection.  
The resulting control sample has 225 galaxies, slightly fewer than $4\times$ the CPG sample size. These are all the galaxies we could find with a similar $m_{\rm F444W}$ and photometric redshift distribution.}

The control sample was also analyzed with the same two-dimensional light-profile and SED-analysis as the CPG sample, providing robust number statistics and similar galaxy demographics to assess the reliability of the classifications.

The first spoke of the NEP TDF, observed in 2022 Aug, had a scheduling interruption due to a guide star failure. The second half of that spoke was observed 10 days later, resulting in two sets of diffraction spikes rotated by 10\degr. 
The PSF structure with 12 diffraction spikes could not be modeled for morphological parameters with affordable
computational efficiency, and that left only 132 control galaxies that could be so modeled.  All other parameters were determined for the full control sample of 225 galaxies.
\section{Results \& Discussion}
\label{sec:characterization}

\subsection{Two-Dimensional Light Modeling}
\label{sec:galfit}

\added{We used the two-dimensional fitting algorithm \galfit{} \citep{Peng2002} to model the 4.4~\micron\ light profiles of the CPGs.
We modeled all galaxies with both a double-S\'ersic model and a S\'ersic component plus a PSF component. Comparison of the two models determines  whether CPGs contain true point sources or  a compact stellar bulge. Many galaxies in both the CPG and control samples cannot be completely modeled with only two components, but this modeling approach is adequate to characterize the apparent point-like features. This structural analysis is primarily sensitive to AGN that have similar or higher luminosities to or higher than that of a central bulge.}

\begin{figure*}
\centering
\includegraphics[width=0.33\textwidth]{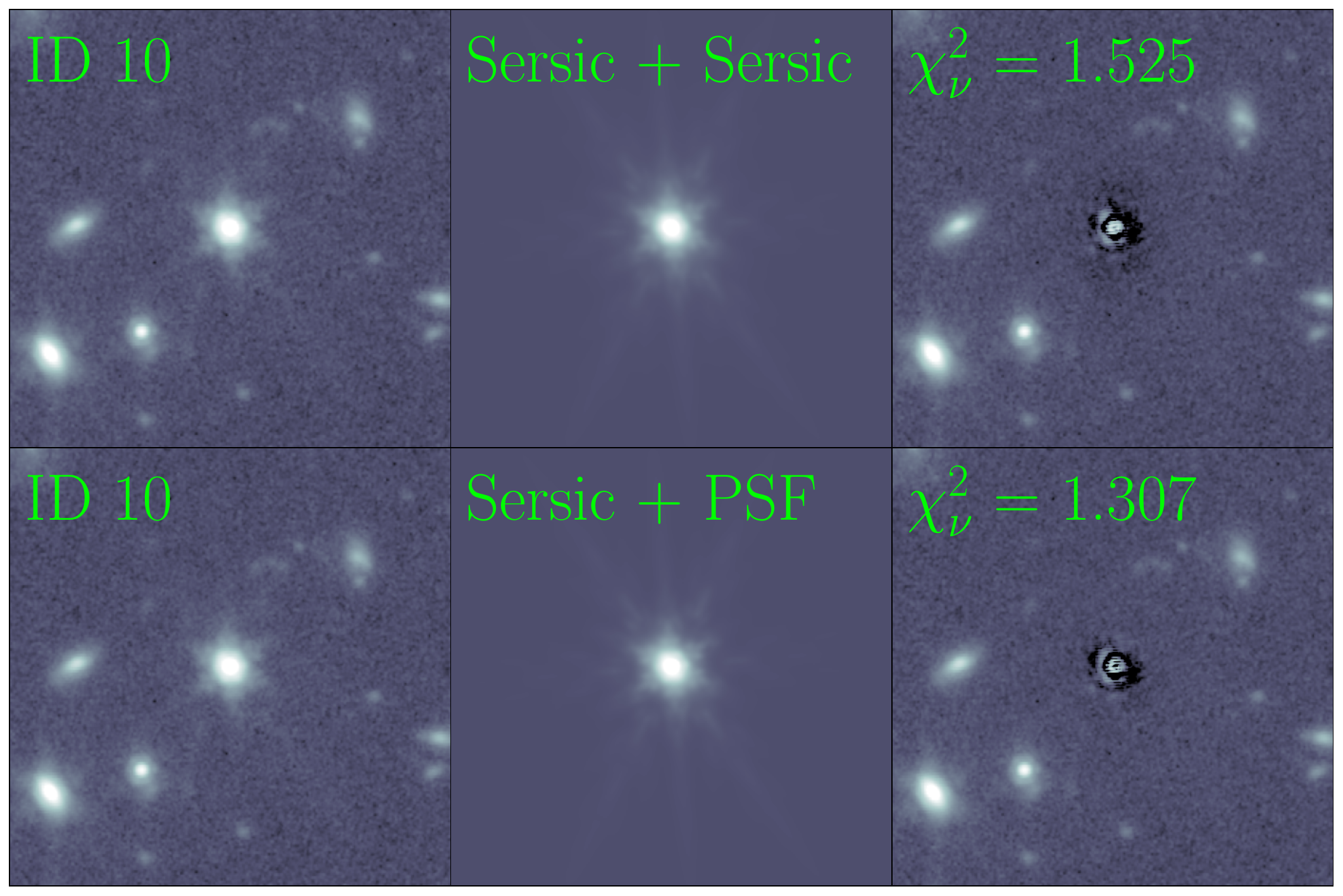}
\includegraphics[width=0.33\textwidth]{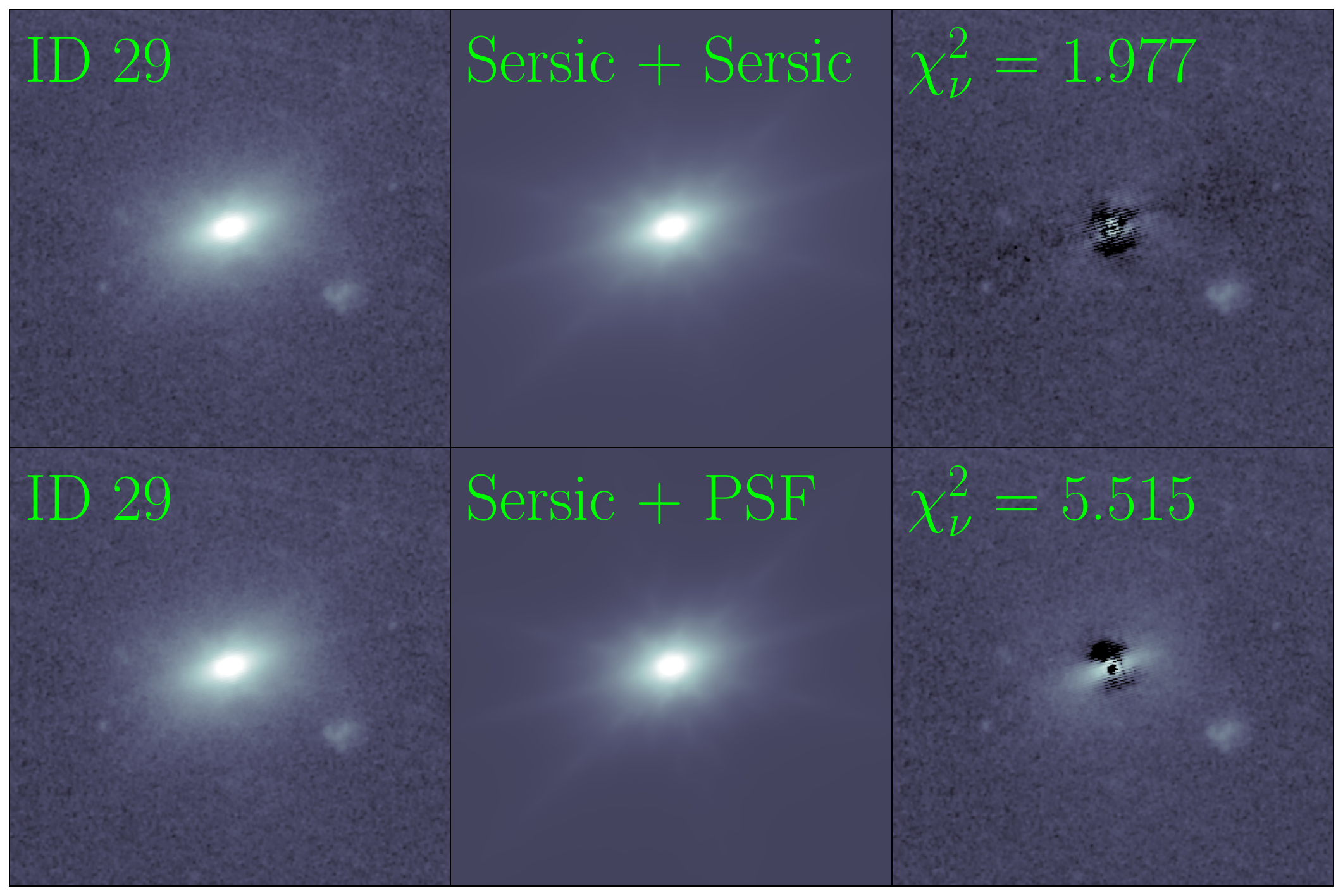}
\includegraphics[width=0.33\textwidth]{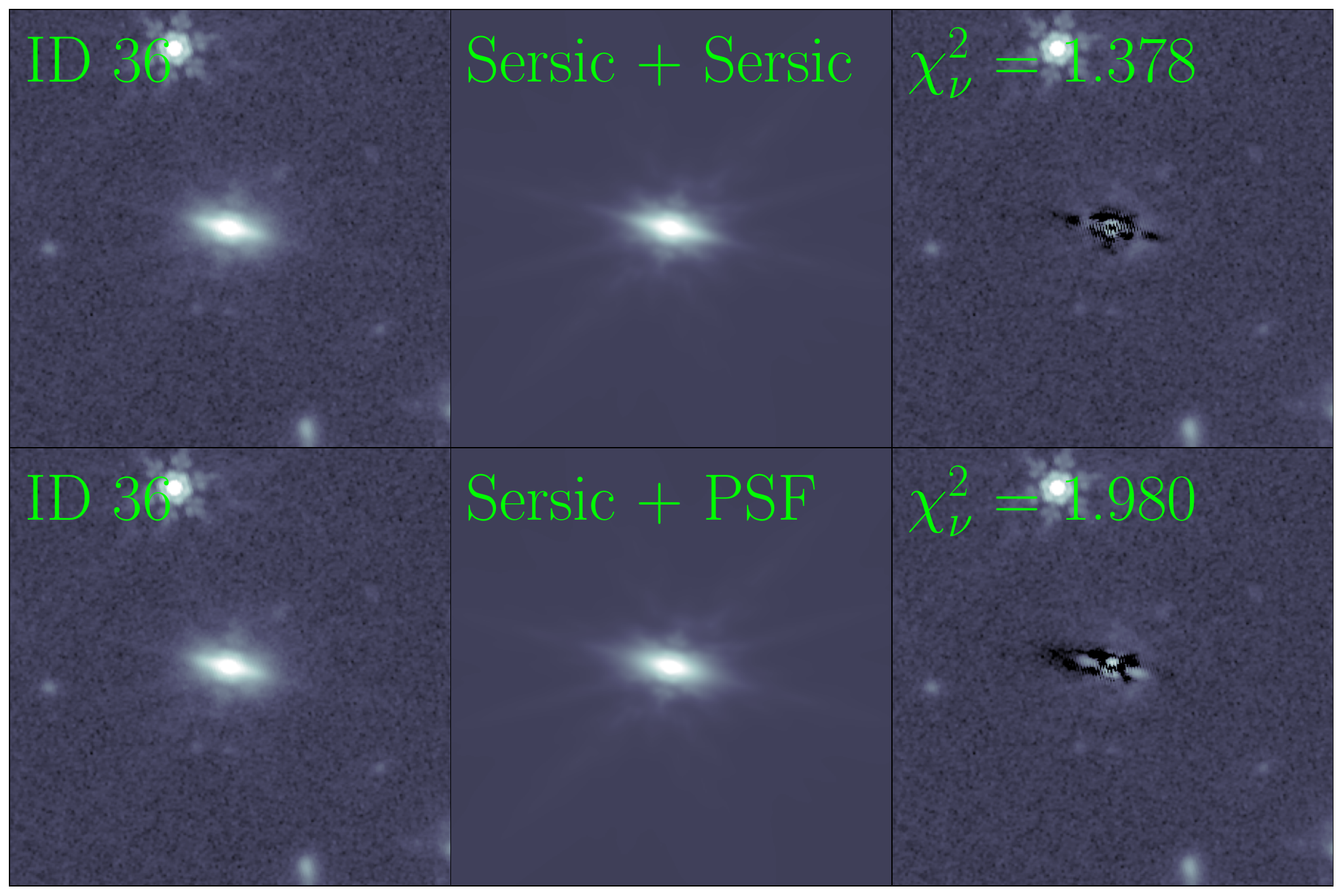}
\includegraphics[width=0.33\textwidth]{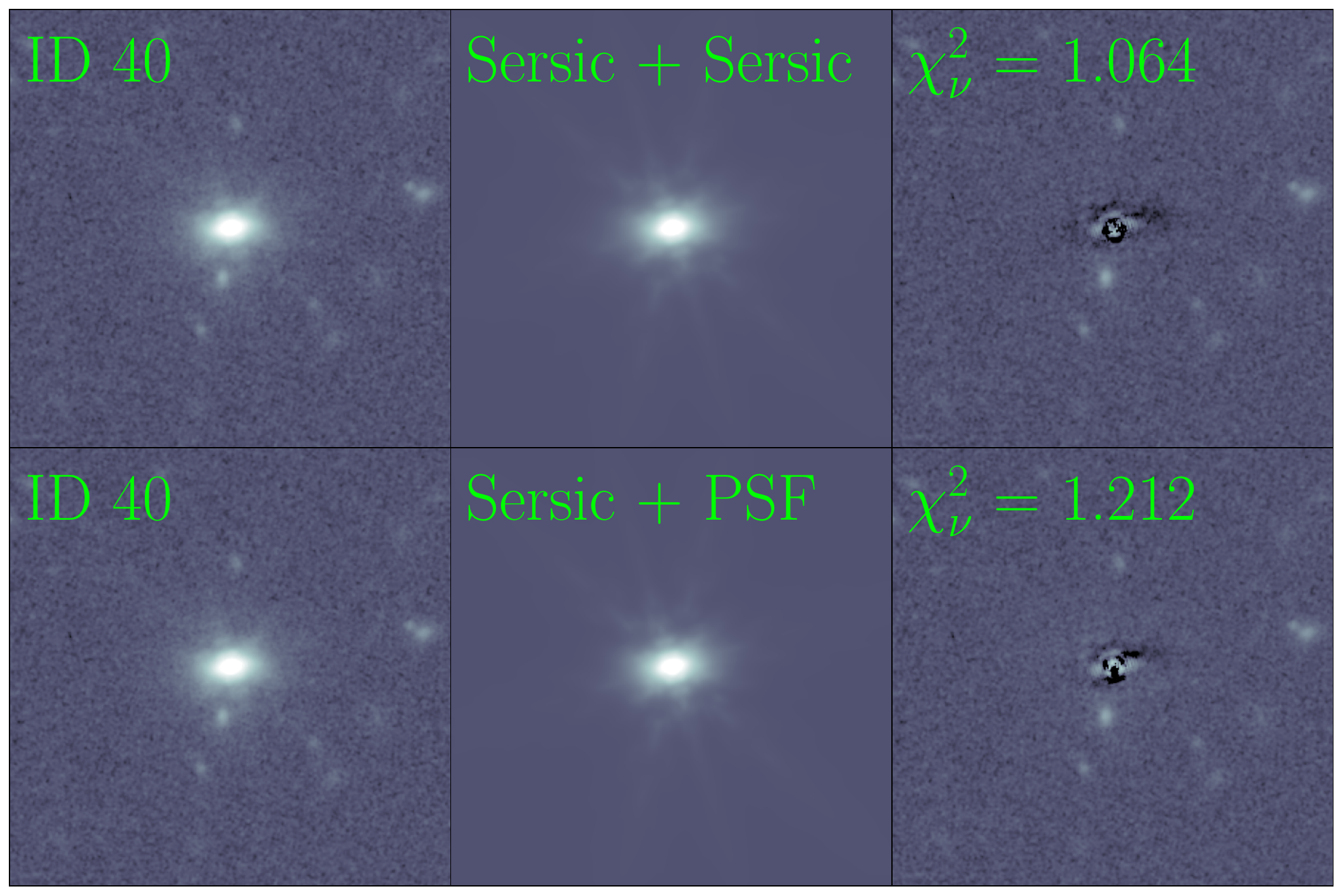}
\includegraphics[width=0.33\textwidth]{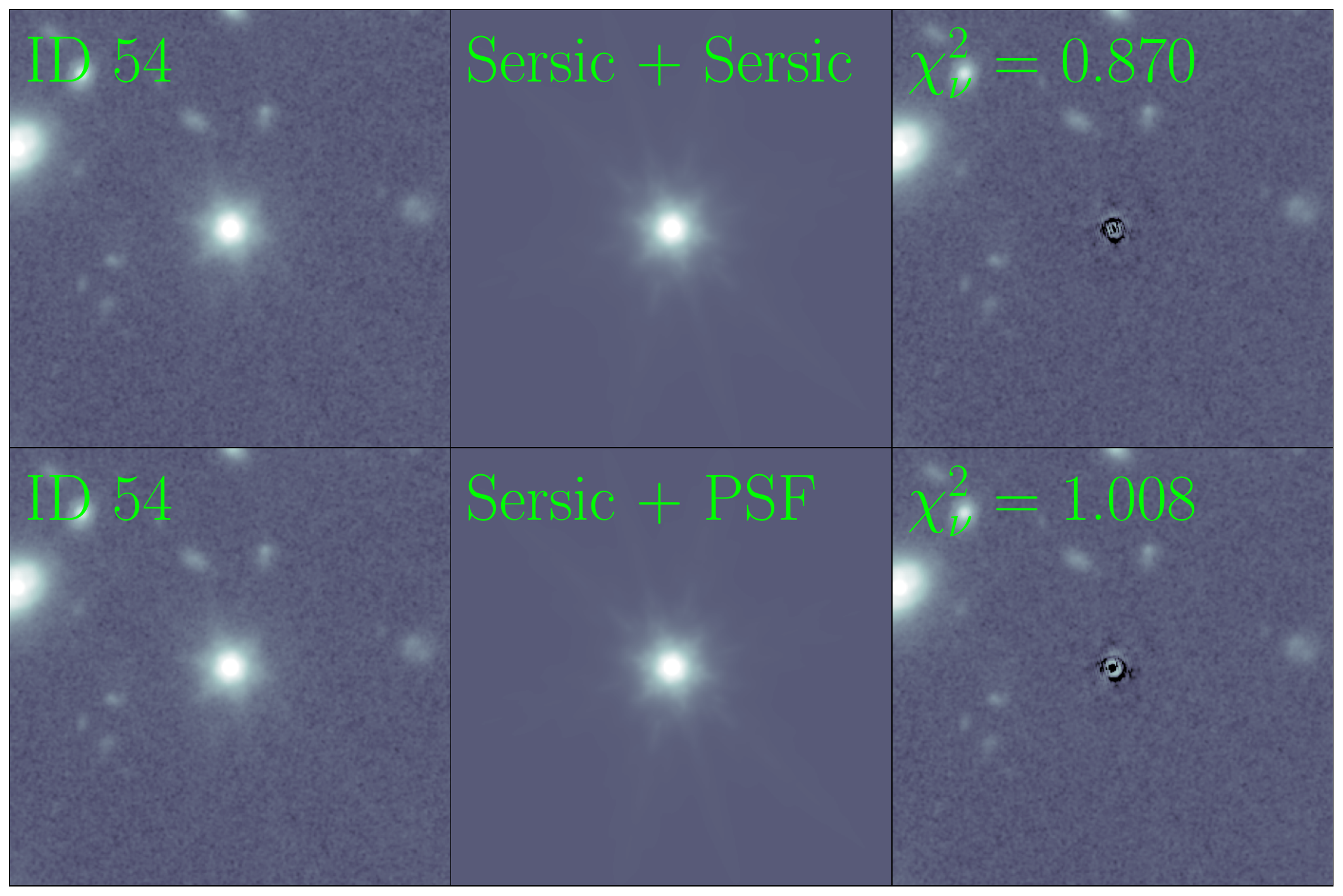}
\includegraphics[width=0.33\textwidth]{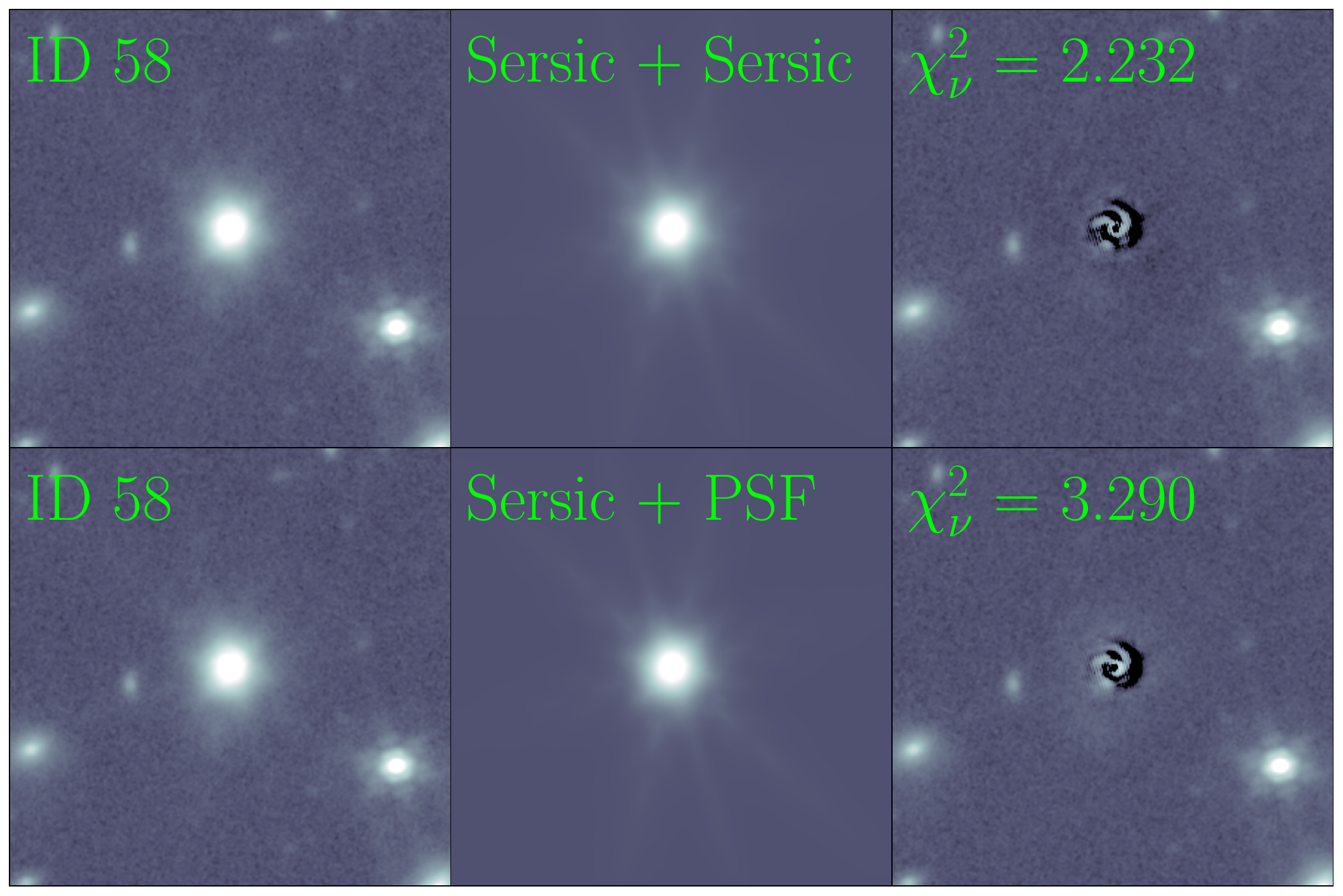}
\caption{\added{\galfit{} image output  for six  CPGs. For each object, there are six subpanels; the top three subpanels visualize the S\'ersic+S\'ersic fit, and the bottom three visualize the S\'ersic+PSF fit. Each row  shows the image (left), model (middle), and residual image (right). Green text identifies the CPG ID, the two-component fit, and the goodness of fit statistic, $\chi^2_{\nu}$, for the fit. The color map indicates negative flux in black and positive flux in white. }}
\label{fig:residuals}
\end{figure*}

\added{
The S\'ersic profile is a robust representation of a variety of galaxy types because of its flexibility in profile characteristics. The S\'ersic profile is parameterized as
\begin{equation}
    I(R) = I_e \exp\left\{-b_n \left[\left(\frac{R_m}{R_e}\right)^{1/n}-1\right]\right\}\quad,
\end{equation}
where $I_e$ is the surface brightness at the half-light radius $R_e$, $n$ is the S\'ersic index, $b_n$ is a derived parameter ensuring proper integration at $R_e$ (and is the value at which the Gamma probability distribution function integrates to $0.5$ with a shape parameter of $2n$), and $R_m$ is the two-dimensional modified radius where the profile is being evaluated \citep[\eg][]{Robotham2017}. For our purposes, the S\'ersic profile is assumed to generally model the nucleus and extended stellar disc of a galaxy in a double-S\'ersic component fit. For two-component fits with a S\'ersic component and a PSF component, the S\'ersic profile generally models the bulk of the extended galaxy, while the PSF component models an unresolved point source in the galaxy nucleus. Thus a \galfit{} analysis classifies the core type of the CPGs.}

\added{For both our CPG and control sample, we used a simulated 4.4~\micron\ PSF from \texttt{WebbPSF} \citep{webbpsf} based on 300$\times$300 pixel (9\arcsec$\times$9\arcsec)  image stamps of each galaxy. We masked out neighboring objects through a \sextractor-generated segmentation map and estimated the sky background using the uncertainty image from the \JWST\ pipeline. A note at the end of Table~\ref{table:galfitTable} gives the constraints on the \galfit{} parameters.}
\added{Figure~\ref{fig:residuals} visualizes \galfit{} output for six CPGs. 
}

\begin{figure}[b]
    \centering
\includegraphics[width=0.95\linewidth]{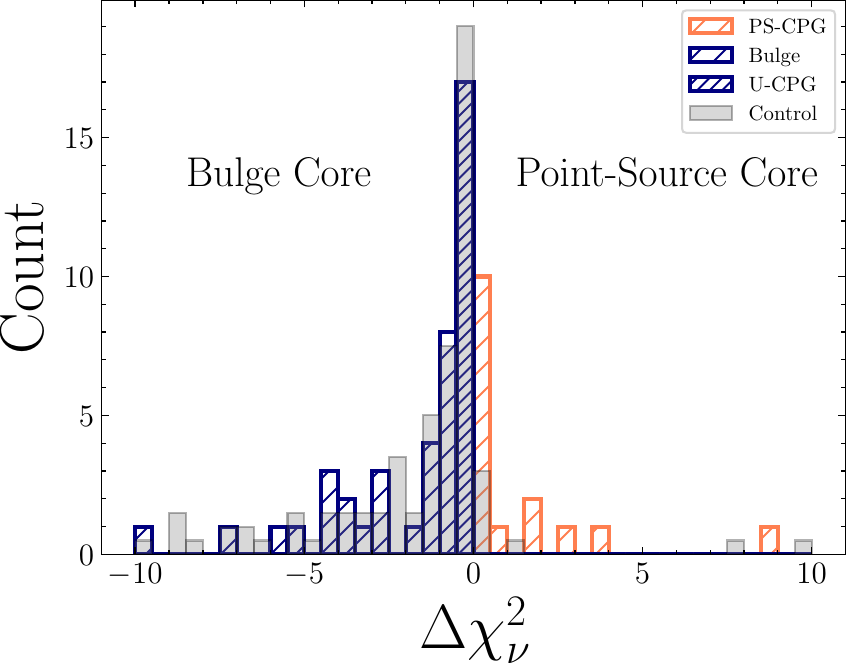}
    \caption{\added{Histogram of the goodness of fit statistic ($\chi^2_{\nu}$) between the \galfit{} procedure that models both the CPG and control sample with a S\'ersic+S\'ersic two-component fit and a S\'ersic+PSF two-component fit. $\Delta\chi^2_{\nu}\equiv \chi^2_{\nu}[\hbox{S\'ersic}+\hbox{S\'ersic}]--\chi^2_{\nu}[\hbox{S\'ersic}+\rm PSF]$, indicating a preference for a bulge core (negative values) or a point-source core (positive values). The hashed histogram identifies CPGs, with distinction being made between general CPGs (navy) and PS-CPGs (orange) to further classify and characterize the CPGs core type. The navy histogram contains hashed and double-hashed bins, identifying Bulge cores and undetermined CPGs core types, respectively. The grey histogram represents the control sample and is normalized to the CPG sample size.}}
    \label{fig:deltachi2}
\end{figure}

\added{Figure \ref{fig:deltachi2} compares the goodness-of-fit results for the single- and double-S\'ersic fits. Most CPGs prefer the {double}-S\'ersic fit, but 16  are best fit with the {single}-S\'ersic component plus a PSF component. This is not the case for the control-sample galaxies. Many galaxies in both samples are equally well fit by both models with a preference for a double-S\'ersic fit in borderline cases. The preference is consistent with the double-S\'ersic having more free parameters. Both the CPG and control sample demonstrate a similar bulge-core frequency.}
\added{Hereafter, we refer to CPGs with a point-source core classification from this two-component \galfit{} procedure as ``PS-CPGs'', borderline cases as Undetermined-CPGs (``U-CPGs''), and bulge core classified CPGs are referred to as Bulge-CPGs (``Bulge''). This distinction is particularly useful when using SED techniques to infer physical properties of the CPGs with likely point-source features in their cores.}

\added{
A number of CPGs are not completely fit with two-components. There are definitely galaxies within our CPG sample that were visually selected simply for having ultraluminous,  compact nuclei, thus having steep S\'ersic indices and showing faint features of the PSF\null. Consequently, these Bulge-CPGs are likely galaxies not hosting an observable AGN at our observed wavelengths and are more suggestive of being a compact stellar bulge or nuclear starburst \citep[\eg][]{Bruce2016}. It is reasonable to expect that the CPG sample would show a mix of true point-source and stellar-bulge cores. More comprehensive \galfit{} modeling would clarify the point-source presence in galaxies that are not well fit with only two components (i.e., U-CPGs). Nevertheless, Figure~\ref{fig:deltachi2} suggests that the CPG sample does indeed have  point-sources within galactic nuclei. Additional constraints from SED fitting are needed to clarify whether an AGN is responsible for these features.}

\subsection{Seyfert Template Fitting with \eazy{}}
\label{sec:eazy}

In order to estimate AGN fractions and derive photometric redshifts, we used \eazy{} to fit the photometry for each of the CPG and control sample galaxies. For this single-component template fitting procedure, we ran \eazy{} twice: one run allowing a single-component fit from the 12 \eazy{} \texttt{tweak\_fsps\_QSF\_12\_v3} templates and a second run fitting the AGN-ATLAS SEDs. The former templates are galaxy templates used for stellar population synthesis and the latter templates \citep{BrownAGNATLAS} are blends of AGN and host-galaxy SEDs covering 0.09--30.0~\micron.  The host/AGN ratios available are powers of two from 0.5 to 64 normalized at 0.6~\micron.

\begin{figure}[t]
    \centering
    \includegraphics[width=0.49\txw]{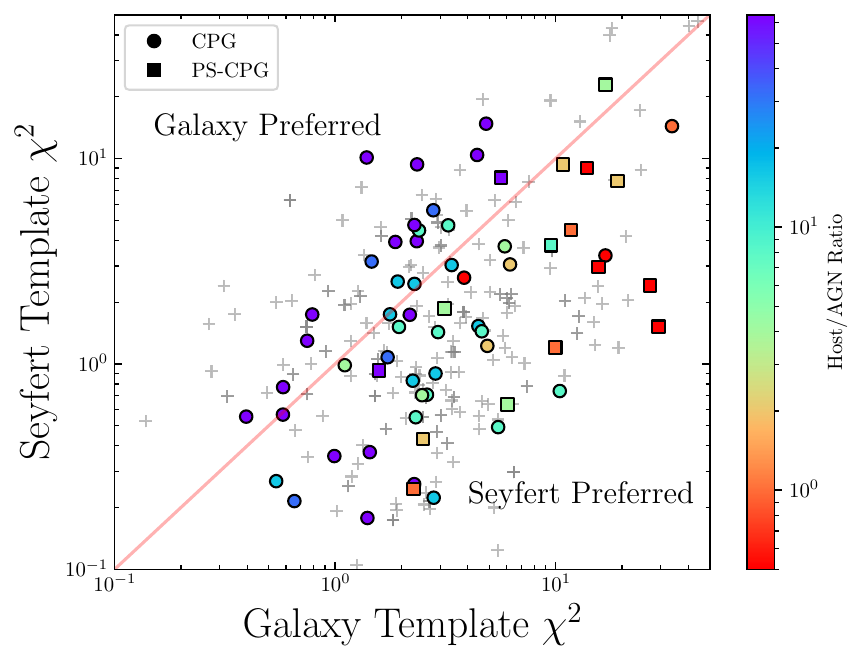}
    \caption{$\chi^2$ of \eazy\ fits with and without an AGN\null. Colored markers represent CPGs with squares identifying PS-CPGs and circles representing the remaining CPGs. Colors indicate each CPGs' best-fit host/AGN ratio as indicated in the color bar. Grey crosses represent   control-sample galaxies. The diagonal line indicates equality, \ie\, an undetermined classification. CPG IDs 1 and 29 are not plotted because their fits failed to converge owing to the core's extreme brightness.}
    \label{fig:template}
\end{figure}

As shown in Figure~\ref{fig:template}, 70\% of the CPGs prefer an AGN-ATLAS template over a standard \eazy{} template. Generally, the fits with the highest AGN contributions to the total SED are still fairly well fit with an \eazy{} template, \ie\, with no AGN at all. Even when an AGN is indicated, for most galaxies in the sample, the host galaxy outshines AGN emission. The fitting does not prove that our sample galaxies are Seyferts. Offering more SED templates is likely just adding more free parameters and resulting in better fits \added{as demonstrated by the control sample, which shows a similar distribution to the CPGs in Figure~\ref{fig:template}.  In fact, 86\% of control-sample fits prefer an AGN-ATLAS template}. Nevertheless, the fitting is consistent with the hypothesis that the morphological selection is finding AGN.

\subsection{SED Parameter Estimation with \cigale{}}
\label{sec:cigale}

\begin{figure*}[t]
    \centering
    \includegraphics[width=.97\txw]{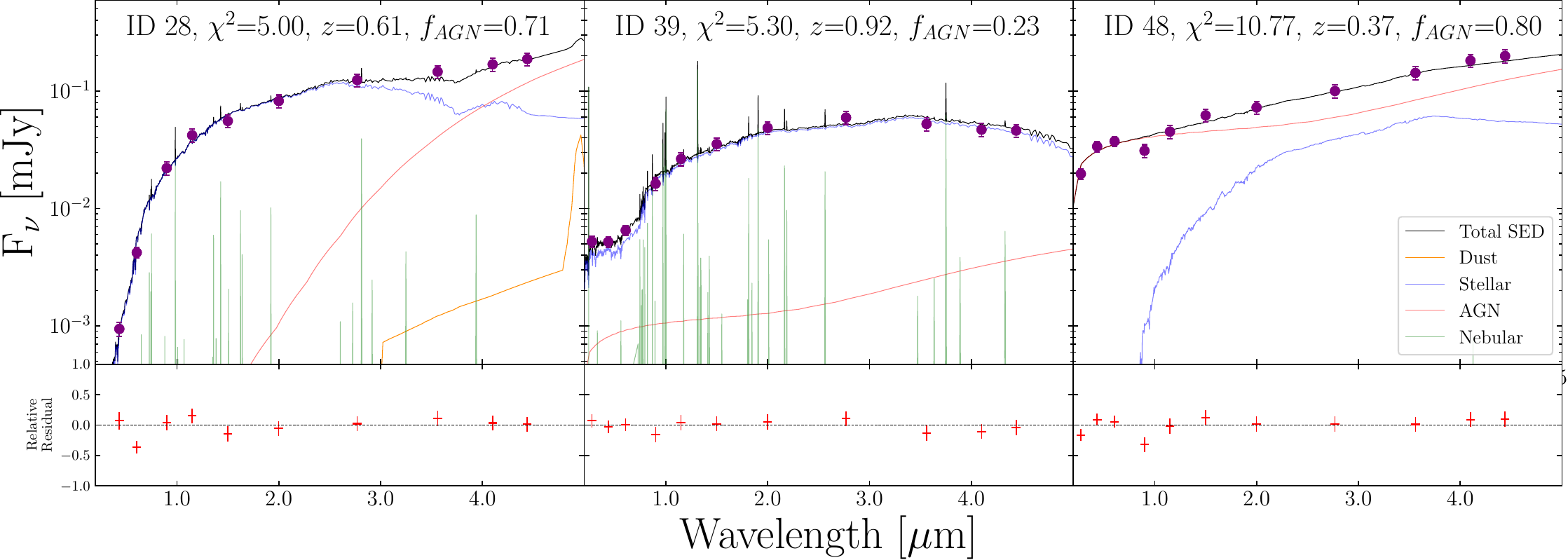}
    \caption{Example \cigale\ model fits for three CPG SEDs. Fit residuals are shown below each SED panel. The CPG ID\added{, $\chi^2$, photometric redshift, and $f_{\rm AGN}$ are} indicated  in each panel. The legend identifies model components included in the SED fitting. The online journal includes a figure set showing all 66 \cigale{} fits.}
    \label{fig:26215SED}
\end{figure*}

To infer the galaxies' physical properties, we used \cigale{}, an SED-fitting code relying on energy balance between the ultraviolet and infrared. 
The fits are good, as expected with so many free parameters, with \added{94\% of the fits having $\chi^2_{\nu} \leq 1.0$}. The online version of this paper includes a Figure Set of all 66 \cigale\ fits for our CPG sample, and Figure~\ref{fig:26215SED}  shows three examples. 

\figsetstart
\figsetnum{6}
\figsettitle{CIGALE SEDs for CPGs}

A \cigale{} output parameter of immediate interest is $f_{\rm AGN}$, the fractional contribution of AGN emission to the bolometric luminosity. We tuned \cigale{} to compute $f_{\rm AGN}$ based on AGN emission from 0.1 to 30~\micron\  \citep[following][their Eq.~1]{Assef2013} and the NIRCam photometry. 

\added{Most interestingly, the PS-CPG median $f_{\rm AGN}=0.44\pm0.12$} \added{while the control sample's $f_{\rm AGN}=0.24\pm0.09$}. The formal statistical separation is only at the 1$\sigma$ level, though our result suggests the PS-CPG sample has measurable AGN emission in the infrared. $f_{\rm AGN}$ can be better constrained with additional  photometry in future work, particularly in the sub-mm \citep[see, \eg][]{Ciesla2015}.

\begin{figure}[b]
    \centering
    \includegraphics[width=.45\txw]{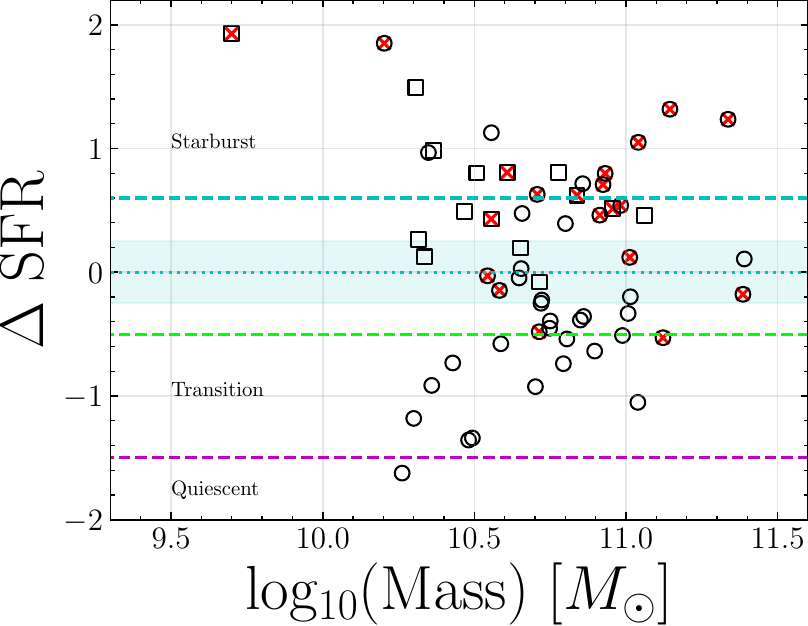}
    \caption{Star-formation rates from \cigale{} relative to the respective star-formation main sequence. The $y$-axis shows $\rm \Delta SFR \equiv \log_{10}(SFR)-\log_{10}(SFR_{\rm MS}$) with SFR$_{\rm MS}$ defined by \citet[their Eq.~28]{Speagle2014}. The horizontal dotted line and shaded region mark the main sequence and the $\pm$0.2 dex range. The upper cyan, the middle green, and the lower magenta dashed lines mark 0.6 dex above the main sequence, commonly taken as the starburst boundary \citep[\eg][]{Rodighiero2011}, the lower boundary to the main sequence, and the upper boundary to the quiescent region \citep[\eg][]{Renzini2015}, respectively. Circles and squares identify CPGs and PS-CPGs, respectively, and markers with $\times$ symbols indicate sources with VLA 3~GHz matches.}
    \label{fig:SFRMS}
\end{figure}

\cigale{} also provides estimated stellar masses and star-formation rates (SFRs), which provide context for the CPGs. Typical galaxies fall on the star-formation main sequence, which depends on both  redshift and stellar mass \citep[Eq.~28]{Speagle2014}. 
As Figure~\ref{fig:SFRMS} shows, the majority of our sample lie near the star-formation main sequence, and $\sim$31\% lie at or above the starburst boundary given by \citet{Rodighiero2011}. An independent SFR estimate comes from radio observations \citep[\eg][]{Condon1992,Tabatabaei2017}. Positions for 24 CPGs (identified in Table~\ref{table:catalog}) match sources in the VLA 3~GHz catalog \citep{Hyun2023}.  Generally, massive and starburst CPGs are the ones having radio detections. This could be attributed to either the synchrotron radiation produced from remnants of Type~II supernovae \citep[\eg][]{Chevalier1982}, \ie\, star formation, or radio emission from the AGN \citep[\eg][]{Kellermann1987}, or both.  \citet[their Fig.~11]{Willner2023} similarly showed that few radio sources in the TDF are in the starburst range. 

\begin{figure*}
    \centering
    \includegraphics[width=0.95\txw]{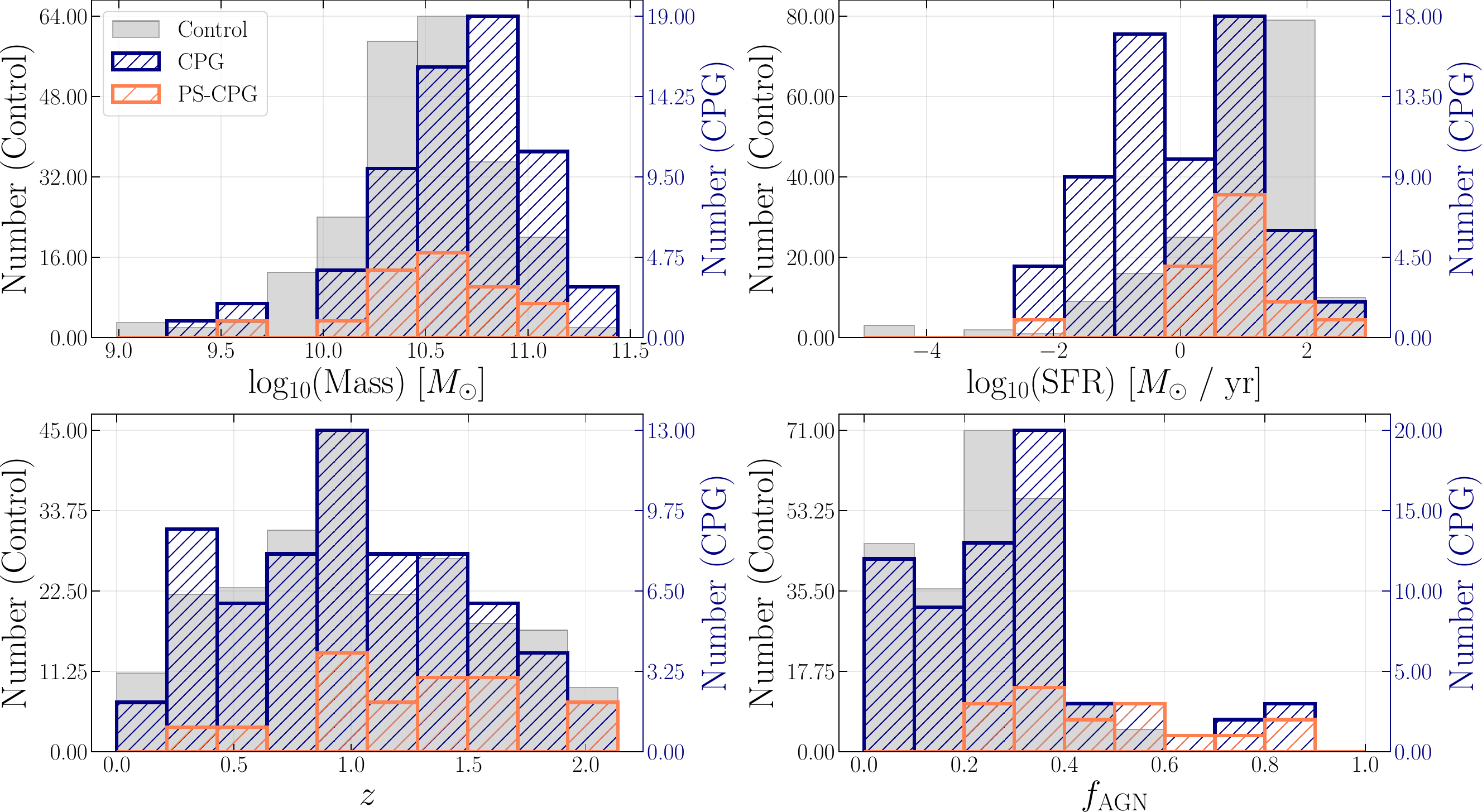}
    \caption{\added{\cigale{}-inferred physical parameters  from SED fitting 0.2--4.4\,$\mu$m observations. All panels are histograms of the CPG sample in navy against the control sample in gray. PS-CPGs are plotted in orange and are included in the CPG distribution. From top left, the panels show the inferred stellar mass, star-formation rates, photometric redshifts, and $f_{\rm AGN}$.}}
    \label{fig:fourhist}
\end{figure*}

\added{Figure~\ref{fig:fourhist} compares the inferred physical properties for the CPG and control samples.  In short, the CPGs are massive galaxies with lower SFRs compared to the control sample. The PS-CPGs tend to have slightly lower inferred stellar masses and higher SFRs than other CPGs. \citet[][]{Rosario2013} showed that AGN are more likely to be hosted by a star-forming galaxy owing to the need for cold gas to fuel both AGN and star formation, though X-ray AGN can be present in similar mass and redshift non-active galaxies. While  the CPG and control samples show a similar distribution of $f_{\rm AGN}$, the PS-CPGs dominate the upper quartile of CPG $f_{\rm AGN}$ values with  $\langle f_{\rm AGN}\rangle= 0.47$. \citet[][]{Ciesla2015} emphasized that small $f_{\rm AGN}$ values from \cigale{} are not well-constrained absent sub-mm photometry, and therefore the values of $f_{\rm AGN} \la 0.2$ may be overestimated. Therefore, our \cigale{} inferences do not necessarily preclude AGN presence for the rest of our CPGs, though they do suggest unique AGN activity in our PS-CPGs.}

\added{Finally,  \cigale{} identified a visual selection bias towards massive galaxies in the CPG sample because only bright galaxies could be examined for morphology. Often it was the luminous core that qualified a galaxy for our visual sample.} Regardless, PS-CPGs' active star formation, normal stellar mass, and frequent radio loudness all suggest an AGN giving rise to a near-infrared point-source is present.

\subsection{($m_{\rm F356W}-m_{\rm F444W}$) Colors}
\label{sec:Color}

\begin{figure}[b]
    \centering
    \includegraphics[width=.485\textwidth]{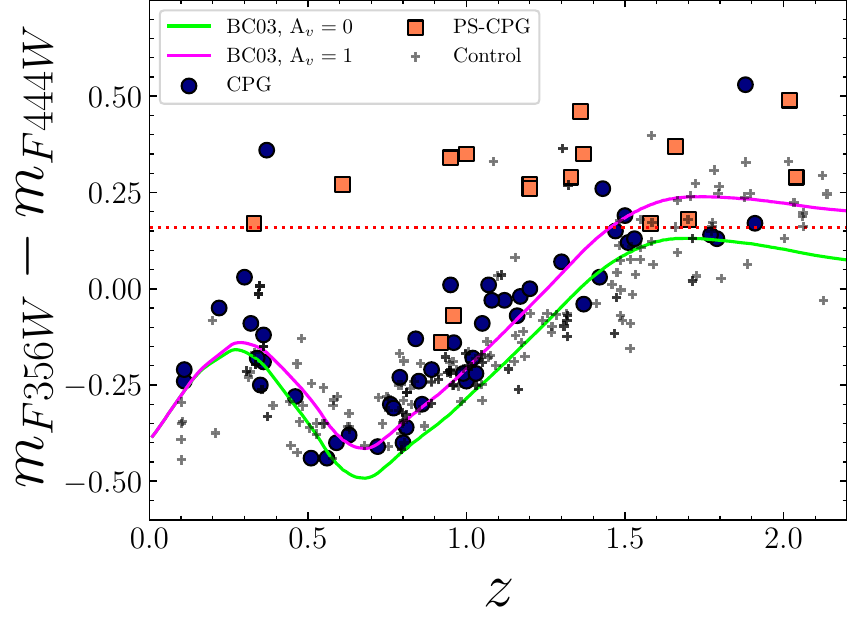}
    \caption{($m_{\rm F356W,AB}-m_{\rm F444W,AB}$) color  versus \cigale-computed photometric redshift. Navy blue circles and orange squares identify CPGs and PS-CPGs, respectively. Plus signs represent galaxies in the control sample. The green  and  magenta lines show the colors of a galaxy that formed at $z=5$ and has $A_{V}=0$ and $A_{V}=1$, respectively (\citealt{Calzetti2000} extinction law). The galaxy model  is a \citet{Bruzual2003} stellar population with Salpeter IMF, solar metallicity, and $\tau=0.5$~Gyr exponentially declining star-formation history. The horizontal dotted line at 0.16 shows the $(m_{\rm F356W,Vega}-m_{\rm F444W,Vega})=0.8$ AGN color-selection criterion in AB mag\added{, replicating (\textsl{W1}--\textsl{W2}) AGN colors \citep[]{Assef2013}.}}
    \label{fig:color_v_z}
\end{figure}

Another standard AGN selection method is red ($m_{\rm F356W}-m_{\rm F444W}$) color \citep[\eg][]{Stern2005}. The NIRCam F356W and F444W filters are close to the WISE W1 and W2 filters \citep{Wright2010} and can replicate the $\rm ({\textsl{W1}}-{\textsl{W2}})>0.80$ (Vega magnitudes) AGN color selection \citep[\eg][]{Stern2012}. There are 19 CPGs with ($m_{\rm F356W,Vega}-m_{\rm F444W,Vega}) > 0.80$ as shown in Figure~\ref{fig:color_v_z}. (Table~\ref{table:catalog} gives the CPG colors in 
AB-mag.) Nearly all remaining CPGs and control-sample galaxies have integrated colors consistent with a pure stellar population that formed early and evolved passively. These colors do not preclude a modest AGN or star-formation contribution, especially for higher redshift CPGs. \added{However, PS-CPGs demonstrate a strong preference for colors consistent with an AGN \added{(i.e., ($m_{\rm F356W}-m_{\rm F444W}) > 0.16$~mag)}, further suggesting that the near-infrared point-source feature comes from a visible AGN\null. This is most likely because dust obscuration decreases at the longer wavelengths \citep{Kim2019}, resulting in redder colors because an AGN is shining through the galaxy. This is consistent with the point-source feature often dominating the reddest wavelengths of our observations \added{(Figure~\ref{fig:RGB_fig})}.}

\section{Summary \& Future Prospects}
\label{sec:discussion}

The superb angular resolution of \JWST/NIRCam reveals galaxies (``CPGs'') with point-like cores at the redder wavelengths. 
Based on \galfit\ analysis, 16/66 CPGs have a point-source core. This is a much higher frequency of point-source cores than in a control sample.

Best-fit SEDs for the CPGs in the \JWST\ NEP--TDF field suggest that the CPGs are massive and luminous, and in most cases, the extended galaxy outshines the unresolved core. The SEDs are well-characterized with AGN components, though the CPG sample is not AGN-dominated.
\added{However, the core classification from \galfit{} suggests that at least 14 PS-CPGs host IR-luminous AGN for the following reasons: 
\begin{itemize}
\item 14/16 are classified as AGN via their $m_{\rm F356W}-m_{\rm F444W}$ colors;
\item \galfit{} prefers  a point-source nucleus over a compact stellar bulge in the nucleus;
\item \cigale\ finds $\langle f_{\rm AGN}\rangle= 0.47$, higher than for the control sample; and
\item All display the point-source signatures in their images that motivated this work (Figure~\ref{fig:RGB_fig}).
\end{itemize}}

Photometry at additional wavelengths and spectroscopy would aid in classifying these objects, confirming AGN presence, constraining host-galaxy parameters, and characterizing the mechanisms driving the pointlike galaxy cores for the entire CPG sample. \added{More robust \galfit{} analysis could clarify which are either PS-CPGs or Bulge-CPGs (e.g., CPG ID 48 was classified as a Bulge-CPG despite  displaying an obvious point-source signature).} Additionally, cross correlation of our CPGs with X-ray detections  will confirm their AGN nature and measure the nuclear luminosity and absorption. This work has identified AGN only at 4.4\,$\mu$m. Future work could extend this morphological classification  to shorter wavelength \JWST\ data to improve the angular resolution and make it more broadly applicable---despite the larger effects from dust at the shorter wavelengths---and to test at what wavelength an automated morphological approach would break down.

The initial visual sample selection revives the classical approach to  AGN identification via a bright, star-like nucleus. This simple selection method is possible because of \JWST's superb angular resolution at long wavelengths, where extinction is much lower than in visible light. Taking advantage of the near-infrared surface brightness and morphology of our CPG sample made it possible to quantify and automate the morphological approach. 
Future simulations and modeling can refine this procedure to constrain where the PS-CPG separation from brighter galaxies and stars can confidently exist.
Even as a prototype, this automated sample selection of brighter galaxies with point-source features streamlines the search and analysis of AGN in current and future \JWST/NIRCam imaging. 

\begin{acknowledgements}
ROIII dedicates this work to the glory of God who makes all things possible.
ROIII acknowledges support from an undergraduate Arizona NASA Space Grant, Cooperative Agreement 80NSSC20M0041. RAW, SHC, and RAJ acknowledge support from NASA JWST Interdisciplinary Scientist grants NAG5-12460, NNX14AN10G, and 80NSSC18K0200 from GSFC\null. This work is based on observations made with the NASA/ESA/CSA James Webb Space Telescope. The data were obtained from the Mikulski Archive for Space Telescopes (MAST) at the Space Telescope Science Institute, which is operated by the Association of Universities for Research in Astronomy, Inc., under NASA contract NAS 5-03127 for JWST\null. These observations are associated with JWST programs 1176, 2736, and 2738. AZ acknowledges support by Grant No.\ 2020750 from the United States--Israel Binational Science Foundation (BSF) and Grant No.\ 2109066 from the United States National Science Foundation (NSF); by the Ministry of Science \& Technology, Israel; and by the Israel Science Foundation Grant No.\ 864/23. HBH and SNM also acknowledge support from NASA JWST Interdisciplinary Scientist grant 21-SMDSS21-0013.
\end{acknowledgements}

All the \JWST\ data used in this paper can be found in MAST: \dataset[10.17909/jtd6-af15]{http://dx.doi.org/10.17909/jtd6-af15}

\software{\astropy{} \citep{Astropy2013,astropy2018, Astropy2022}; \cigale{} \citep{cigale}; \sextractor{} \citep{sextractor}; \eazy{} \citep{eaZy}; \added{\galfit{} \citep{Peng2002}; \texttt{WebbPSF} \citep{webbpsf}}}

\facilities{James Webb Space Telescope (\JWST/NIRCam); Hubble Space Telescope, Very Large Array.}


\bibliographystyle{aasjournal}
\bibliography{references}{}

\begin{thebibliography}{}
\expandafter\ifx\csname natexlab\endcsname\relax\def\natexlab#1{#1}\fi
\providecommand{\url}[1]{\href{#1}{#1}}
\providecommand{\dodoi}[1]{doi:~\href{http://doi.org/#1}{\nolinkurl{#1}}}
\providecommand{\doeprint}[1]{\href{http://ascl.net/#1}{\nolinkurl{http://ascl.net/#1}}}
\providecommand{\doarXiv}[1]{\href{https://arxiv.org/abs/#1}{\nolinkurl{https://arxiv.org/abs/#1}}}

\bibitem[{{Antonucci} \& {Miller}(1985)}]{Antonucci1985}
{Antonucci}, R.~R.~J., \& {Miller}, J.~S. 1985, \apj, 297, 621,
  \dodoi{10.1086/163559}

\bibitem[{{Assef} {et~al.}(2013){Assef}, {Stern}, {Kochanek}, {Blain},
  {Brodwin}, {Brown}, {Donoso}, {Eisenhardt}, {Jannuzi}, {Jarrett}, {Stanford},
  {Tsai}, {Wu}, \& {Yan}}]{Assef2013}
{Assef}, R.~J., {Stern}, D., {Kochanek}, C.~S., {et~al.} 2013, \apj, 772, 26,
  \dodoi{10.1088/0004-637X/772/1/26}

\bibitem[{{Astropy Collaboration} {et~al.}(2013){Astropy Collaboration},
  {Robitaille}, {Tollerud}, {Greenfield}, {Droettboom}, {Bray}, {Aldcroft},
  {Davis}, {Ginsburg}, {Price-Whelan}, {Kerzendorf}, {Conley}, {Crighton},
  {Barbary}, {Muna}, {Ferguson}, {Grollier}, {Parikh}, {Nair}, {Unther},
  {Deil}, {Woillez}, {Conseil}, {Kramer}, {Turner}, {Singer}, {Fox}, {Weaver},
  {Zabalza}, {Edwards}, {Azalee Bostroem}, {Burke}, {Casey}, {Crawford},
  {Dencheva}, {Ely}, {Jenness}, {Labrie}, {Lim}, {Pierfederici}, {Pontzen},
  {Ptak}, {Refsdal}, {Servillat}, \& {Streicher}}]{Astropy2013}
{Astropy Collaboration}, {Robitaille}, T.~P., {Tollerud}, E.~J., {et~al.} 2013,
  \aap, 558, A33, \dodoi{10.1051/0004-6361/201322068}

\bibitem[{{Astropy Collaboration} {et~al.}(2018){Astropy Collaboration},
  {Price-Whelan}, {Sip{\H{o}}cz}, {G{\"u}nther}, {Lim}, {Crawford}, {Conseil},
  {Shupe}, {Craig}, {Dencheva}, {Ginsburg}, {VanderPlas}, {Bradley},
  {P{\'e}rez-Su{\'a}rez}, {de Val-Borro}, {Aldcroft}, {Cruz}, {Robitaille},
  {Tollerud}, {Ardelean}, {Babej}, {Bach}, {Bachetti}, {Bakanov}, {Bamford},
  {Barentsen}, {Barmby}, {Baumbach}, {Berry}, {Biscani}, {Boquien}, {Bostroem},
  {Bouma}, {Brammer}, {Bray}, {Breytenbach}, {Buddelmeijer}, {Burke},
  {Calderone}, {Cano Rodr{\'\i}guez}, {Cara}, {Cardoso}, {Cheedella}, {Copin},
  {Corrales}, {Crichton}, {D'Avella}, {Deil}, {Depagne}, {Dietrich}, {Donath},
  {Droettboom}, {Earl}, {Erben}, {Fabbro}, {Ferreira}, {Finethy}, {Fox},
  {Garrison}, {Gibbons}, {Goldstein}, {Gommers}, {Greco}, {Greenfield},
  {Groener}, {Grollier}, {Hagen}, {Hirst}, {Homeier}, {Horton}, {Hosseinzadeh},
  {Hu}, {Hunkeler}, {Ivezi{\'c}}, {Jain}, {Jenness}, {Kanarek}, {Kendrew},
  {Kern}, {Kerzendorf}, {Khvalko}, {King}, {Kirkby}, {Kulkarni}, {Kumar},
  {Lee}, {Lenz}, {Littlefair}, {Ma}, {Macleod}, {Mastropietro}, {McCully},
  {Montagnac}, {Morris}, {Mueller}, {Mumford}, {Muna}, {Murphy}, {Nelson},
  {Nguyen}, {Ninan}, {N{\"o}the}, {Ogaz}, {Oh}, {Parejko}, {Parley}, {Pascual},
  {Patil}, {Patil}, {Plunkett}, {Prochaska}, {Rastogi}, {Reddy Janga},
  {Sabater}, {Sakurikar}, {Seifert}, {Sherbert}, {Sherwood-Taylor}, {Shih},
  {Sick}, {Silbiger}, {Singanamalla}, {Singer}, {Sladen}, {Sooley},
  {Sornarajah}, {Streicher}, {Teuben}, {Thomas}, {Tremblay}, {Turner},
  {Terr{\'o}n}, {van Kerkwijk}, {de la Vega}, {Watkins}, {Weaver}, {Whitmore},
  {Woillez}, {Zabalza}, \& {Astropy Contributors}}]{astropy2018}
{Astropy Collaboration}, {Price-Whelan}, A.~M., {Sip{\H{o}}cz}, B.~M., {et~al.}
  2018, \aj, 156, 123, \dodoi{10.3847/1538-3881/aabc4f}

\bibitem[{{Astropy Collaboration} {et~al.}(2022){Astropy Collaboration},
  {Price-Whelan}, {Lim}, {Earl}, {Starkman}, {Bradley}, {Shupe}, {Patil},
  {Corrales}, {Brasseur}, {N{\"o}the}, {Donath}, {Tollerud}, {Morris},
  {Ginsburg}, {Vaher}, {Weaver}, {Tocknell}, {Jamieson}, {van Kerkwijk},
  {Robitaille}, {Merry}, {Bachetti}, {G{\"u}nther}, {Aldcroft},
  {Alvarado-Montes}, {Archibald}, {B{\'o}di}, {Bapat}, {Barentsen},
  {Baz{\'a}n}, {Biswas}, {Boquien}, {Burke}, {Cara}, {Cara}, {Conroy},
  {Conseil}, {Craig}, {Cross}, {Cruz}, {D'Eugenio}, {Dencheva}, {Devillepoix},
  {Dietrich}, {Eigenbrot}, {Erben}, {Ferreira}, {Foreman-Mackey}, {Fox},
  {Freij}, {Garg}, {Geda}, {Glattly}, {Gondhalekar}, {Gordon}, {Grant},
  {Greenfield}, {Groener}, {Guest}, {Gurovich}, {Handberg}, {Hart},
  {Hatfield-Dodds}, {Homeier}, {Hosseinzadeh}, {Jenness}, {Jones}, {Joseph},
  {Kalmbach}, {Karamehmetoglu}, {Ka{\l}uszy{\'n}ski}, {Kelley}, {Kern},
  {Kerzendorf}, {Koch}, {Kulumani}, {Lee}, {Ly}, {Ma}, {MacBride}, {Maljaars},
  {Muna}, {Murphy}, {Norman}, {O'Steen}, {Oman}, {Pacifici}, {Pascual},
  {Pascual-Granado}, {Patil}, {Perren}, {Pickering}, {Rastogi}, {Roulston},
  {Ryan}, {Rykoff}, {Sabater}, {Sakurikar}, {Salgado}, {Sanghi}, {Saunders},
  {Savchenko}, {Schwardt}, {Seifert-Eckert}, {Shih}, {Jain}, {Shukla}, {Sick},
  {Simpson}, {Singanamalla}, {Singer}, {Singhal}, {Sinha}, {Sip{\H{o}}cz},
  {Spitler}, {Stansby}, {Streicher}, {{\v{S}}umak}, {Swinbank}, {Taranu},
  {Tewary}, {Tremblay}, {de Val-Borro}, {Van Kooten}, {Vasovi{\'c}}, {Verma},
  {de Miranda Cardoso}, {Williams}, {Wilson}, {Winkel}, {Wood-Vasey}, {Xue},
  {Yoachim}, {Zhang}, {Zonca}, \& {Astropy Project Contributors}}]{Astropy2022}
{Astropy Collaboration}, {Price-Whelan}, A.~M., {Lim}, P.~L., {et~al.} 2022,
  \apj, 935, 167, \dodoi{10.3847/1538-4357/ac7c74}

\bibitem[{{Bertin} \& {Arnouts}(1996)}]{sextractor}
{Bertin}, E., \& {Arnouts}, S. 1996, A\&AS, 117, 393,
  \dodoi{10.1051/aas:1996164}

\bibitem[{Bollati {et~al.}(2023)Bollati, Lupi, Dotti, \&
  Haardt}]{bollati2023connection}
Bollati, F., Lupi, A., Dotti, M., \& Haardt, F. 2023, On the connection between
  AGN radiative feedback and massive black hole spin.
\newblock \doarXiv{2311.07576}

\bibitem[{{Boquien} {et~al.}(2019){Boquien}, {Burgarella}, {Roehlly}, {Buat},
  {Ciesla}, {Corre}, {Inoue}, \& {Salas}}]{cigale}
{Boquien}, M., {Burgarella}, D., {Roehlly}, Y., {et~al.} 2019, \aap, 622, A103,
  \dodoi{10.1051/0004-6361/201834156}

\bibitem[{{Brammer} {et~al.}(2008){Brammer}, {van Dokkum}, \& {Coppi}}]{eaZy}
{Brammer}, G.~B., {van Dokkum}, P.~G., \& {Coppi}, P. 2008, \apj, 686, 1503,
  \dodoi{10.1086/591786}

\bibitem[{{Brown} {et~al.}(2019){Brown}, {Duncan}, {Landt}, {Kirk}, {Ricci},
  {Kamraj}, {Salvato}, \& {Ananna}}]{BrownAGNATLAS}
{Brown}, M.~J.~I., {Duncan}, K.~J., {Landt}, H., {et~al.} 2019, \mnras, 489,
  3351, \dodoi{10.1093/mnras/stz2324}

\bibitem[{{Bruce} {et~al.}(2016){Bruce}, {Dunlop}, {Mortlock}, {Kocevski},
  {McGrath}, \& {Rosario}}]{Bruce2016}
{Bruce}, V.~A., {Dunlop}, J.~S., {Mortlock}, A., {et~al.} 2016, \mnras, 458,
  2391, \dodoi{10.1093/mnras/stw467}

\bibitem[{{Bruzual} \& {Charlot}(2003)}]{Bruzual2003}
{Bruzual}, G., \& {Charlot}, S. 2003, \mnras, 344, 1000,
  \dodoi{10.1046/j.1365-8711.2003.06897.x}

\bibitem[{{Burbidge} {et~al.}(1963){Burbidge}, {Burbidge}, \&
  {Prendergast}}]{Burbidge1963}
{Burbidge}, E.~M., {Burbidge}, G.~R., \& {Prendergast}, K.~H. 1963, \apj, 138,
  375, \dodoi{10.1086/147652}

\bibitem[{{Burke} {et~al.}(2024){Burke}, {Liu}, \& {Shen}}]{Burke2024}
{Burke}, C.~J., {Liu}, X., \& {Shen}, Y. 2024, \mnras, 527, 5356,
  \dodoi{10.1093/mnras/stad3592}

\bibitem[{{Cales} \& {Brotherton}(2015)}]{Cales2015}
{Cales}, S.~L., \& {Brotherton}, M.~S. 2015, \mnras, 449, 2374,
  \dodoi{10.1093/mnras/stv370}

\bibitem[{{Calzetti} {et~al.}(2000){Calzetti}, {Armus}, {Bohlin}, {Kinney},
  {Koornneef}, \& {Storchi-Bergmann}}]{Calzetti2000}
{Calzetti}, D., {Armus}, L., {Bohlin}, R.~C., {et~al.} 2000, \apj, 533, 682,
  \dodoi{10.1086/308692}

\bibitem[{{Chevalier}(1982)}]{Chevalier1982}
{Chevalier}, R.~A. 1982, \apj, 259, 302, \dodoi{10.1086/160167}

\bibitem[{{Ciesla} {et~al.}(2015){Ciesla}, {Charmandaris}, {Georgakakis},
  {Bernhard}, {Mitchell}, {Buat}, {Elbaz}, {LeFloc'h}, {Lacey}, {Magdis}, \&
  {Xilouris}}]{Ciesla2015}
{Ciesla}, L., {Charmandaris}, V., {Georgakakis}, A., {et~al.} 2015, \aap, 576,
  A10, \dodoi{10.1051/0004-6361/201425252}

\bibitem[{{Coe} {et~al.}(2012){Coe}, {Umetsu}, {Zitrin}, {Donahue},
  {Medezinski}, {Postman}, {Carrasco}, {Anguita}, {Geller}, {Rines},
  {Diaferio}, {Kurtz}, {Bradley}, {Koekemoer}, {Zheng}, {Nonino}, {Molino},
  {Mahdavi}, {Lemze}, {Infante}, {Ogaz}, {Melchior}, {Host}, {Ford}, {Grillo},
  {Rosati}, {Jim{\'e}nez-Teja}, {Moustakas}, {Broadhurst}, {Ascaso}, {Lahav},
  {Bartelmann}, {Ben{\'\i}tez}, {Bouwens}, {Graur}, {Graves}, {Jha}, {Jouvel},
  {Kelson}, {Moustakas}, {Maoz}, {Meneghetti}, {Merten}, {Riess}, {Rodney}, \&
  {Seitz}}]{Coe2012}
{Coe}, D., {Umetsu}, K., {Zitrin}, A., {et~al.} 2012, \apj, 757, 22,
  \dodoi{10.1088/0004-637X/757/1/22}

\bibitem[{{Condon}(1992)}]{Condon1992}
{Condon}, J.~J. 1992, \araa, 30, 575,
  \dodoi{10.1146/annurev.aa.30.090192.003043}

\bibitem[{{Conselice}(2003)}]{Conselice2003}
{Conselice}, C.~J. 2003, \apjs, 147, 1, \dodoi{10.1086/375001}

\bibitem[{Costa-Souza {et~al.}(2023)Costa-Souza, Riffel, Dors, Riffel, \&
  da~Rocha-Poppe}]{costasouza2023spatially}
Costa-Souza, J.~H., Riffel, R.~A., Dors, O.~L., Riffel, R., \& da~Rocha-Poppe,
  P.~C. 2023, Spatially resolved observations of the peculiar galaxy NGC 232:
  AGN winds and stellar populations.
\newblock \doarXiv{2310.15842}

\bibitem[{{Dale} {et~al.}(2014){Dale}, {Helou}, {Magdis}, {Armus},
  {D{\'\i}az-Santos}, \& {Shi}}]{Dale2014}
{Dale}, D.~A., {Helou}, G., {Magdis}, G.~E., {et~al.} 2014, \apj, 784, 83,
  \dodoi{10.1088/0004-637X/784/1/83}

\bibitem[{{Davies} {et~al.}(2024){Davies}, {Belli}, {Park}, {Mendel},
  {Johnson}, {Conroy}, {Benton}, {Bugiani}, {Emami}, {Leja}, {Li}, {Maheson},
  {Mathews}, {Naidu}, {Nelson}, {Tacchella}, {Terrazas}, \&
  {Weinberger}}]{Davies2024}
{Davies}, R.~L., {Belli}, S., {Park}, M., {et~al.} 2024, \mnras, 528, 4976,
  \dodoi{10.1093/mnras/stae327}

\bibitem[{{Elvis} {et~al.}(1978){Elvis}, {Maccacaro}, {Wilson}, {Ward},
  {Penston}, {Fosbury}, \& {Perola}}]{Elvis1978}
{Elvis}, M., {Maccacaro}, T., {Wilson}, A.~S., {et~al.} 1978, \mnras, 183, 129,
  \dodoi{10.1093/mnras/183.2.129}

\bibitem[{{Fanaroff} \& {Riley}(1974)}]{Fanaroff1974}
{Fanaroff}, B.~L., \& {Riley}, J.~M. 1974, \mnras, 167, 31P,
  \dodoi{10.1093/mnras/167.1.31P}

\bibitem[{{Furtak} {et~al.}(2023){Furtak}, {Labb{\'e}}, {Zitrin}, {Greene},
  {Dayal}, {Chemerynska}, {Kokorev}, {Miller}, {Goulding}, {Bezanson},
  {Brammer}, {Cutler}, {Leja}, {Pan}, {Price}, {Wang}, {Weaver}, {Whitaker},
  {Atek}, {Bogd{\'a}n}, {Charlot}, {Curtis-Lake}, {van Dokkum}, {Endsley},
  {Fudamoto}, {Fujimoto}, {de Graaff}, {Glazebrook}, {Juneau}, {Marchesini},
  {Maseda}, {Nelson}, {Oesch}, {Plat}, {Setton}, {Stark}, \&
  {Williams}}]{FurtakColors2023}
{Furtak}, L.~J., {Labb{\'e}}, I., {Zitrin}, A., {et~al.} 2023, arXiv e-prints,
  arXiv:2308.05735, \dodoi{10.48550/arXiv.2308.05735}

\bibitem[{{Glikman} {et~al.}(2012){Glikman}, {Lacy}, {Urrutia}, {Djorgovski},
  \& {Mahabal}}]{Glikman2012}
{Glikman}, E., {Lacy}, M., {Urrutia}, T., {Djorgovski}, G., \& {Mahabal}, A.
  2012, in American Astronomical Society Meeting Abstracts, Vol. 219, American
  Astronomical Society Meeting Abstracts \#219, 209.03

\bibitem[{{Hickox} \& {Alexander}(2018)}]{Hickox2018}
{Hickox}, R.~C., \& {Alexander}, D.~M. 2018, \araa, 56, 625,
  \dodoi{10.1146/annurev-astro-081817-051803}

\bibitem[{{Hwang} {et~al.}(2021){Hwang}, {Wang}, {Chang}, {Lim}, {Chen}, {Gao},
  {Dunlop}, {Gao}, {Ho}, {Hwang}, {Koprowski}, {Micha{\l}owski}, {Peng},
  {Shim}, {Simpson}, \& {Toba}}]{Hwang2021}
{Hwang}, Y.-H., {Wang}, W.-H., {Chang}, Y.-Y., {et~al.} 2021, \apj, 913, 6,
  \dodoi{10.3847/1538-4357/abf11a}

\bibitem[{{Hyun} {et~al.}(2023){Hyun}, {Im}, {Smail}, {Cotton}, {Birkin},
  {Kikuta}, {Shim}, {Willmer}, {Condon}, {Windhorst}, {Cohen}, {Jansen}, {Ly},
  {Matsuda}, {Fazio}, {Swinbank}, \& {Yan}}]{Hyun2023}
{Hyun}, M., {Im}, M., {Smail}, I.~R., {et~al.} 2023, \apjs, 264, 19,
  \dodoi{10.3847/1538-4365/ac9bf4}

\bibitem[{{Inoue}(2011)}]{Inoue2011}
{Inoue}, A.~K. 2011, \mnras, 415, 2920,
  \dodoi{10.1111/j.1365-2966.2011.18906.x}

\bibitem[{Jansen \& Windhorst(2018)}]{Jansen_2018}
Jansen, R.~A., \& Windhorst, R.~A. 2018, \pasp, 130, 124001,
  \dodoi{10.1088/1538-3873/aae476}

\bibitem[{{Juod{\v{z}}balis} {et~al.}(2023){Juod{\v{z}}balis}, {Conselice},
  {Singh}, {Adams}, {Ormerod}, {Harvey}, {Austin}, {Volonteri}, {Cohen},
  {Jansen}, {Summers}, {Windhorst}, {D'Silva}, {Koekemoer}, {Coe}, {Driver},
  {Frye}, {Grogin}, {Marshall}, {Nonino}, {Pirzkal}, {Robotham}, {Ryan},
  {Ortiz}, {Tompkins}, {Willmer}, \& {Yan}}]{Joudz2023}
{Juod{\v{z}}balis}, I., {Conselice}, C.~J., {Singh}, M., {et~al.} 2023, \mnras,
  525, 1353, \dodoi{10.1093/mnras/stad2396}

\bibitem[{{Kellermann}(1987)}]{Kellermann1987}
{Kellermann}, K.~I. 1987, in Observational Evidence of Activity in Galaxies,
  ed. E.~E. {Khachikian}, K.~J. {Fricke}, \& J.~{Melnick}, Vol. 121, 273

\bibitem[{{Kim} {et~al.}(2019){Kim}, {Jansen}, {Windhorst}, {Cohen}, \&
  {McCabe}}]{Kim2019}
{Kim}, D., {Jansen}, R.~A., {Windhorst}, R.~A., {Cohen}, S.~H., \& {McCabe},
  T.~J. 2019, \apj, 884, 21, \dodoi{10.3847/1538-4357/ab385c}

\bibitem[{Li {et~al.}(2023)Li, Conselice, Adams, Trussler, Austin, Harvey,
  Ferreira, Caruana, Ormerod, \& Juodžbalis}]{li2023epochs}
Li, Q., Conselice, C.~J., Adams, N., {et~al.} 2023, EPOCHS VIII. An Insight
  into MIRI-selected Galaxies in SMACS-0723 and the Benefits of Deep MIRI
  Photometry in Revealing AGN and the Dusty Universe.
\newblock \doarXiv{2309.06932}

\bibitem[{{Lyu} {et~al.}(2023){Lyu}, {Alberts}, {Rieke}, {Shivaei},
  {Perez-Gonzalez}, {Sun}, {Hainline}, {Baum}, {Bonaventura}, {Bunker},
  {Egami}, {Eisenstein}, {Florian}, {Ji}, {Johnson}, {Morrison}, {Rieke},
  {Robertson}, {Rujopakarn}, {Tacchella}, {Scholtz}, \& {Willmer}}]{Lyu2023}
{Lyu}, J., {Alberts}, S., {Rieke}, G.~H., {et~al.} 2023, arXiv e-prints,
  arXiv:2310.12330, \dodoi{10.48550/arXiv.2310.12330}

\bibitem[{{Mandal} {et~al.}(2018){Mandal}, {Rakshit}, {Kurian}, {Stalin},
  {Mathew}, {Hoenig}, {Gandhi}, {Sagar}, \& {Pandge}}]{Mandal2018}
{Mandal}, A.~K., {Rakshit}, S., {Kurian}, K.~S., {et~al.} 2018, \mnras, 475,
  5330, \dodoi{10.1093/mnras/sty200}

\bibitem[{{Masini} {et~al.}(2020){Masini}, {Hickox}, {Carroll}, {Aird},
  {Alexander}, {Assef}, {Bower}, {Brodwin}, {Brown}, {Chatterjee}, {Chen},
  {Dey}, {DiPompeo}, {Duncan}, {Eisenhardt}, {Forman}, {Gonzalez}, {Goulding},
  {Hainline}, {Jannuzi}, {Jones}, {Kochanek}, {Kraft}, {Lee}, {Miller},
  {Mullaney}, {Myers}, {Ptak}, {Stanford}, {Stern}, {Vikhlinin}, {Wake}, \&
  {Murray}}]{Masini2020}
{Masini}, A., {Hickox}, R.~C., {Carroll}, C.~M., {et~al.} 2020, \apjs, 251, 2,
  \dodoi{10.3847/1538-4365/abb607}

\bibitem[{{Mehdipour} {et~al.}(2024){Mehdipour}, {Kriss}, {Kaastra},
  {Costantini}, {Gu}, {Landt}, {Mao}, \& {Rogantini}}]{Mehdipour2024}
{Mehdipour}, M., {Kriss}, G.~A., {Kaastra}, J.~S., {et~al.} 2024, \apj, 962,
  155, \dodoi{10.3847/1538-4357/ad1bcb}

\bibitem[{{O'Brien} {et~al.}(2024){O'Brien}, {Jansen}, {Grogin}, {Cohen},
  {Smith}, {Silver}, {Maksym}, {Windhorst}, {Carleton}, {Koekemoer}, {Hathi},
  {Willmer}, {Frye}, {Alpaslan}, {Ashby}, {Ashcraft}, {Bonoli}, {Brisken},
  {Cappelluti}, {Civano}, {Conselice}, {Dhillon}, {Driver}, {Duncan}, {Dupke},
  {Elvis}, {Fazio}, {Finkelstein}, {Gim}, {Griffiths}, {Hammel}, {Hyun}, {Im},
  {Jones}, {Kim}, {Ladjelate}, {Larson}, {Malhotra}, {Marshall}, {Milam},
  {Pierel}, {Rhoads}, {Rodney}, {R{\"o}ttgering}, {Rutkowski}, {Ryan}, {Ward},
  {White}, {van Weeren}, {Zhao}, {Summers}, {D'Silva}, {Ortiz}, {Robotham},
  {Coe}, {Nonino}, {Pirzkal}, {Yan}, \& {Acharya}}]{OBrien2024}
{O'Brien}, R., {Jansen}, R.~A., {Grogin}, N.~A., {et~al.} 2024, arXiv e-prints,
  arXiv:2401.04944, \dodoi{10.48550/arXiv.2401.04944}

\bibitem[{{Ohsuga} \& {Umemura}(2001)}]{Ohsuga2001}
{Ohsuga}, K., \& {Umemura}, M. 2001, \apj, 559, 157, \dodoi{10.1086/322398}

\bibitem[{{Oke} \& {Gunn}(1983)}]{Oke_1983}
{Oke}, J.~B., \& {Gunn}, J.~E. 1983, \apj, 266, 713, \dodoi{10.1086/160817}

\bibitem[{{Osterbrock}(1993)}]{Osterbrock1993}
{Osterbrock}, D.~E. 1993, \rmxaa, 26, 65

\bibitem[{{Padovani} {et~al.}(2017){Padovani}, {Alexander}, {Assef}, {De
  Marco}, {Giommi}, {Hickox}, {Richards}, {Smol{\v{c}}i{\'c}},
  {Hatziminaoglou}, {Mainieri}, \& {Salvato}}]{Padovani2017}
{Padovani}, P., {Alexander}, D.~M., {Assef}, R.~J., {et~al.} 2017, \aapr, 25,
  2, \dodoi{10.1007/s00159-017-0102-9}

\bibitem[{{Peng} {et~al.}(2002){Peng}, {Ho}, {Impey}, \& {Rix}}]{Peng2002}
{Peng}, C.~Y., {Ho}, L.~C., {Impey}, C.~D., \& {Rix}, H.-W. 2002, \aj, 124,
  266, \dodoi{10.1086/340952}

\bibitem[{{Perrin} {et~al.}(2012){Perrin}, {Soummer}, {Elliott}, {Lallo}, \&
  {Sivaramakrishnan}}]{webbpsf}
{Perrin}, M.~D., {Soummer}, R., {Elliott}, E.~M., {Lallo}, M.~D., \&
  {Sivaramakrishnan}, A. 2012, in Society of Photo-Optical Instrumentation
  Engineers (SPIE) Conference Series, Vol. 8442, Space Telescopes and
  Instrumentation 2012: Optical, Infrared, and Millimeter Wave, ed. M.~C.
  {Clampin}, G.~G. {Fazio}, H.~A. {MacEwen}, \& J.~{Oschmann}, Jacobus~M.,
  84423D, \dodoi{10.1117/12.925230}

\bibitem[{{Planck Collaboration} {et~al.}(2016){Planck Collaboration}, {Ade},
  {Aghanim}, {Arnaud}, {Ashdown}, {Aumont}, {Baccigalupi}, {Banday},
  {Barreiro}, {Bartlett}, {Bartolo}, {Battaner}, {Battye}, {Benabed},
  {Beno{\^\i}t}, {Benoit-L{\'e}vy}, {Bernard}, {Bersanelli}, {Bielewicz},
  {Bock}, {Bonaldi}, {Bonavera}, {Bond}, {Borrill}, {Bouchet}, {Boulanger},
  {Bucher}, {Burigana}, {Butler}, {Calabrese}, {Cardoso}, {Catalano},
  {Challinor}, {Chamballu}, {Chary}, {Chiang}, {Chluba}, {Christensen},
  {Church}, {Clements}, {Colombi}, {Colombo}, {Combet}, {Coulais}, {Crill},
  {Curto}, {Cuttaia}, {Danese}, {Davies}, {Davis}, {de Bernardis}, {de Rosa},
  {de Zotti}, {Delabrouille}, {D{\'e}sert}, {Di Valentino}, {Dickinson},
  {Diego}, {Dolag}, {Dole}, {Donzelli}, {Dor{\'e}}, {Douspis}, {Ducout},
  {Dunkley}, {Dupac}, {Efstathiou}, {Elsner}, {En{\ss}lin}, {Eriksen},
  {Farhang}, {Fergusson}, {Finelli}, {Forni}, {Frailis}, {Fraisse},
  {Franceschi}, {Frejsel}, {Galeotta}, {Galli}, {Ganga}, {Gauthier}, {Gerbino},
  {Ghosh}, {Giard}, {Giraud-H{\'e}raud}, {Giusarma}, {Gjerl{\o}w},
  {Gonz{\'a}lez-Nuevo}, {G{\'o}rski}, {Gratton}, {Gregorio}, {Gruppuso},
  {Gudmundsson}, {Hamann}, {Hansen}, {Hanson}, {Harrison}, {Helou},
  {Henrot-Versill{\'e}}, {Hern{\'a}ndez-Monteagudo}, {Herranz}, {Hildebrandt},
  {Hivon}, {Hobson}, {Holmes}, {Hornstrup}, {Hovest}, {Huang}, {Huffenberger},
  {Hurier}, {Jaffe}, {Jaffe}, {Jones}, {Juvela}, {Keih{\"a}nen}, {Keskitalo},
  {Kisner}, {Kneissl}, {Knoche}, {Knox}, {Kunz}, {Kurki-Suonio}, {Lagache},
  {L{\"a}hteenm{\"a}ki}, {Lamarre}, {Lasenby}, {Lattanzi}, {Lawrence}, {Leahy},
  {Leonardi}, {Lesgourgues}, {Levrier}, {Lewis}, {Liguori}, {Lilje},
  {Linden-V{\o}rnle}, {L{\'o}pez-Caniego}, {Lubin}, {Mac{\'\i}as-P{\'e}rez},
  {Maggio}, {Maino}, {Mandolesi}, {Mangilli}, {Marchini}, {Maris}, {Martin},
  {Martinelli}, {Mart{\'\i}nez-Gonz{\'a}lez}, {Masi}, {Matarrese}, {McGehee},
  {Meinhold}, {Melchiorri}, {Melin}, {Mendes}, {Mennella}, {Migliaccio},
  {Millea}, {Mitra}, {Miville-Desch{\^e}nes}, {Moneti}, {Montier}, {Morgante},
  {Mortlock}, {Moss}, {Munshi}, {Murphy}, {Naselsky}, {Nati}, {Natoli},
  {Netterfield}, {N{\o}rgaard-Nielsen}, {Noviello}, {Novikov}, {Novikov},
  {Oxborrow}, {Paci}, {Pagano}, {Pajot}, {Paladini}, {Paoletti}, {Partridge},
  {Pasian}, {Patanchon}, {Pearson}, {Perdereau}, {Perotto}, {Perrotta},
  {Pettorino}, {Piacentini}, {Piat}, {Pierpaoli}, {Pietrobon}, {Plaszczynski},
  {Pointecouteau}, {Polenta}, {Popa}, {Pratt}, {Pr{\'e}zeau}, {Prunet},
  {Puget}, {Rachen}, {Reach}, {Rebolo}, {Reinecke}, {Remazeilles}, {Renault},
  {Renzi}, {Ristorcelli}, {Rocha}, {Rosset}, {Rossetti}, {Roudier},
  {Rouill{\'e} d'Orfeuil}, {Rowan-Robinson}, {Rubi{\~n}o-Mart{\'\i}n},
  {Rusholme}, {Said}, {Salvatelli}, {Salvati}, {Sandri}, {Santos},
  {Savelainen}, {Savini}, {Scott}, {Seiffert}, {Serra}, {Shellard}, {Spencer},
  {Spinelli}, {Stolyarov}, {Stompor}, {Sudiwala}, {Sunyaev}, {Sutton},
  {Suur-Uski}, {Sygnet}, {Tauber}, {Terenzi}, {Toffolatti}, {Tomasi},
  {Tristram}, {Trombetti}, {Tucci}, {Tuovinen}, {T{\"u}rler}, {Umana},
  {Valenziano}, {Valiviita}, {Van Tent}, {Vielva}, {Villa}, {Wade}, {Wandelt},
  {Wehus}, {White}, {White}, {Wilkinson}, {Yvon}, {Zacchei}, \&
  {Zonca}}]{PlanckCollaboration2016}
{Planck Collaboration}, {Ade}, P.~A.~R., {Aghanim}, N., {et~al.} 2016, \aap,
  594, A13, \dodoi{10.1051/0004-6361/201525830}

\bibitem[{{Planck Collaboration} {et~al.}(2020){Planck Collaboration},
  {Aghanim}, {Akrami}, {Ashdown}, {Aumont}, {Baccigalupi}, {Ballardini},
  {Banday}, {Barreiro}, {Bartolo}, {Basak}, {Battye}, {Benabed}, {Bernard},
  {Bersanelli}, {Bielewicz}, {Bock}, {Bond}, {Borrill}, {Bouchet}, {Boulanger},
  {Bucher}, {Burigana}, {Butler}, {Calabrese}, {Cardoso}, {Carron},
  {Challinor}, {Chiang}, {Chluba}, {Colombo}, {Combet}, {Contreras}, {Crill},
  {Cuttaia}, {de Bernardis}, {de Zotti}, {Delabrouille}, {Delouis}, {Di
  Valentino}, {Diego}, {Dor{\'e}}, {Douspis}, {Ducout}, {Dupac}, {Dusini},
  {Efstathiou}, {Elsner}, {En{\ss}lin}, {Eriksen}, {Fantaye}, {Farhang},
  {Fergusson}, {Fernandez-Cobos}, {Finelli}, {Forastieri}, {Frailis},
  {Fraisse}, {Franceschi}, {Frolov}, {Galeotta}, {Galli}, {Ganga},
  {G{\'e}nova-Santos}, {Gerbino}, {Ghosh}, {Gonz{\'a}lez-Nuevo}, {G{\'o}rski},
  {Gratton}, {Gruppuso}, {Gudmundsson}, {Hamann}, {Handley}, {Hansen},
  {Herranz}, {Hildebrandt}, {Hivon}, {Huang}, {Jaffe}, {Jones}, {Karakci},
  {Keih{\"a}nen}, {Keskitalo}, {Kiiveri}, {Kim}, {Kisner}, {Knox},
  {Krachmalnicoff}, {Kunz}, {Kurki-Suonio}, {Lagache}, {Lamarre}, {Lasenby},
  {Lattanzi}, {Lawrence}, {Le Jeune}, {Lemos}, {Lesgourgues}, {Levrier},
  {Lewis}, {Liguori}, {Lilje}, {Lilley}, {Lindholm}, {L{\'o}pez-Caniego},
  {Lubin}, {Ma}, {Mac{\'\i}as-P{\'e}rez}, {Maggio}, {Maino}, {Mandolesi},
  {Mangilli}, {Marcos-Caballero}, {Maris}, {Martin}, {Martinelli},
  {Mart{\'\i}nez-Gonz{\'a}lez}, {Matarrese}, {Mauri}, {McEwen}, {Meinhold},
  {Melchiorri}, {Mennella}, {Migliaccio}, {Millea}, {Mitra},
  {Miville-Desch{\^e}nes}, {Molinari}, {Montier}, {Morgante}, {Moss}, {Natoli},
  {N{\o}rgaard-Nielsen}, {Pagano}, {Paoletti}, {Partridge}, {Patanchon},
  {Peiris}, {Perrotta}, {Pettorino}, {Piacentini}, {Polastri}, {Polenta},
  {Puget}, {Rachen}, {Reinecke}, {Remazeilles}, {Renzi}, {Rocha}, {Rosset},
  {Roudier}, {Rubi{\~n}o-Mart{\'\i}n}, {Ruiz-Granados}, {Salvati}, {Sandri},
  {Savelainen}, {Scott}, {Shellard}, {Sirignano}, {Sirri}, {Spencer},
  {Sunyaev}, {Suur-Uski}, {Tauber}, {Tavagnacco}, {Tenti}, {Toffolatti},
  {Tomasi}, {Trombetti}, {Valenziano}, {Valiviita}, {Van Tent}, {Vibert},
  {Vielva}, {Villa}, {Vittorio}, {Wandelt}, {Wehus}, {White}, {White},
  {Zacchei}, \& {Zonca}}]{PlanckCollaboration2018}
{Planck Collaboration}, {Aghanim}, N., {Akrami}, Y., {et~al.} 2020, \aap, 641,
  A6, \dodoi{10.1051/0004-6361/201833910}

\bibitem[{{Pouliasis} {et~al.}(2019){Pouliasis}, {Georgantopoulos}, {Bonanos},
  {Yang}, {Sokolovsky}, {Hatzidimitriou}, {Mountrichas}, {Gavras},
  {Charmandaris}, {Bellas-Velidis}, {Spetsieri}, \&
  {Tsinganos}}]{Pouliasis2019}
{Pouliasis}, E., {Georgantopoulos}, I., {Bonanos}, A.~Z., {et~al.} 2019,
  \mnras, 487, 4285, \dodoi{10.1093/mnras/stz1483}

\bibitem[{{Renzini} \& {Peng}(2015)}]{Renzini2015}
{Renzini}, A., \& {Peng}, Y.-j. 2015, \apjl, 801, L29,
  \dodoi{10.1088/2041-8205/801/2/L29}

\bibitem[{{Richards} {et~al.}(2003){Richards}, {Hall}, {Vanden Berk},
  {Strauss}, {Schneider}, {Weinstein}, {Reichard}, {York}, {Knapp}, {Fan},
  {Ivezi{\'c}}, {Brinkmann}, {Budav{\'a}ri}, {Csabai}, \&
  {Nichol}}]{Richards2003}
{Richards}, G.~T., {Hall}, P.~B., {Vanden Berk}, D.~E., {et~al.} 2003, \aj,
  126, 1131, \dodoi{10.1086/377014}

\bibitem[{{Rigby} {et~al.}(2023){Rigby}, {Perrin}, {McElwain}, {Kimble},
  {Friedman}, {Lallo}, {Doyon}, {Feinberg}, {Ferruit}, {Glasse}, {Rieke},
  {Rieke}, {Wright}, {Willott}, {Colon}, {Milam}, {Neff}, {Stark}, {Valenti},
  {Abell}, {Abney}, {Abul-Huda}, {Acton}, {Adams}, {Adler}, {Aguilar}, {Ahmed},
  {Albert}, {Alberts}, {Aldridge}, {Allen}, {Altenburg},
  {{\'A}lvarez-M{\'a}rquez}, {Alves de Oliveira}, {Andersen}, {Anderson},
  {Anderson}, {Argyriou}, {Armstrong}, {Arribas}, {Artigau}, {Arvai},
  {Atkinson}, {Bacon}, {Bair}, {Banks}, {Barrientes}, {Barringer}, {Bartosik},
  {Bast}, {Baudoz}, {Beatty}, {Bechtold}, {Beck}, {Bergeron}, {Bergkoetter},
  {Bhatawdekar}, {Birkmann}, {Blazek}, {Blome}, {Boccaletti}, {B{\"o}ker},
  {Boia}, {Bonaventura}, {Bond}, {Bosley}, {Boucarut}, {Bourque}, {Bouwman},
  {Bower}, {Bowers}, {Boyer}, {Bradley}, {Brady}, {Braun}, {Breda},
  {Bresnahan}, {Bright}, {Britt}, {Bromenschenkel}, {Brooks}, {Brooks},
  {Brown}, {Brown}, {Brown}, {Bunker}, {Burger}, {Bushouse}, {Cale}, {Cameron},
  {Cameron}, {Canipe}, {Caplinger}, {Caputo}, {Cara}, {Carey}, {Carniani},
  {Carrasquilla}, {Carruthers}, {Case}, {Catherine}, {Chance}, {Chapman},
  {Charlot}, {Charlow}, {Chayer}, {Chen}, {Cherinka}, {Chichester}, {Chilton},
  {Chonis}, {Clampin}, {Clark}, {Clark}, {Coe}, {Coleman}, {Comber}, {Comeau},
  {Connolly}, {Cooper}, {Cooper}, {Coppock}, {Correnti}, {Cossou}, {Coulais},
  {Coyle}, {Cracraft}, {Curti}, {Cuturic}, {Davis}, {Davis}, {Dean}, {DeLisa},
  {deMeester}, {Dencheva}, {Dencheva}, {DePasquale}, {Deschenes}, {Hunor
  Detre}, {Diaz}, {Dicken}, {DiFelice}, {Dillman}, {Dixon}, {Doggett},
  {Donaldson}, {Douglas}, {DuPrie}, {Dupuis}, {Durning}, {Easmin}, {Eck},
  {Edeani}, {Egami}, {Ehrenwinkler}, {Eisenhamer}, {Eisenhower}, {Elie},
  {Elliott}, {Elliott}, {Ellis}, {Engesser}, {Espinoza}, {Etienne}, {Etxaluze},
  {Falini}, {Feeney}, {Ferry}, {Filippazzo}, {Fincham}, {Fix}, {Flagey},
  {Florian}, {Flynn}, {Fontanella}, {Ford}, {Forshay}, {Fox}, {Franz}, {Fu},
  {Fullerton}, {Galkin}, {Galyer}, {Garc{\'\i}a Mar{\'\i}n}, {Gardner},
  {Gardner}, {Garland}, {Garrett}, {Gasman}, {Gaspar}, {Gaudreau}, {Gauthier},
  {Geers}, {Geithner}, {Gennaro}, {Giardino}, {Girard}, {Giuliano},
  {Glassmire}, {Glauser}, {Glazer}, {Godfrey}, {Golimowski}, {Gollnitz},
  {Gong}, {Gonzaga}, {Gordon}, {Gordon}, {Goudfrooij}, {Greene}, {Greenhouse},
  {Grimaldi}, {Groebner}, {Grundy}, {Guillard}, {Gutman}, {Ha}, {Haderlein},
  {Hagedorn}, {Hainline}, {Haley}, {Hami}, {Hamilton}, {Hammel}, {Hansen},
  {Harkins}, {Harr}, {Hart}, {Hart}, {Hartig}, {Hashimoto}, {Haskins},
  {Hathaway}, {Havey}, {Hayden}, {Hecht}, {Heller-Boyer}, {Henriques}, {Henry},
  {Hermann}, {Hernandez}, {Hesman}, {Hicks}, {Hilbert}, {Hines}, {Hoffman},
  {Holfeltz}, {Holler}, {Hoppa}, {Hott}, {Howard}, {Howard}, {Hunter},
  {Hunter}, {Hurst}, {Husemann}, {Hustak}, {Ilinca Ignat}, {Illingworth},
  {Irish}, {Jackson}, {Jahromi}, {Jakobsen}, {James}, {James}, {Januszewski},
  {Jenkins}, {Jirdeh}, {Johnson}, {Johnson}, {Jones}, {Jones}, {Jones},
  {Jones}, {Jordan}, {Jordan}, {Jurczyk}, {Jurling}, {Kaleida}, {Kalmanson},
  {Kammerer}, {Kang}, {Kao}, {Karakla}, {Kavanagh}, {Kelly}, {Kendrew},
  {Kennedy}, {Kenny}, {Keski-kuha}, {Keyes}, {Kidwell}, {Kinzel}, {Kirk},
  {Kirkpatrick}, {Kirshenblat}, {Klaassen}, {Knapp}, {Knight}, {Knollenberg},
  {Koehler}, {Koekemoer}, {Kovacs}, {Kulp}, {Kumari}, {Kyprianou}, {La Massa},
  {Labador}, {Labiano}, {Lagage}, {Lajoie}, {Lallo}, {Lam}, {Lamb}, {Lambros},
  {Lampenfield}, {Langston}, {Larson}, {Law}, {Lawrence}, {Lee}, {Leisenring},
  {Lepo}, {Leveille}, {Levenson}, {Levine}, {Levy}, {Lewis}, {Lewis},
  {Libralato}, {Lightsey}, {Link}, {Liu}, {Lo}, {Lockwood}, {Logue}, {Long},
  {Long}, {Loomis}, {Lopez-Caniego}, {Lorenzo Alvarez}, {Love-Pruitt}, {Lucy},
  {Luetzgendorf}, {Maghami}, {Maiolino}, {Major}, {Malla}, {Malumuth},
  {Manjavacas}, {Mannfolk}, {Marrione}, {Marston}, {Martel}, {Maschmann},
  {Masci}, {Masciarelli}, {Maszkiewicz}, {Mather}, {McKenzie}, {McLean},
  {McMaster}, {Melbourne}, {Mel{\'e}ndez}, {Menzel}, {Merz}, {Meyett}, {Meza},
  {Miskey}, {Misselt}, {Moller}, {Morrison}, {Morse}, {Moseley}, {Mosier},
  {Mountain}, {Mueckay}, {Mueller}, {Mullally}, {Murphy}, {Murray}, {Murray},
  {Mustelier}, {Muzerolle}, {Mycroft}, {Myers}, {Myrick}, {Nanavati}, {Nance},
  {Nayak}, {Naylor}, {Nelan}, {Nickson}, {Nielson}, {Nieto-Santisteban},
  {Nikolov}, {Noriega-Crespo}, {O'Shaughnessy}, {O'Sullivan}, {Ochs}, {Ogle},
  {Oleszczuk}, {Olmsted}, {Osborne}, {Ottens}, {Owens}, {Pacifici}, {Pagan},
  {Page}, {Park}, {Parrish}, {Patapis}, {Paul}, {Pauly}, {Pavlovsky}, {Pedder},
  {Peek}, {Pena-Guerrero}, {Penanen}, {Perez}, {Perna}, {Perriello},
  {Phillips}, {Pietraszkiewicz}, {Pinaud}, {Pirzkal}, {Pitman}, {Piwowar},
  {Platais}, {Player}, {Plesha}, {Pollizi}, {Polster}, {Pontoppidan},
  {Porterfield}, {Proffitt}, {Pueyo}, {Pulliam}, {Quirt}, {Quispe Neira},
  {Ramos Alarcon}, {Ramsay}, {Rapp}, {Rapp}, {Rauscher}, {Ravindranath},
  {Rawle}, {Regan}, {Reichard}, {Reis}, {Ressler}, {Rest}, {Reynolds}, {Rhue},
  {Richon}, {Rickman}, {Ridgaway}, {Ritchie}, {Rix}, {Robberto}, {Robinson},
  {Robinson}, {Robinson}, {Rock}, {Rodriguez}, {Rodriguez Del Pino}, {Roellig},
  {Rohrbach}, {Roman}, {Romelfanger}, {Rose}, {Roteliuk}, {Roth}, {Rothwell},
  {Rowlands}, {Roy}, {Royer}, {Royle}, {Rui}, {Rumler}, {Runnels}, {Russ},
  {Rustamkulov}, {Ryden}, {Ryer}, {Sabata}, {Sabatke}, {Sabbi}, {Samuelson},
  {Sapp}, {Sappington}, {Sargent}, {Sauer}, {Scheithauer}, {Schlawin},
  {Schlitz}, {Schmitz}, {Schneider}, {Schreiber}, {Schulze}, {Schwab}, {Scott},
  {Sembach}, {Shanahan}, {Shaughnessy}, {Shaw}, {Shawger}, {Shay}, {Sheehan},
  {Shen}, {Sherman}, {Shiao}, {Shih}, {Shivaei}, {Sienkiewicz}, {Sing},
  {Sirianni}, {Sivaramakrishnan}, {Skipper}, {Sloan}, {Slocum}, {Slowinski},
  {Smith}, {Smith}, {Smith}, {Smith}, {Snyder}, {Soh}, {Sohn}, {Soto},
  {Spencer}, {Stallcup}, {Stansberry}, {Starr}, {Starr}, {Stewart},
  {Stiavelli}, {Straughn}, {Strickland}, {Stys}, {Summers}, {Sun}, {Sunnquist},
  {Swade}, {Swam}, {Swaters}, {Swoish}, {Taylor}, {Taylor}, {Te Plate}, {Tea},
  {Teague}, {Telfer}, {Temim}, {Thatte}, {Thompson}, {Thompson}, {Thomson},
  {Tikkanen}, {Tippet}, {Todd}, {Toolan}, {Tran}, {Trejo}, {Truong},
  {Tsukamoto}, {Tustain}, {Tyra}, {Ubeda}, {Underwood}, {Uzzo}, {Van Campen},
  {Vandal}, {Vandenbussche}, {Vila}, {Volk}, {Wahlgren}, {Waldman}, {Walker},
  {Wander}, {Warfield}, {Warner}, {Wasiak}, {Watkins}, {Weaver}, {Weilert},
  {Weiser}, {Weiss}, {Weissman}, {Welty}, {West}, {Wheate}, {Wheatley},
  {Wheeler}, {White}, {Whiteaker}, {Whitehouse}, {Whiteleather}, {Whitman},
  {Williams}, {Willmer}, {Willoughby}, {Wilson}, {Wirth}, {Wislowski}, {Wolf},
  {Wolfe}, {Wolff}, {Workman}, {Wright}, {Wu}, {Wu}, {Wymer}, {Yates},
  {Yeager}, {Yeates}, {Yerger}, {Yoon}, {Young}, {Yu}, {Zak}, {Zeidler},
  {Zhou}, {Zielinski}, {Zincke}, \& {Zonak}}]{rigby2023}
{Rigby}, J., {Perrin}, M., {McElwain}, M., {et~al.} 2023, \pasp, 135, 048001,
  \dodoi{10.1088/1538-3873/acb293}

\bibitem[{{Robotham} {et~al.}(2023){Robotham}, {D'Silva}, {Windhorst},
  {Jansen}, {Summers}, {Driver}, {Wilmer}, \& {Bellstedt}}]{Robotham_2023}
{Robotham}, A.~S.~G., {D'Silva}, J.~C.~J., {Windhorst}, R.~A., {et~al.} 2023,
  \pasp, 135, 085003, \dodoi{10.1088/1538-3873/acea42}

\bibitem[{{Robotham} {et~al.}(2017){Robotham}, {Taranu}, {Tobar}, {Moffett}, \&
  {Driver}}]{Robotham2017}
{Robotham}, A.~S.~G., {Taranu}, D.~S., {Tobar}, R., {Moffett}, A., \& {Driver},
  S.~P. 2017, \mnras, 466, 1513, \dodoi{10.1093/mnras/stw3039}

\bibitem[{{Rodighiero} {et~al.}(2011){Rodighiero}, {Daddi}, {Baronchelli},
  {Cimatti}, {Renzini}, {Aussel}, {Popesso}, {Lutz}, {Andreani}, {Berta},
  {Cava}, {Elbaz}, {Feltre}, {Fontana}, {F{\"o}rster Schreiber},
  {Franceschini}, {Genzel}, {Grazian}, {Gruppioni}, {Ilbert}, {Le Floch},
  {Magdis}, {Magliocchetti}, {Magnelli}, {Maiolino}, {McCracken}, {Nordon},
  {Poglitsch}, {Santini}, {Pozzi}, {Riguccini}, {Tacconi}, {Wuyts}, \&
  {Zamorani}}]{Rodighiero2011}
{Rodighiero}, G., {Daddi}, E., {Baronchelli}, I., {et~al.} 2011, \apjl, 739,
  L40, \dodoi{10.1088/2041-8205/739/2/L40}

\bibitem[{{Rosario} {et~al.}(2013){Rosario}, {Santini}, {Lutz}, {Netzer},
  {Bauer}, {Berta}, {Magnelli}, {Popesso}, {Alexander}, {Brandt}, {Genzel},
  {Maiolino}, {Mullaney}, {Nordon}, {Saintonge}, {Tacconi}, \&
  {Wuyts}}]{Rosario2013}
{Rosario}, D.~J., {Santini}, P., {Lutz}, D., {et~al.} 2013, \apj, 771, 63,
  \dodoi{10.1088/0004-637X/771/1/63}

\bibitem[{Rutkowski {et~al.}(2013)Rutkowski, Hegel, Kim, Tamura, \&
  Windhorst}]{Rutkowski2013}
Rutkowski, M.~J., Hegel, P.~R., Kim, H., Tamura, K., \& Windhorst, R.~A. 2013,
  Astron. J., 146, 11, \dodoi{10.1088/0004-6256/146/1/11}

\bibitem[{{Sampaio} {et~al.}(2023){Sampaio}, {Arag{\'o}n-Salamanca},
  {Merrifield}, {de Carvalho}, {Zhou}, \& {Ferreras}}]{Sampaio2023}
{Sampaio}, V.~M., {Arag{\'o}n-Salamanca}, A., {Merrifield}, M.~R., {et~al.}
  2023, \mnras, 524, 5327, \dodoi{10.1093/mnras/stad2211}

\bibitem[{{Seyfert}(1943)}]{Seyfert1943}
{Seyfert}, C.~K. 1943, \apj, 97, 28, \dodoi{10.1086/144488}

\bibitem[{{Speagle} {et~al.}(2014){Speagle}, {Steinhardt}, {Capak}, \&
  {Silverman}}]{Speagle2014}
{Speagle}, J.~S., {Steinhardt}, C.~L., {Capak}, P.~L., \& {Silverman}, J.~D.
  2014, \apjs, 214, 15, \dodoi{10.1088/0067-0049/214/2/15}

\bibitem[{{Stalevski} {et~al.}(2016){Stalevski}, {Ricci}, {Ueda}, {Lira},
  {Fritz}, \& {Baes}}]{Skirtor2016}
{Stalevski}, M., {Ricci}, C., {Ueda}, Y., {et~al.} 2016, \mnras, 458, 2288,
  \dodoi{10.1093/mnras/stw444}

\bibitem[{{Stern} {et~al.}(2005){Stern}, {Eisenhardt}, {Gorjian}, {Kochanek},
  {Caldwell}, {Eisenstein}, {Brodwin}, {Brown}, {Cool}, {Dey}, {Green},
  {Jannuzi}, {Murray}, {Pahre}, \& {Willner}}]{Stern2005}
{Stern}, D., {Eisenhardt}, P., {Gorjian}, V., {et~al.} 2005, \apj, 631, 163,
  \dodoi{10.1086/432523}

\bibitem[{{Stern} {et~al.}(2012){Stern}, {Assef}, {Benford}, {Blain}, {Cutri},
  {Dey}, {Eisenhardt}, {Griffith}, {Jarrett}, {Lake}, {Masci}, {Petty},
  {Stanford}, {Tsai}, {Wright}, {Yan}, {Harrison}, \& {Madsen}}]{Stern2012}
{Stern}, D., {Assef}, R.~J., {Benford}, D.~J., {et~al.} 2012, \apj, 753, 30,
  \dodoi{10.1088/0004-637X/753/1/30}

\bibitem[{{Tabatabaei} {et~al.}(2017){Tabatabaei}, {Schinnerer}, {Krause},
  {Dumas}, {Meidt}, {Damas-Segovia}, {Beck}, {Murphy}, {Mulcahy}, {Groves},
  {Bolatto}, {Dale}, {Galametz}, {Sandstrom}, {Boquien}, {Calzetti},
  {Kennicutt}, {Hunt}, {De Looze}, \& {Pellegrini}}]{Tabatabaei2017}
{Tabatabaei}, F.~S., {Schinnerer}, E., {Krause}, M., {et~al.} 2017, \apj, 836,
  185, \dodoi{10.3847/1538-4357/836/2/185}

\bibitem[{{Trump} {et~al.}(2013){Trump}, {Hsu}, {Fang}, {Faber}, {Koo}, \&
  {Kocevski}}]{Trump2013}
{Trump}, J.~R., {Hsu}, A.~D., {Fang}, J.~J., {et~al.} 2013, \apj, 763, 133,
  \dodoi{10.1088/0004-637X/763/2/133}

\bibitem[{{Wang} {et~al.}(2019){Wang}, {Xu}, {Wang}, {Zhang}, {Zheng}, \&
  {Wei}}]{Wang2019}
{Wang}, J., {Xu}, D.~W., {Wang}, Y., {et~al.} 2019, \apj, 887, 15,
  \dodoi{10.3847/1538-4357/ab4d90}

\bibitem[{{Wilkes} {et~al.}(2002){Wilkes}, {Schmidt}, {Cutri}, {Ghosh},
  {Hines}, {Nelson}, \& {Smith}}]{Wilkes2002}
{Wilkes}, B.~J., {Schmidt}, G.~D., {Cutri}, R.~M., {et~al.} 2002, \apjl, 564,
  L65, \dodoi{10.1086/338908}

\bibitem[{{Willner} {et~al.}(2023){Willner}, {Gim}, {del Carmen Polletta},
  {Cohen}, {Willmer}, {Zhao}, {D'Silva}, {Jansen}, {Koekemoer}, {Summers},
  {Windhorst}, {Coe}, {Conselice}, {Driver}, {Frye}, {Grogin}, {Marshall},
  {Nonino}, {Ortiz}, {Pirzkal}, {Robotham}, {Rutkowski}, {Ryan}, {Tompkins},
  {Yan}, {Hammel}, {Milam}, {Adams}, {Beacom}, {Bhatawdekar}, {Cheng},
  {Civano}, {Cotton}, {Hyun}, {Kikuta}, {Nyland}, {Peters}, {Petric},
  {R{\"o}ttgering}, {Shimwell}, \& {Yun}}]{Willner2023}
{Willner}, S.~P., {Gim}, H.~B., {del Carmen Polletta}, M., {et~al.} 2023, \apj,
  958, 176, \dodoi{10.3847/1538-4357/acfdfb}

\bibitem[{{Windhorst} \& {Cohen}(2010)}]{Cohen2010}
{Windhorst}, R.~A., \& {Cohen}, S.~H. 2010, in American Institute of Physics
  Conference Series, Vol. 1294, First Stars and Galaxies: Challenges for the
  Next Decade, ed. D.~J. {Whalen}, V.~{Bromm}, \& N.~{Yoshida} (AIP), 225--233,
  \dodoi{10.1063/1.3518858}

\bibitem[{{Windhorst} {et~al.}(2023){Windhorst}, {Cohen}, {Jansen}, {Summers},
  {Tompkins}, {Conselice}, {Driver}, {Yan}, {Coe}, {Frye}, {Grogin},
  {Koekemoer}, {Marshall}, {O'Brien}, {Pirzkal}, {Robotham}, {Ryan}, {Willmer},
  {Carleton}, {Diego}, {Keel}, {Porto}, {Redshaw}, {Scheller}, {Wilkins},
  {Willner}, {Zitrin}, {Adams}, {Austin}, {Arendt}, {Beacom}, {Bhatawdekar},
  {Bradley}, {Broadhurst}, {Cheng}, {Civano}, {Dai}, {Dole}, {D'Silva},
  {Duncan}, {Fazio}, {Ferrami}, {Ferreira}, {Finkelstein}, {Furtak}, {Gim},
  {Griffiths}, {Hammel}, {Harrington}, {Hathi}, {Holwerda}, {Honor}, {Huang},
  {Hyun}, {Im}, {Joshi}, {Kamieneski}, {Kelly}, {Larson}, {Li}, {Lim}, {Ma},
  {Maksym}, {Manzoni}, {Meena}, {Milam}, {Nonino}, {Pascale}, {Petric},
  {Pierel}, {del Carmen Polletta}, {R{\"o}ttgering}, {Rutkowski}, {Smail},
  {Straughn}, {Strolger}, {Swirbul}, {Trussler}, {Wang}, {Welch}, {B. Wyithe},
  {Yun}, {Zackrisson}, {Zhang}, \& {Zhao}}]{Windhorst2023}
{Windhorst}, R.~A., {Cohen}, S.~H., {Jansen}, R.~A., {et~al.} 2023, \aj, 165,
  13, \dodoi{10.3847/1538-3881/aca163}

\bibitem[{{Wright} {et~al.}(2010){Wright}, {Eisenhardt}, {Mainzer}, {Ressler},
  {Cutri}, {Jarrett}, {Kirkpatrick}, {Padgett}, {McMillan}, {Skrutskie},
  {Stanford}, {Cohen}, {Walker}, {Mather}, {Leisawitz}, {Gautier}, {McLean},
  {Benford}, {Lonsdale}, {Blain}, {Mendez}, {Irace}, {Duval}, {Liu}, {Royer},
  {Heinrichsen}, {Howard}, {Shannon}, {Kendall}, {Walsh}, {Larsen}, {Cardon},
  {Schick}, {Schwalm}, {Abid}, {Fabinsky}, {Naes}, \& {Tsai}}]{Wright2010}
{Wright}, E.~L., {Eisenhardt}, P. R.~M., {Mainzer}, A.~K., {et~al.} 2010, \aj,
  140, 1868, \dodoi{10.1088/0004-6256/140/6/1868}

\bibitem[{{Yang} {et~al.}(2023){Yang}, {Caputi}, {Papovich}, {Arrabal Haro},
  {Bagley}, {Behroozi}, {Bell}, {Bisigello}, {Buat}, {Burgarella}, {Cheng},
  {Cleri}, {Dav{\'e}}, {Dickinson}, {Elbaz}, {Ferguson}, {Finkelstein},
  {Grogin}, {Hathi}, {Hirschmann}, {Holwerda}, {Huertas-Company}, {Hutchison},
  {Iani}, {Kartaltepe}, {Kirkpatrick}, {Kocevski}, {Koekemoer}, {Kokorev},
  {Larson}, {Lucas}, {P{\'e}rez-Gonz{\'a}lez}, {Rinaldi}, {Shen}, {Trump}, {de
  la Vega}, {Yung}, \& {Zavala}}]{CeersYang2023}
{Yang}, G., {Caputi}, K.~I., {Papovich}, C., {et~al.} 2023, \apjl, 950, L5,
  \dodoi{10.3847/2041-8213/acd639}

\bibitem[{{York} {et~al.}(2000){York}, {Adelman}, {Anderson}, {Anderson},
  {Annis}, {Bahcall}, {Bakken}, {Barkhouser}, {Bastian}, {Berman}, {Boroski},
  {Bracker}, {Briegel}, {Briggs}, {Brinkmann}, {Brunner}, {Burles}, {Carey},
  {Carr}, {Castander}, {Chen}, {Colestock}, {Connolly}, {Crocker}, {Csabai},
  {Czarapata}, {Davis}, {Doi}, {Dombeck}, {Eisenstein}, {Ellman}, {Elms},
  {Evans}, {Fan}, {Federwitz}, {Fiscelli}, {Friedman}, {Frieman}, {Fukugita},
  {Gillespie}, {Gunn}, {Gurbani}, {de Haas}, {Haldeman}, {Harris}, {Hayes},
  {Heckman}, {Hennessy}, {Hindsley}, {Holm}, {Holmgren}, {Huang}, {Hull},
  {Husby}, {Ichikawa}, {Ichikawa}, {Ivezi{\'c}}, {Kent}, {Kim}, {Kinney},
  {Klaene}, {Kleinman}, {Kleinman}, {Knapp}, {Korienek}, {Kron}, {Kunszt},
  {Lamb}, {Lee}, {Leger}, {Limmongkol}, {Lindenmeyer}, {Long}, {Loomis},
  {Loveday}, {Lucinio}, {Lupton}, {MacKinnon}, {Mannery}, {Mantsch}, {Margon},
  {McGehee}, {McKay}, {Meiksin}, {Merelli}, {Monet}, {Munn}, {Narayanan},
  {Nash}, {Neilsen}, {Neswold}, {Newberg}, {Nichol}, {Nicinski}, {Nonino},
  {Okada}, {Okamura}, {Ostriker}, {Owen}, {Pauls}, {Peoples}, {Peterson},
  {Petravick}, {Pier}, {Pope}, {Pordes}, {Prosapio}, {Rechenmacher}, {Quinn},
  {Richards}, {Richmond}, {Rivetta}, {Rockosi}, {Ruthmansdorfer}, {Sandford},
  {Schlegel}, {Schneider}, {Sekiguchi}, {Sergey}, {Shimasaku}, {Siegmund},
  {Smee}, {Smith}, {Snedden}, {Stone}, {Stoughton}, {Strauss}, {Stubbs},
  {SubbaRao}, {Szalay}, {Szapudi}, {Szokoly}, {Thakar}, {Tremonti}, {Tucker},
  {Uomoto}, {Vanden Berk}, {Vogeley}, {Waddell}, {Wang}, {Watanabe},
  {Weinberg}, {Yanny}, {Yasuda}, \& {SDSS Collaboration}}]{York2000}
{York}, D.~G., {Adelman}, J., {Anderson}, John~E., J., {et~al.} 2000, \aj, 120,
  1579, \dodoi{10.1086/301513}

\bibitem[{{Zhao} {et~al.}(2021){Zhao}, {Civano}, {Fornasini}, {Alexander},
  {Cappelluti}, {Chen}, {Cohen}, {Elvis}, {Gandhi}, {Grogin}, {Hickox},
  {Jansen}, {Koekemoer}, {Lanzuisi}, {Maksym}, {Masini}, {Rosario}, {Ward},
  {Willmer}, \& {Windhorst}}]{Zhao2021}
{Zhao}, X., {Civano}, F., {Fornasini}, F.~M., {et~al.} 2021, \mnras, 508, 5176,
  \dodoi{10.1093/mnras/stab2885}

\end{thebibliography}


\appendix

\vspace*{-1.50cm}

\startlongtable
\begin{deluxetable*}{ccccccccccccccccCc}
\tablewidth{0pt} 
\tabletypesize{\ssmall}
\tablecaption{Catalog of  morphologically identified CPGs in the \JWST\ NEP--TDF\null.}
\tablenum{1}
\tablehead{\colhead{ID} & \colhead{R.A.} & \colhead{Decl.} & \colhead{F275W} & \colhead{F435W} & \colhead{F606W} & \colhead{F090W} & \colhead{F115W} & \colhead{F150W} & \colhead{F200W} & \colhead{F277W} & \colhead{F356W} & \colhead{F410M} & \colhead{F444W} & \colhead{$z$} & \colhead{$f_{\rm AGN}$} & \colhead{color} & \colhead{H23}\\ \colhead{} & \colhead{J2000} & \colhead{J2000} & \multicolumn{11}{c}{----------------------------------------------------------------------AB mag--------------------------------------------------------------------} & \colhead{} & \colhead{} & \colhead{AB mag} & \colhead{} \\
\colhead{(1)} & \colhead{(2)} & \colhead{(3)} & \colhead{(4)} & \colhead{(5)} & \colhead{(6)} & \colhead{(7)} & \colhead{(8)} & \colhead{(9)} & \colhead{(10)} & \colhead{(11)} & \colhead{(12)} & \colhead{(13)} & \colhead{(14)} & \colhead{(15)} & \colhead{(16)} & \colhead{(17)} & \colhead{(18)}}
\startdata
1 & 260.671752 & +65.7116883 & \nd & 19.99 & 18.75 & \nd & \nd & \nd & \nd & 16.69 & 17.07 & 17.05 & 17.32 & 0.35 & 0.13 & -0.25 & 194 \\
2 & 260.756207 & +65.7131767 & 27.05 & 24.95 & 24.54 & 23.11 & 22.16 & 21.61 & 21.44 & 20.92 & 20.72 & 20.46 & 20.37 & 1.37 & 0.5 & 0.35 & 314 \\
3 & 260.694024 & +65.7169286 & \nd & 24.02 & 22.12 & 20.74 & 20.14 & 19.63 & 19.27 & 19.15 & 19.66 & 19.78 & 19.78 & 0.36 & 0.07 & -0.12 & \nd \\
4 & 260.751425 & +65.735378 & 23.24 & 21.03 & 19.96 & 19.21 & 18.9 & 18.71 & 18.61 & 19.13 & 19.5 & 19.63 & 19.74 & 0.11 & 0.09 & -0.24 & 308 \\
5 & 260.690857 & +65.7380851 & 25.43 & 25.29 & 25.12 & 22.9 & 21.76 & 21.22 & 20.8 & 20.29 & 20.15 & 20.06 & 20.08 & 1.3 & 0.22 & 0.07 & 218 \\
6 & 260.70125 & +65.7761802 & 25.7 & 25.88 & 24.91 & 22.79 & 22.02 & 21.52 & 21.14 & 20.59 & 20.26 & 19.99 & 19.91 & 1.0 & 0.83 & 0.35 & 232 \\
7 & 260.697887 & +65.7779331 & 25.88 & 26.0 & 25.52 & 24.22 & 22.84 & 21.98 & 21.31 & 20.89 & 20.57 & 20.34 & 20.2 & 1.66 & 0.33 & 0.37 & \nd \\
8 & 260.659521 & +65.7845551 & 25.76 & 26.86 & 23.03 & 20.74 & 20.18 & 19.78 & 19.41 & 19.18 & 19.19 & 19.32 & 19.55 & 0.81 & 0.21 & -0.36 & \nd \\
9 & 260.695583 & +65.7843522 & 24.9 & 25.23 & 24.2 & 23.31 & 22.8 & 22.43 & 21.92 & 21.55 & 21.22 & 21.04 & 20.88 & 0.95 & 0.57 & 0.34 & 223 \\
10 & 260.746297 & +65.7842212 & \nd & 26.62 & 26.31 & 25.07 & 23.94 & 22.74 & 22.13 & 21.58 & 21.11 & 20.94 & 20.82 & 2.04 & 0.51 & 0.29 & 300 \\
11 & 260.719022 & +65.7862604 & 27.13 & 25.89 & 24.95 & 22.53 & 21.59 & 21.15 & 20.81 & 20.48 & 20.34 & 20.28 & 20.36 & 1.17 & 0.35 & -0.02 & \nd \\
12 & 260.773037 & +65.7856466 & 25.74 & 24.97 & 23.01 & 21.27 & 20.79 & 20.45 & 20.14 & 19.98 & 20.16 & 20.52 & 20.6 & 0.51 & 0.07 & -0.44 & \nd \\
13 & 260.641043 & +65.787557 & 25.94 & \nd & 25.79 & 23.04 & 22.02 & 21.45 & 21.02 & 20.65 & 20.45 & 20.46 & 20.52 & 1.16 & 0.34 & -0.07 & \nd \\
14 & 260.53697 & +65.7952973 & 23.06 & 20.8 & 20.8 & 20.97 & 20.8 & 20.73 & 20.53 & 20.58 & 20.27 & 19.95 & 19.78 & 2.02 & 0.84 & 0.49 & \nd \\
15 & 260.733993 & +65.7985085 & \nd & 21.49 & 20.1 & 19.04 & 18.64 & 18.29 & 18.06 & 18.17 & 18.59 & 18.66 & 18.56 & 0.3 & 0.13 & 0.03 & 283 \\
16 & 260.758634 & +65.800809 & \nd & \nd & \nd & 18.11 & 17.84 & 17.67 & 17.52 & 17.94 & 18.06 & 17.95 & 18.11 & 0.22 & 0.1 & -0.05 & 319 \\
17 & 260.639149 & +65.799387 & 24.94 & 24.6 & 22.76 & 20.0 & 19.53 & 19.21 & 18.87 & 18.67 & 18.55 & 18.66 & 18.79 & 1.0 & 0.21 & -0.24 & 141 \\
18 & 260.507214 & +65.799489 & \nd & 27.3 & 24.45 & 22.3 & 21.64 & 21.13 & 20.64 & 20.31 & 20.43 & 20.58 & 20.73 & 0.76 & 0.2 & -0.3 & \nd \\
19 & 260.684968 & +65.7998337 & 27.55 & 26.5 & 24.42 & 22.16 & 21.56 & 21.09 & 20.69 & 20.42 & 20.52 & 20.74 & 20.92 & 0.8 & 0.13 & -0.4 & \nd \\
20 & 260.722957 & +65.8043552 & 23.72 & 23.04 & 20.84 & 19.41 & 18.95 & 18.56 & 18.3 & 18.28 & 18.71 & 18.93 & 18.99 & 0.46 & 0.03 & -0.28 & \nd \\
21 & 260.473516 & +65.8030315 & 26.88 & 26.22 & 23.28 & 21.35 & 20.85 & 20.5 & 20.19 & 19.98 & 20.22 & 20.46 & 20.6 & 0.63 & 0.08 & -0.38 & \nd \\
22 & 260.536763 & +65.8052802 & 26.52 & 26.09 & 24.88 & 22.7 & 21.89 & 21.33 & 20.9 & 20.59 & 20.46 & 20.56 & 20.6 & 0.96 & 0.29 & -0.14 & 64 \\
23 & 260.456228 & +65.8076758 & \nd & 26.97 & 24.91 & 22.62 & 21.55 & 21.08 & 20.7 & 20.35 & 20.26 & 20.24 & 20.29 & 1.12 & 0.34 & -0.03 & \nd \\
24 & 260.524706 & +65.8099822 & 26.19 & 26.23 & 24.75 & 22.34 & 21.23 & 20.74 & 20.36 & 20.0 & 19.89 & 19.82 & 19.89 & 1.2 & 0.33 & 0.0 & \nd \\
25 & 260.483404 & +65.8118176 & \nd & 25.87 & 24.21 & 23.52 & 22.87 & 22.22 & 21.37 & 20.82 & 21.15 & 21.03 & 20.88 & 1.2 & 0.31 & 0.27 & \nd \\
26 & 260.486571 & +65.8162335 & 24.16 & 23.89 & 22.01 & 20.15 & 19.64 & 19.24 & 18.91 & 18.71 & 18.99 & 19.34 & 19.39 & 0.59 & 0.05 & -0.4 & \nd \\
27 & 260.531751 & +65.8156273 & 25.67 & 25.37 & 24.46 & 22.6 & 21.85 & 21.12 & 20.57 & 20.01 & 19.76 & 19.72 & 19.79 & 1.08 & 0.32 & -0.03 & 63 \\
28 & 260.896667 & +65.8174626 & \nd & 23.96 & 22.34 & 20.54 & 19.84 & 19.54 & 19.11 & 18.67 & 18.49 & 18.33 & 18.22 & 0.61 & 0.71 & 0.27 & 497 \\
29 & 260.905043 & +65.8175627 & \nd & 22.77 & 20.92 & \nd & \nd & \nd & \nd & 18.72 & 19.31 & 19.45 & 19.5 & 0.36 & 0.04 & -0.19 & \nd \\
30 & 260.81488 & +65.821168 & \nd & 25.38 & 26.26 & 24.78 & 23.62 & 22.99 & 22.6 & 22.18 & 22.04 & 21.96 & 21.87 & 1.58 & 0.32 & 0.17 & \nd \\
31 & 260.851755 & +65.8249769 & 24.8 & 24.36 & 21.99 & 20.05 & 19.6 & 19.27 & 18.96 & 18.74 & 18.9 & 19.16 & 19.31 & 0.72 & 0.1 & -0.41 & \nd \\
32 & 260.898138 & +65.8218193 & \nd & 27.19 & 26.36 & 23.92 & 22.53 & 21.99 & 21.69 & 21.33 & 21.25 & 21.14 & 21.1 & 1.47 & 0.36 & 0.15 & \nd \\
33 & 260.423208 & +65.8239109 & 26.69 & 25.69 & 24.67 & 22.44 & 21.63 & 21.18 & 20.82 & 20.44 & 20.33 & 20.25 & 20.32 & 1.07 & 0.37 & 0.01 & \nd \\
34 & 260.768573 & +65.8261999 & 28.36 & 26.89 & 25.35 & 23.12 & 21.65 & 21.05 & 20.73 & 20.37 & 20.18 & 20.08 & 20.06 & 1.51 & 0.38 & 0.12 & \nd \\
35 & 260.538451 & +65.8275573 & \nd & 24.48 & 23.78 & 22.05 & 21.51 & 21.19 & 20.85 & 20.53 & 20.42 & 20.44 & 20.49 & 0.96 & 0.29 & -0.07 & \nd \\
36 & 260.942294 & +65.8276848 & 28.26 & 26.06 & 24.22 & 21.87 & 21.19 & 20.73 & 20.34 & 20.03 & 20.09 & 20.21 & 20.39 & 0.86 & 0.19 & -0.3 & \nd \\
37 & 260.924987 & +65.8298372 & 25.49 & 23.17 & 21.07 & 19.69 & 19.22 & 18.86 & 18.62 & 18.74 & 19.31 & 19.45 & 19.49 & 0.34 & 0.02 & -0.18 & \nd \\
38 & 260.919468 & +65.8313406 & 27.59 & 23.49 & 21.48 & 20.38 & 20.03 & 19.72 & 19.52 & 19.56 & 19.92 & 19.88 & 19.75 & 0.33 & 0.25 & 0.17 & 528 \\
39 & 260.96505 & +65.8346423 & 22.1 & 22.11 & 21.87 & 20.87 & 20.34 & 20.03 & 19.68 & 19.46 & 19.6 & 19.72 & 19.74 & 0.92 & 0.23 & -0.14 & \nd \\
40 & 260.661331 & +65.832686 & 25.39 & 27.82 & 26.32 & 23.97 & 22.52 & 21.86 & 21.42 & 21.01 & 20.81 & 20.67 & 20.68 & 1.53 & 0.39 & 0.13 & \nd \\
41 & 260.886193 & +65.8338751 & 26.57 & 24.65 & 23.77 & 22.27 & 21.59 & 21.08 & 20.64 & 20.28 & 20.32 & 20.4 & 20.55 & 0.79 & 0.11 & -0.23 & \nd \\
42 & 260.498867 & +65.8337582 & 23.85 & 23.2 & 23.06 & 22.56 & 22.07 & 21.71 & 21.42 & 21.13 & 21.0 & 20.86 & 20.71 & 1.33 & 0.44 & 0.29 & \nd \\
43 & 260.66199 & +65.8414 & \nd & 24.6 & 23.69 & 21.45 & 20.29 & 19.84 & 19.51 & 19.14 & 18.98 & 18.86 & 18.95 & 1.42 & 0.36 & 0.03 & \nd \\
44 & 260.673587 & +65.837855 & 26.21 & 25.48 & 24.68 & 22.42 & 21.72 & 21.2 & 20.8 & 20.5 & 20.36 & 20.49 & 20.58 & 0.99 & 0.15 & -0.22 & \nd \\
45 & 260.894292 & +65.8416492 & 22.68 & 20.75 & 19.61 & 18.77 & 18.35 & 18.06 & 17.87 & 18.37 & 18.64 & 18.73 & 18.85 & 0.11 & 0.07 & -0.21 & 492 \\
46 & 260.679864 & +65.8446726 & \nd & 23.57 & 22.84 & 21.2 & 20.56 & 20.11 & 19.69 & 19.36 & 19.25 & 19.34 & 19.46 & 0.89 & 0.18 & -0.21 & 205 \\
47 & 260.74563 & +65.8423048 & \nd & 24.82 & 23.48 & 21.31 & 20.67 & 20.19 & 19.69 & 19.31 & 19.2 & 19.26 & 19.44 & 0.85 & 0.2 & -0.24 & 299 \\
48 & 260.63956 & +65.8449596 & 20.66 & 20.08 & 19.97 & 20.17 & 19.76 & 19.42 & 19.25 & 18.9 & 18.51 & 18.25 & 18.15 & 0.37 & 0.8 & 0.36 & 142 \\
49 & 260.626602 & +65.8521721 & 24.96 & 26.08 & 24.25 & 21.51 & 20.75 & 20.23 & 19.82 & 19.52 & 19.39 & 19.52 & 19.61 & 1.03 & 0.24 & -0.22 & 124 \\
50 & 260.756515 & +65.849021 & 26.22 & 25.48 & 24.08 & 21.63 & 21.0 & 20.64 & 20.34 & 20.1 & 19.98 & 20.08 & 20.16 & 1.02 & 0.24 & -0.18 & \nd \\
51 & 260.692874 & +65.8619061 & 23.94 & 21.71 & 20.01 & 18.89 & 18.41 & 17.98 & 17.64 & 17.74 & 18.15 & 18.2 & 18.24 & 0.32 & 0.08 & -0.09 & 222 \\
52 & 260.624274 & +65.8678777 & 24.78 & 23.49 & 22.43 & 20.75 & 20.16 & 19.66 & 19.15 & 18.79 & 18.56 & 18.59 & 18.69 & 0.84 & 0.33 & -0.13 & 121 \\
53 & 260.681038 & +65.8692821 & \nd & 24.39 & 23.35 & 22.22 & 21.58 & 21.01 & 20.43 & 19.94 & 19.5 & 19.19 & 18.97 & 1.88 & 0.72 & 0.53 & 207 \\
54 & 260.701071 & +65.876123 & \nd & 25.98 & 23.78 & 21.86 & 21.33 & 20.93 & 20.62 & 20.39 & 20.61 & 20.94 & 21.05 & 0.56 & 0.07 & -0.44 & \nd \\
55 & 260.73803 & +65.9028757 & \nd & 27.06 & 26.17 & 24.04 & 22.76 & 22.16 & 21.71 & 21.23 & 20.97 & 20.73 & 20.51 & 1.36 & 0.66 & 0.46 & \nd \\
56 & 260.708784 & +65.876954 & 26.69 & \nd & 25.62 & 24.01 & 22.44 & 21.85 & 21.55 & 21.26 & 21.15 & 21.1 & 21.02 & 1.79 & 0.36 & 0.13 & \nd \\
57 & 260.651265 & +65.8935901 & 27.95 & 27.81 & 25.81 & 23.5 & 22.6 & 22.07 & 21.68 & 21.37 & 21.24 & 21.27 & 21.33 & 1.05 & 0.34 & -0.09 & \nd \\
58 & 260.668029 & +65.8778179 & 24.97 & 27.77 & 25.18 & 23.65 & 22.46 & 21.76 & 21.16 & 20.62 & 20.29 & 20.15 & 20.03 & 1.43 & 0.4 & 0.26 & 187 \\
59 & 260.665718 & +65.8772567 & 25.42 & 27.52 & 26.35 & 24.78 & 23.24 & 22.31 & 21.92 & 21.59 & 21.42 & 21.32 & 21.25 & 1.91 & 0.38 & 0.17 & \nd \\
60 & 260.690274 & +65.8996399 & \nd & \nd & \nd & 22.37 & 21.4 & 20.99 & 20.67 & 20.29 & 20.24 & 20.23 & 20.28 & 1.37 & 0.3 & -0.04 & \nd \\
61 & 260.668589 & +65.8807234 & 25.36 & 23.86 & 23.31 & 21.78 & 21.16 & 20.74 & 20.43 & 20.04 & 19.88 & 19.83 & 19.87 & 0.95 & 0.28 & 0.01 & \nd \\
62 & 260.709682 & +65.8805701 & 25.76 & 25.99 & 25.45 & 24.22 & 23.03 & 22.44 & 21.95 & 21.43 & 21.23 & 21.05 & 21.04 & 1.5 & 0.33 & 0.19 & \nd \\
63 & 260.62903 & +65.8741014 & \nd & 25.8 & 25.72 & 24.68 & 23.15 & 22.42 & 21.91 & 21.4 & 21.17 & 21.0 & 21.03 & 1.77 & 0.32 & 0.14 & \nd \\
64 & 260.649498 & +65.8709509 & 26.83 & 24.95 & 24.46 & 22.85 & 21.98 & 21.72 & 21.58 & 21.39 & 21.21 & 21.02 & 20.95 & 1.2 & 0.44 & 0.26 & \nd \\
65 & 260.739964 & +65.9251014 & 25.13 & 24.63 & 23.57 & 21.89 & 21.15 & 20.62 & 20.04 & 19.59 & 19.68 & 19.93 & 19.99 & 0.77 & 0.11 & -0.31 & 292 \\
66 & 260.672783 & +65.9111158 & \nd & \nd & \nd & 24.73 & 23.24 & 22.45 & 21.93 & 21.51 & 21.38 & 21.29 & 21.2 & 1.7 & 0.37 & 0.18 & \nd \\
\enddata
\tabletypesize{\small}
\tablecomments{\small Columns show ID number, J2000 positions, AB mag in three \HST\ filters and eight \JWST/NIRCam filters, photometric redshift ($z$) as measured by \cigale{}, $f_{\rm AGN}$ from \cigale{},  ($m_{\rm F356W}-m_{\rm F444W}$) color, and counterpart ID in the \citet{Hyun2023} (H23) VLA 3 GHz source catalog.
\added{The  \cigale{} fits incorporated a stellar continuum component from \citet{Bruzual2003} models, a dust component from \citet{Dale2014} templates, a \citet{Calzetti2000} dust attenuation law, nebular emission templates from \citet{Inoue2011}, and a clumpy two-phase torus model  from \citet{Skirtor2016}.}}
\label{table:catalog}
\end{deluxetable*}

\startlongtable
\added{\begin{deluxetable*}{cccccccccccccc}
\label{table:galfitTable}
\tablecaption{Catalog of \galfit{} output parameters for two-component fitting of our visual sample.
}
\tablenum{2}
\tablehead{\colhead{} & \multicolumn{7}{c}{-------------------------- S\'ersic + S\'ersic -------------------------} & \multicolumn{5}{c}{-------------------- S\'ersic + PSF ----------------} & \colhead{} \\ \colhead{ID} & \colhead{mag} & \colhead{R$_e$} & \colhead{n} & \colhead{mag} & \colhead{R$_e$} & \colhead{n} & \colhead{$\chi^2_{\nu}$} & \colhead{mag} & \colhead{R$_e$} & \colhead{n} & \colhead{mag} & \colhead{$\chi^2_{\nu}$} & \colhead{Core Type} \\ \multicolumn{1}{c}{$\quad$} & \multicolumn{3}{c}{---------S\'ersic 1---------} & \multicolumn{3}{c}{---------S\'ersic 2---------} & \multicolumn{1}{c}{$\quad$}
  & \multicolumn{3}{c}{---------S\'ersic 1---------} & \multicolumn{1}{c}{PSF 1} \\ \colhead{(1)} & \colhead{(2)} & \colhead{(3)} & \colhead{(4)} & \colhead{(5)} & \colhead{(6)} & \colhead{(7)} & \colhead{(8)} & \colhead{(9)} & \colhead{(10)} & \colhead{(11)} & \colhead{(12)} & \colhead{(13)} & \colhead{(14)}}
\startdata
1 & 16.52 & 4.00 & 6.00 & 16.52 & 25.00 & 6.00 & 109.26 & 17.68 & 30.00 & 2.21 & 20.68 & 185.59 & Bulge \\
2 & 21.02 & 4.00 & 1.11 & 21.02 & 4.00 & 6.00 & 1.57 & 20.41 & 4.00 & 2.25 & 22.80 & 1.51 & Point-Source \\
3 & 20.34 & 5.66 & 3.05 & 20.34 & 16.28 & 0.83 & 2.06 & 19.79 & 10.94 & 1.96 & 22.79 & 3.12 & Bulge \\
4 & 20.97 & 4.00 & 3.95 & 20.97 & 16.05 & 3.65 & 1.93 & 19.68 & 11.88 & 4.01 & 22.68 & 2.22 & Undetermined \\
5 & 20.80 & 6.20 & 0.74 & 20.80 & 7.63 & 4.74 & 1.87 & 20.15 & 5.25 & 2.56 & 23.15 & 5.82 & Bulge \\
6 & 20.46 & 4.00 & 2.29 & 20.46 & 4.00 & 1.95 & 3.65 & 19.91 & 4.00 & 1.87 & 22.79 & 3.45 & Point-Source \\
7 & 20.83 & 4.00 & 6.00 & 20.83 & 4.00 & 0.99 & 8.06 & 20.49 & 4.00 & 1.31 & 21.59 & 5.32 & Point-Source \\
8 & 20.19 & 23.88 & 0.77 & 20.19 & 4.00 & 1.04 & 1.72 & 19.53 & 11.04 & 3.07 & 22.53 & 4.77 & Bulge \\
9 & 21.21 & 4.00 & 1.59 & 21.21 & 4.00 & 6.00 & 3.01 & 21.07 & 4.00 & 1.47 & 22.72 & 2.68 & Point-Source \\
10 & 21.36 & 4.00 & 1.29 & 21.36 & 4.00 & 4.71 & 1.53 & 20.83 & 4.00 & 1.86 & 23.33 & 1.31 & Point-Source \\
11 & 20.92 & 4.00 & 2.66 & 20.92 & 4.41 & 6.00 & 1.86 & 20.36 & 4.00 & 4.08 & 23.36 & 2.07 & Undetermined \\
12 & 21.77 & 16.92 & 0.70 & 21.77 & 4.05 & 1.76 & 1.47 & 20.59 & 6.71 & 2.81 & 23.59 & 2.28 & Bulge \\
13 & 21.33 & 12.39 & 4.04 & 21.33 & 4.00 & 1.25 & 1.12 & 20.54 & 4.42 & 2.86 & 23.54 & 1.51 & Undetermined \\
14 & 20.49 & 4.00 & 6.00 & 20.49 & 4.00 & 0.89 & 9.08 & 20.34 & 4.00 & 1.09 & 20.72 & 5.27 & Point-Source \\
15 & 18.88 & 24.11 & 1.12 & 18.88 & 4.00 & 0.70 & 20.71 & 18.61 & 18.97 & 2.26 & 21.45 & 25.07 & Bulge \\
16 & 19.00 & 6.61 & 0.70 & 19.00 & 18.44 & 2.37 & 12.37 & 18.36 & 7.21 & 3.43 & 21.36 & 93.77 & Bulge \\
17 & 18.86 & 15.75 & 3.64 & 18.86 & 4.00 & 2.06 & 3.99 & 18.78 & 10.15 & 4.45 & 21.78 & 13.56 & Bulge \\
18 & 22.53 & 4.00 & 0.70 & 22.53 & 10.02 & 1.75 & 1.84 & 20.69 & 8.37 & 2.08 & 23.69 & 2.27 & Undetermined \\
19 & 21.17 & 7.42 & 1.23 & 21.17 & 4.00 & 6.00 & 1.09 & 20.91 & 6.57 & 1.73 & 23.91 & 1.16 & Undetermined \\
20 & 19.64 & 12.88 & 5.65 & 19.64 & 25.00 & 6.00 & 16.59 & 19.16 & 17.39 & 5.44 & 22.16 & 35.62 & Bulge \\
21 & 23.06 & 4.00 & 0.85 & 23.06 & 8.98 & 2.14 & 1.14 & 20.59 & 8.68 & 1.85 & 23.59 & 1.18 & Undetermined \\
22 & 21.53 & 23.52 & 6.00 & 21.53 & 4.25 & 2.54 & 1.75 & 20.62 & 5.11 & 3.05 & 23.62 & 2.23 & Undetermined \\
23 & 21.55 & 15.25 & 1.29 & 21.55 & 4.00 & 2.74 & 2.18 & 20.31 & 5.72 & 2.63 & 23.31 & 2.91 & Bulge \\
24 & 20.87 & 14.47 & 6.00 & 20.87 & 5.66 & 2.08 & 1.61 & 19.92 & 5.94 & 2.69 & 22.92 & 2.96 & Bulge \\
25 & 21.38 & 4.00 & 6.00 & 21.38 & 4.00 & 6.00 & 4.58 & 21.13 & 4.00 & 6.00 & 22.10 & 2.67 & Point-Source \\
26 & 21.42 & 28.29 & 0.70 & 21.42 & 11.88 & 3.56 & 18.16 & 19.53 & 14.50 & 2.52 & 22.53 & 19.77 & Bulge \\
27 & 20.23 & 8.61 & 0.70 & 20.23 & 7.46 & 0.70 & 8.42 & 19.85 & 8.16 & 0.70 & 22.85 & 11.20 & Bulge \\
28 & 18.83 & 4.00 & 3.51 & 18.83 & 4.00 & 6.00 & 55.99 & 18.53 & 4.00 & 4.61 & 19.94 & 47.26 & Point-Source \\
29 & 20.02 & 25.01 & 5.32 & 20.02 & 9.24 & 2.84 & 1.98 & 19.61 & 10.83 & 3.04 & 22.61 & 5.51 & Bulge \\
30 & 22.37 & 30.00 & 6.00 & 22.37 & 4.00 & 6.00 & 3.58 & 21.87 & 4.00 & 6.00 & 24.33 & 1.81 & Point-Source \\
31 & 21.87 & 8.48 & 6.00 & 21.87 & 16.45 & 4.41 & 3.45 & 19.40 & 13.62 & 3.15 & 22.40 & 8.98 & Bulge \\
32 & 21.82 & 4.00 & 6.00 & 21.82 & 4.00 & 1.08 & 1.35 & 21.23 & 4.00 & 1.51 & 22.89 & 1.38 & Undetermined \\
33 & 21.23 & 4.00 & 6.00 & 21.23 & 4.00 & 1.71 & 4.20 & 20.34 & 4.00 & 2.89 & 23.14 & 4.74 & Bulge \\
34 & 20.62 & 8.30 & 1.37 & 20.62 & 4.00 & 1.96 & 1.87 & 20.05 & 6.54 & 1.56 & 23.05 & 2.26 & Undetermined \\
35 & 21.58 & 4.00 & 6.00 & 21.58 & 4.00 & 6.00 & 2.21 & 20.43 & 4.00 & 6.00 & 23.36 & 1.86 & Point-Source \\
36 & 20.90 & 4.09 & 2.18 & 20.90 & 12.93 & 0.70 & 1.38 & 20.40 & 7.94 & 1.62 & 23.40 & 1.98 & Bulge \\
37 & 21.20 & 15.53 & 0.70 & 21.20 & 8.04 & 4.29 & 1.95 & 19.50 & 9.69 & 2.38 & 22.50 & 2.95 & Bulge \\
38 & 20.23 & 6.44 & 6.00 & 20.23 & 4.00 & 6.00 & 3.18 & 19.67 & 5.48 & 6.00 & 22.41 & 3.01 & Point-Source \\
39 & 20.12 & 4.00 & 4.67 & 20.12 & 4.00 & 6.00 & 7.25 & 19.77 & 4.00 & 4.33 & 22.60 & 6.88 & Point-Source \\
40 & 21.17 & 5.26 & 1.88 & 21.17 & 7.63 & 3.53 & 1.06 & 20.69 & 6.11 & 2.07 & 23.69 & 1.21 & Undetermined \\
41 & 20.89 & 5.45 & 1.44 & 20.89 & 25.00 & 6.00 & 5.83 & 20.77 & 4.66 & 2.20 & 23.77 & 8.68 & Bulge \\
42 & 21.48 & 4.00 & 3.76 & 21.48 & 4.00 & 6.00 & 1.62 & 20.68 & 4.00 & 3.95 & 23.20 & 1.37 & Point-Source \\
43 & 19.18 & 30.00 & 3.03 & 19.18 & 4.00 & 0.79 & 4.60 & 19.00 & 10.64 & 5.95 & 22.00 & 18.97 & Bulge \\
44 & 21.43 & 16.03 & 0.70 & 21.43 & 4.00 & 2.40 & 1.23 & 20.53 & 8.94 & 2.26 & 23.49 & 1.75 & Bulge \\
45 & 18.94 & 10.41 & 4.57 & 18.94 & 25.00 & 0.70 & 9.74 & 18.97 & 12.67 & 2.71 & 21.97 & 14.97 & Bulge \\
46 & 20.18 & 5.41 & 2.09 & 20.18 & 25.00 & 6.00 & 17.06 & 19.74 & 7.16 & 2.65 & 22.74 & 18.40 & Bulge \\
47 & 20.02 & 4.00 & 0.72 & 20.02 & 25.00 & 0.70 & 4.56 & 19.51 & 8.61 & 3.02 & 22.51 & 15.58 & Bulge \\
48 & 20.29 & 25.12 & 0.70 & 20.29 & 4.00 & 6.00 & 13.22 & 18.51 & 9.80 & 3.64 & 19.44 & 17.46 & Bulge \\
49 & 19.97 & 30.00 & 6.00 & 19.97 & 7.14 & 2.63 & 3.54 & 19.72 & 9.36 & 3.44 & 22.72 & 6.11 & Bulge \\
50 & 20.31 & 6.99 & 2.59 & 20.31 & 5.79 & 3.55 & 1.26 & 20.16 & 7.40 & 2.16 & 23.16 & 1.53 & Undetermined \\
51 & 20.23 & 5.27 & 0.87 & 20.23 & 25.00 & 2.28 & 31.77 & 18.41 & 19.72 & 2.32 & 21.41 & 60.68 & Bulge \\
52 & 19.69 & 7.67 & 5.57 & 19.69 & 15.23 & 0.70 & 41.44 & 18.87 & 13.86 & 0.72 & 21.04 & 45.58 & Bulge \\
53 & 21.26 & 25.59 & 0.70 & 21.26 & 4.00 & 6.00 & 36.44 & 19.29 & 9.68 & 2.85 & 20.20 & 48.88 & Bulge \\
54 & 21.49 & 4.00 & 2.27 & 21.49 & 4.55 & 4.61 & 0.87 & 21.01 & 4.32 & 2.54 & 24.01 & 1.01 & Undetermined \\
55 & 21.34 & 4.00 & 2.87 & 21.34 & 4.00 & 4.28 & 2.68 & 20.81 & 4.00 & 1.68 & 21.84 & 1.83 & Point-Source \\
56 & 21.93 & 6.44 & 3.60 & 21.93 & 4.00 & 2.13 & 1.04 & 21.02 & 4.37 & 2.05 & 24.02 & 1.08 & Undetermined \\
57 & 22.03 & 6.92 & 6.00 & 22.03 & 4.00 & 1.39 & 1.12 & 21.36 & 4.00 & 3.01 & 24.36 & 1.34 & Undetermined \\
58 & 20.68 & 8.08 & 0.70 & 20.68 & 6.17 & 2.97 & 2.23 & 20.07 & 7.39 & 1.06 & 23.07 & 3.29 & Bulge \\
59 & 21.93 & 4.00 & 6.00 & 21.93 & 4.00 & 0.84 & 1.95 & 21.26 & 4.00 & 2.87 & 23.71 & 2.58 & Bulge \\
60 & 20.36 & 5.24 & 3.39 & 20.36 & 11.66 & 0.70 & 1.23 & 20.27 & 6.74 & 2.34 & 23.27 & 1.38 & Undetermined \\
61 & 20.44 & 15.60 & 0.70 & 20.44 & 4.94 & 1.31 & 7.62 & 19.91 & 10.78 & 1.59 & 22.54 & 14.69 & Bulge \\
62 & 21.81 & 4.00 & 6.00 & 21.81 & 4.00 & 0.89 & 1.06 & 21.03 & 4.00 & 1.58 & 23.98 & 1.27 & Undetermined \\
63 & 21.80 & 9.66 & 6.00 & 21.80 & 4.00 & 2.85 & 1.49 & 21.06 & 4.51 & 3.42 & 24.06 & 1.60 & Undetermined \\
64 & 21.93 & 4.00 & 6.00 & 21.93 & 4.00 & 2.15 & 4.23 & 21.08 & 4.00 & 2.32 & 22.84 & 3.95 & Point-Source \\
65 & 20.17 & 11.32 & 1.78 & 20.17 & 12.22 & 1.73 & 2.94 & 20.02 & 12.41 & 1.28 & 23.02 & 3.55 & Bulge \\
66 & 21.24 & 4.00 & 1.74 & 21.24 & 4.00 & 6.00 & 1.40 & 21.25 & 4.00 & 1.60 & 23.95 & 1.35 & Point-Source \\
\enddata
\tablecomments{Columns show ID number, AB mag, half-light radius, S\'ersic index, and $\chi^2_{\nu}$ for respective components in two-component fits with \galfit{}. The galaxy core-type classification is the final column, which is determined by which fit ---double-S\'ersic or S\'ersic  + PSF--- has $\chi^2_{\nu}$ closest to 1 .
Constraints on \galfit{} parameters are as follows for each two-component fit: 
S\'ersic+S\'ersic: $\rm mag\pm3$ of $\rm \texttt{MAG\_AUTO}_{F444W}$ (via \sextractor{}), $0.7 \leq n \leq 6.0$, $R_{e} \geq 4$ pixels (=\,0\farcs12). 
S\'ersic+PSF: $\rm mag_{\hbox{S\'ersic}}\pm3$ of $\rm \texttt{MAG\_AUTO}_{F444W}$ (via \sextractor{}), $\frac{1}{3} \leq \frac{\rm mag_{\hbox{\footnotesize S\'ersic}}}{\rm mag_{PSF}} \leq 3$, $0.7 \leq n \leq 6.0$, $R_{e} \geq 4$ pixels (=\,0\farcs12).}
\end{deluxetable*}}


\end{document}